\documentclass[aps,fleqn,superscriptaddress,groupedaddress,
  nofootinbib,preprintnumbers,nobalancelastpage]{revtex4}
\usepackage[pdfborder={0 0 0}]{hyperref}
\usepackage{pgf}
\usepackage{mciteplus}
\pdfoutput=1
\usepackage{amsmath}
\usepackage{amssymb}
\usepackage{array}
\usepackage{calc}
\usepackage{longtable}
\usepackage{multirow}
\usepackage{lineno}
\usepackage{pstricks}
\usepackage{graphicx}
\graphicspath{{fhc_graphics/}}
\DeclareGraphicsExtensions{.pdf,.png,.jpg}
\usepackage{xspace}
\usepackage{siunitx}
\usepackage[caption=false]{subfig}

\numberwithin{equation}{section}
\usepackage{calc}
\newcolumntype{z}{>{\raggedleft\arraybackslash}m{.1085\textwidth}}
\newcolumntype{n}{>{\raggedright\arraybackslash}m{.06\textwidth}}

\numberwithin{equation}{section}

\newcommand{\MCatNLO}{M\protect\scalebox{0.8}{C}@N\protect\scalebox{0.8}{LO}\xspace}

\newcommand{\MENLOPS}{ME\protect\scalebox{0.8}{NLO}PS\xspace}

\newcommand{\Herwig}{H\protect\scalebox{0.8}{ERWIG}\xspace}

\newcommand{\Blackhat}{B\protect\scalebox{0.8}{LACK}H\protect\scalebox{0.8}{AT}\xspace}
\newcommand{\Pythia}{P\protect\scalebox{0.8}{YTHIA}\xspace}

\newcommand{\Rivet}{R\protect\scalebox{0.8}{IVET}\xspace}

\newcommand{\OpenLoops}{O\protect\scalebox{0.8}{PEN}L\protect\scalebox{0.8}{OOPS}\xspace}

\newcommand{\Collier}{C\protect\scalebox{0.8}{OLLIER}\xspace}
\newcommand{\CutTools}{C\protect\scalebox{0.8}{UT}T\protect\scalebox{0.8}{OOLS}\xspace}

\newcommand{\FastJet}{F\protect\scalebox{0.8}{ASTJET}\xspace}
\newcommand{\Sherpa}{S\protect\scalebox{0.8}{HERPA}\xspace}
\newcommand{\DIRE}{D\protect\scalebox{0.8}{IRE}\xspace}

\newcommand{\Comix}{C\protect\scalebox{0.8}{OMIX}\xspace}

\newcommand{\Amegic}{A\protect\scalebox{0.8}{MEGIC++}\xspace}

\newcommand{\ATLAS}{ATLAS\xspace}
\newcommand{\CMS}{CMS\xspace}
\newcommand{\RunII}{Run II\xspace}

\long\def\symbolfootnote[#1]#2{\begingroup%
\def\thefootnote{\fnsymbol{footnote}}\footnote[#1]{#2}\endgroup}

\newcommand{\order}{\mathcal{O}}

\newcommand{\bea}{\begin{eqnarray}}
\newcommand{\eea}{\end{eqnarray}}
\newcommand{\bi}{\begin{itemize}}
\newcommand{\ei}{\end{itemize}}

\makeatletter
\newcommand\thefontsize[1]{{#1 The current font size is: \f@size pt\par}}
\makeatother

\hypersetup{
  pdfauthor={},
  pdftitle={}
  pdfkeywords={}
}

\begin{document}

\preprint{MCNET-16-11}

\title{Aspects of pQCD at a 100 TeV future hadron collider}

\author{Enrico Bothmann}
\affiliation{II. Physikalisches Institut Universit\"at G\"ottingen, 
  Friedrich-Hund-Platz 1, 37077 G\"ottingen, Germany}
\author{Piero Ferrarese}
\affiliation{II. Physikalisches Institut Universit\"at G\"ottingen, 
  Friedrich-Hund-Platz 1, 37077 G\"ottingen, Germany}
\author{Frank Krauss}
\affiliation{Institute for Particle Physics Phenomenology, 
  Durham University, Durham DH1 3LE, UK}
\author{Silvan Kuttimalai}
\affiliation{Institute for Particle Physics Phenomenology, 
  Durham University, Durham DH1 3LE, UK}
\author{Steffen Schumann}
\affiliation{II. Physikalisches Institut Universit\"at G\"ottingen, 
  Friedrich-Hund-Platz 1, 37077 G\"ottingen, Germany}
\author{Jennifer Thompson}
\affiliation{II. Physikalisches Institut Universit\"at G\"ottingen, 
  Friedrich-Hund-Platz 1, 37077 G\"ottingen, Germany}

\begin{abstract}
  In this publication we consider particle production at a future circular
  hadron collider with 100~TeV centre of mass energy within the Standard
  Model, and in particular their QCD aspects.  Accurate predictions for
  these processes pose severe theoretical challenges related to large
  hierarchies of scales and possible large multiplicities of final state
  particles.  We investigate scaling patterns in multijet-production rates
  allowing to extrapolate predictions to very high final-state multiplicities.
  Furthermore, we consider large-area QCD jets and study the expectation for
  the mean number of subjets to be reconstructed from their constituents and
  confront these with analytical resummed predictions and with the expectation
  for boosted hadronic decays of top-quarks and W-bosons.  We also discuss
  the validity of Higgs-Effective-Field-Theory in making predictions for
  Higgs-boson production in association with jets.  Finally, we consider
  the case of New Physics searches at such a 100 TeV hadron-collider machine
  and discuss the expectations for corresponding Standard-Model background
  processes.
\end{abstract}

\maketitle
\section{Introduction}
The first run at the Large Hadron Collider (LHC-Run I) was a great success.
This is best exemplified with the discovery of the Higgs boson by the \ATLAS
and \CMS collaborations~\cite{Aad:2012tfa,Chatrchyan:2012xdj}.  As a concrete
manifestation of the Brout-Englert-Higgs (BEH) mechanism of electroweak 
symmetry breaking, this boson is responsible for the generation of elementary 
masses~\cite{Englert:1964et,Higgs:1964ia,Higgs:1964pj,Guralnik:1964eu}.  Its
discovery ultimately completed the triumph of the gauge principle as the 
paradigm underlying our highly detailed understanding of all fundamental 
interactions at the particle level apart from gravitation.  It is obvious 
that after this discovery further runs at the LHC will concentrate on further 
studies of this newly found particle.  The determination of its quantum numbers
and couplings to other particles in the Standard Model (SM) through an analysis
of various combinations of production and decay channels will definitely shed
additional light on its true nature.  In particular, a careful analysis of the
pattern of its couplings to other particles will clarify if it is indeed the
Higgs boson in the minimal realisation of the Standard Model or if it opens
the opportunity to study new phenomena beyond it.  Examples for such new
phenomena include:
\begin{itemize}
\item possible new dynamics, in particular involving the electroweak sector of 
  the Standard Model and the symmetry breaking through the Higgs potential.\\
  An integral part of the BEH mechanism is the quartic Higgs potential.  This
  potential most unambigously manifests itself in the self--\-interactions of
  the Higgs boson.  Higgs pair production will allow some initial, rough tests
  of this potential at forthcoming runs of the LHC, and it will start
  constraining potential new operator structures of higher dimension stemming
  from extended sectors at larger scales.  At higher energies, this programme
  can be further extended through more precise tests of the pair production
  or through triple Higgs production processes, which may be susceptible to
  further, additional operators.  In a similar way, multiple gauge boson
  production, especially at high energies, offers ample opportunities to
  study and constrain higher dimensional operators -- a programme which, of
  course, has been pursued for decades now.  
\item unitarity of the Standard Model at the highest energies.\\
  interactions between the Higgs, the gauge and the fermion sector exhibit
  subtle relations in order to guarantee the unitarity of cross sections.  At
  increasingly higher energies these relations can be subjected to increasingly
  stringent tests, in particular in those processes where various final
  states emerge in the fusion of electroweak gauge bosons.  
\item the persistent hierarchy problem and its consequences.  \\
  Especially after the discovery of the Higgs boson with its relatively
  close resemblance to its realisation in the SM, it is staggering
  how large quantum corrections to its mass, which are directly sensitive to
  the ratio of the scale related to a more complete model of nature and the
  actual electroweak symmetry breaking scale, can be absorbed in such a stable
  way.  Many mechanisms have been suggested to cover the apparent void between
  these scales and to stabilise the effective theory that the SM is, with 
  supersymmetry being the most prominent.  Up to now, no realisation of such
  models for New Physics has been discovered, and with only few exceptions
  any hints indicating a discovery have not stood the test of time and have
  disappeared.
\end{itemize}  
Questions like the ones outlined above can be studied by pushing the energy
frontier of hadron colliders.  It is thus not surprising that discussions
already have started concerning such a machine~\cite{Arkani-Hamed:2015vfh}.
As a typical setup, hadronic centre-of-mass energies of 100 TeV in $pp$ 
collisions are assumed with an anticipated luminosity of 1--10~ab$^{-1}$
~\cite{Avetisyan:2013onh,Hinchliffe:2015qma,Arkani-Hamed:2015vfh}.

\begin{figure}
   \begin{center}
      \includegraphics[width=0.85\linewidth]{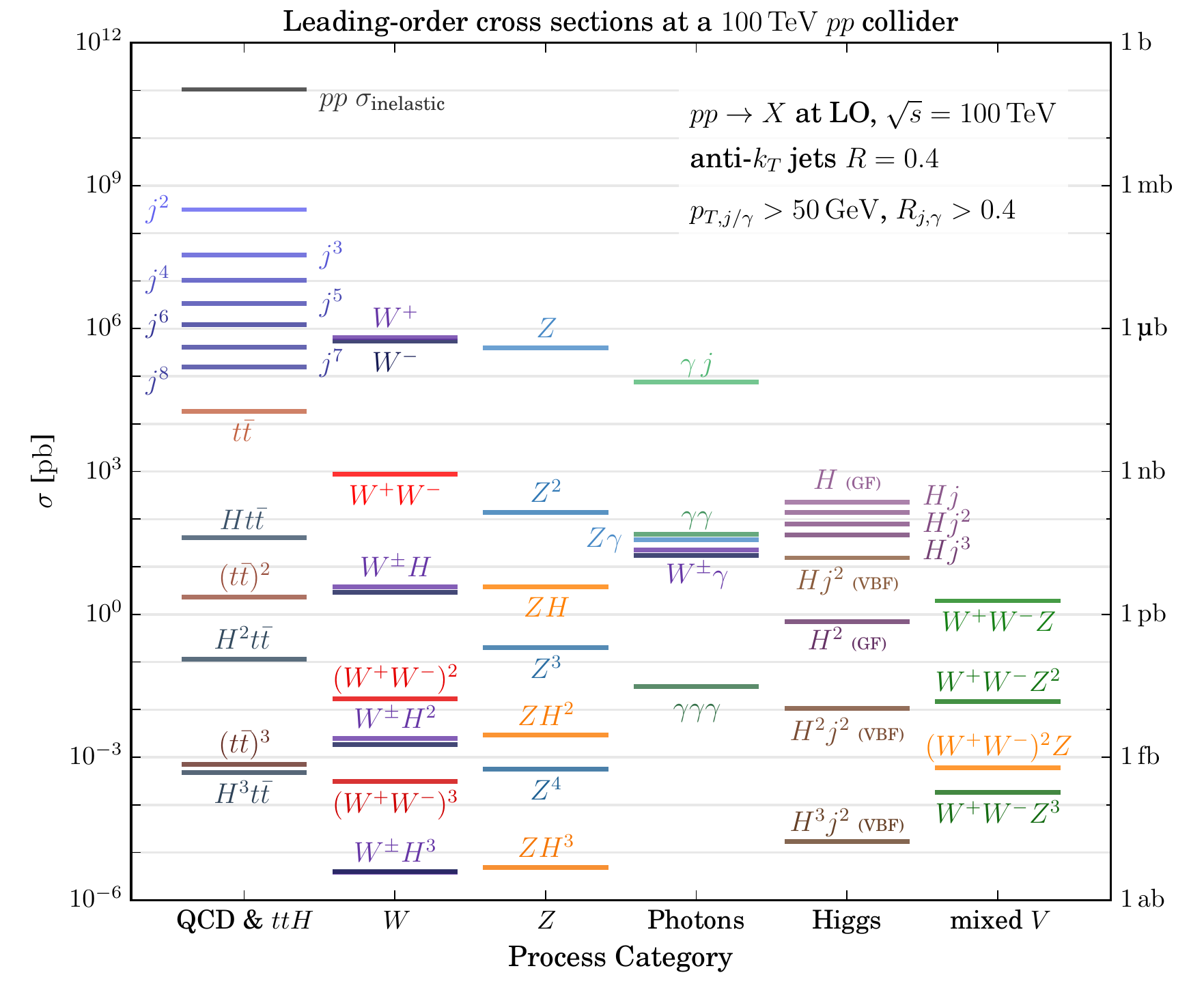}
      \caption{A compilation of LO inclusive production cross sections at the
        100 TeV future $pp$ collider.  The Higgs cross sections labelled
        with ``GF'' refer to Higgs production via gluon fusion, whereas
        ``VBF'' stands for Vector Boson Fusion production. For Gluon Fusion
        Higgs production the top mass effects have been included, see
        Sec.~\ref{sec:loop}}.
   \end{center}
   \label{Fig::XSecs_Summary_Intro}
\end{figure}

In order to gain some intuition about the kind of {\em known} physics that
will be encountered at such a machine, representative cross sections for the
production of relevant high-multiplicity final states are compiled in
Fig.~\ref{Fig::XSecs_Summary_Intro}.
The first thing to notice is the inelastic cross section
at the FCC, being around \SI{105}{mb}~\cite{d'Enterria:2016aur},
which constitutes a \SI{45}{\percent}
raise compared to the LHC ($\sim\SI{72}{mb})$.
To contrast that, we calculated LO cross sections for a multitude
of processes, with cross sections ranging from a few attobarn up to
hundreds of microbarn, across 15 orders of magnitude.
QCD-only processes come with the largest cross-sections
when a jet cut of $p_{T,\text{min}}=\SI{50}{GeV}$ is used,
with dijet production at \SI{315}{\micro\barn}. Also higher jet-multiplicities
have very high cross sections, and only inclusive 7-jet production is less probably than
any other hard process: The inclusive single vector boson production cross
sections come with $\SI{350}{\nano\barn}$--$\SI{600}{\nano\barn}$ and are thus
slightly enhanced compared to 7-jet production.
The least probable cross sections included in Fig.~\ref{Fig::XSecs_Summary_Intro} are those of triple Higgs production in association with a vector boson or
from vector-boson fusion (VBF) with at least two jets.
These cross sections are between 3 and \SI{20}{\atto\barn}. Thus at least
the $H^3j^2$ cross section would still correspond to several events 
at a luminosity of $1$--$10$ $\si{\atto\barn}^{-1}$.

In this paper we want to address some of the challenges of QCD production
processes at a \SI{100}{TeV} hadron collider. 
We first discuss, in Sec.~\ref{sec:scaling}, in some detail the
scaling behaviour in the production of final states involving many jets, using
pure QCD multijet production and vector boson production associated with jets
as typical benchmarks for this part of SM dynamics.  We discuss the scaling
in jet multiplicities between 14 and 100 TeV, and the scaling behaviour in
ratios of multijet cross sections, which differ by one in the jet multiplicity.
The latter is particularly interesting, since increasing collider energies
allow increasingly more hierarchical kinematical situations, which in turn
trigger the transition of the well-known staircase scaling -- also known
as Berends scaling -- to a Poissonian scaling.  The latter is usually 
associated with the onset of practically unconstrained, independent emission
patterns.  We continue our investigations in Sec.~\ref{sec:substructure}
with a related topic, namely the jetty substructure emerging in the production
and decay of heavily boosted unstable particles, such as top quarks or gauge
bosons.  Here, we use analytic results and contrast them with the results
in our simulation.  We change gear in Sec.~\ref{sec:loop}, and quantify finite
mass effects in loop-initiated processes.  Due to the large gluon PDF, such
processes become increasingly important at higher energies, and the large
energies also allow for scales well above the top-mass threshold to be
tested, which in turn leads to jet transverse momentum distributions which
are notably different from expectations driven by effective theories.
Finally, in Sec.~\ref{sec:backgrounds} we discuss the SM backgrounds for
typical signatures used in searches for New Physics.


\subsection{Calculational baseline}

In our studies we use different parts of event simulation tools, which are
available in practically all modern Monte Carlo event 
generators~\cite{Buckley:2011ms}. Their accuracy reflects our detailed 
understanding of the dynamics of the Standard Model, and in particular, 
of QCD.  Their predictions have been confronted with LHC data in many 
studies, for a recent review see Ref.~\cite{Carli:2015qta}.  In particular, 
methods developed to automate the evaluation of QCD one-loop 
corrections~\cite{Ossola:2007ax,vanHameren:2009dr,Hirschi:2011pa,Cullen:2011ac,Cascioli:2011va,Denner:2016kdg},
and the techniques used to interface such exact, fixed--\-order matrix
elements with parton showers~\cite{Catani:2001cc,Frixione:2002ik,Nason:2004rx,Hoeche:2010av,Hoeche:2010kg,Lonnblad:2012ix},
buttress the unprecedented level of accuracy in our simulations.  But of
course, an extrapolation by about an order of magnitude in the centre-of-mass
energy from LHC energies to the 100 TeV scale is plagued with uncertainties
and potential shortcomings:
\begin{itemize}
\item First of all, it is not entirely unlikely that the current PDFs, including
  the gluon, the photon and all quarks up to the bottom quark, must be 
  extended to include also $W$ and $Z$ bosons or even the top--\-quark.  While
  this sounds a bit strange at first, it is worth pointing out that the
  question of whether such objects must be included is synonymous to whether
  partonic scales $\hat{E}$ are probed that yield large logarithms of the
  type $\log(\hat{E}/M)$, where $M$ is the mass of the heavy object.  And,
  similar to the LHC probing the TeV scale and thereby rendering 
  $\log(\SI{1}{TeV}/m_b)$ a possibly large logarithm, a \SI{100}{TeV} 
  machine will probe scales up to around \SI{20}{TeV} and thereby 
  $\log(\SI{20}{TeV}/M)$ with $M\,=\,m_W,\,m_Z,\,m_t$ will be as large.  
\item Vaguely related to this issue is the question concerning the correct 
  factorisation scheme, in particular for processes like jet production with
  jet transverse momenta of 100 GeV or below.  A simple calculation shows that 
  with such processes the PDFs are probed at $x$-ranges of $x\approx 10^{-5}$
  or below, which is typically identified as a kinematical regime where the
  conventional collinear factorisation and its DGLAP scaling might not yield
  correct results, and one would have to resort to the BFKL picture
  or similar. This then would also mean that different factorisation schemes, 
  such as $k_\perp$ factorisation, would have to be invoked. In this 
  publication, however, only conventional collinear factorisation will be used.
\item In a similar way, one could argue that our knowledge of multiple parton
  interactions is fairly limited.  The conventional picture of such multiple 
  scatterings is based on a simple factorisation, which appears to hold
  true at the LHC.
  A similar picture of typically more or less uncorrelated parton--\-parton
  scattering processes is also successfully employed in Monte Carlo simulations 
  programs like \Pythia~\cite{Sjostrand:2014zea}, \Herwig~\cite{Bellm:2015jjp}, 
  or \Sherpa~\cite{Gleisberg:2003xi,Gleisberg:2008ta} to drive what is 
  known as the underlying event. On the other hand, this relatively simple 
  picture cannot be entirely true, and at high energies correlation effects, 
  parton rescattering, the interplay between different scatters etc.\ will 
  become important.
\end{itemize}
These issues render the naive extrapolation of current models a probably too
optimistic procedure, and therefore, in this publication, we concentrate 
on observables expected to be largely insensitive to multiple parton 
scattering.

For the following studies, the \Sherpa event generation framework has been 
used. Proton--proton collisions at centre-of-mass energies of $100$ TeV 
are being considered and, in relevant cases, compared to collisions at LHC \RunII
energies of 14 TeV, to highlight interesting features of energy scaling 
and similar.  If not stated otherwise, jets are reconstructed with the 
anti-$k_T$ algorithm with a radius parameter of $R=0.4$, using the \FastJet 
package~\cite{Cacciari:2008gp,Cacciari:2011ma}.  For matrix element generation
and cross section calculations at leading order \Comix~\cite{Gleisberg:2008fv}
is employed.  The NNPDF3.0 NLO PDF set~\cite{Ball:2014uwa} is used, which also
provides the strong coupling $\alpha_s$.  Renormalisation and factorisation
scales are defined in a process-specific way, they are listed separately in
the respective subsections.  For most distributions, the multijet-merging
technology developed in~\cite{Catani:2001cc,Krauss:2002up,Hoeche:2009rj}
\footnote{
  It is worth noting that other merging techniques exist, like for
  instance those described in Refs.~\cite{Lonnblad:2001iq,Lonnblad:2011xx,
    Lonnblad:2012ng,Mangano:2001xp,Mangano:2006rw,Hamilton:2009ne},
  which however by far and large have been shown to yield comparable results
  at lower energies, see for example Ref.~\cite{Alwall:2007fs}.}
is being employed.  We use the parton shower built on Catani-Seymour subtraction
kernels as proposed in Ref.~\cite{Nagy:2005aa} and implemented in Ref.~\cite{Schumann:2007mg}.
The inclusion of higher-order accuracy in the parton shower simulations is
facilitated by either the \MCatNLO method~\cite{Frixione:2002ik} in the \Sherpa 
version~\cite{Hoeche:2010kg,Hoeche:2011fd,Hoeche:2012ft} or by the multijet 
merging at NLO~\cite{Gehrmann:2012yg,Hoeche:2012yf}\footnote{
  For the matching of NLO matrix elements with parton showers and the
  merging with matrix elements for higher jet multiplicities, other
  methods have been described in Refs.~\cite{Nason:2004rx,Frixione:2007nw,
  Frederix:2012ps,Lonnblad:2012ix}.}.
For the modelling of non-perturbative effects, \Sherpa's in-built
hadronisation~\cite{Winter:2003tt}, underlying event
model~\cite{Alekhin:2005dx}, hadron decays and QED final state
radiation~\cite{Schonherr:2008av} modules have been used.  The Standard Model
input parameters are defined through the $G_\mu$ scheme with
\begin{align}
m_Z &= \SI{91.188}{GeV}\,, &  \Gamma_Z &=\SI{2.49}{GeV}\,, \\
m_W &= \SI{80.419}{GeV}\,, & \Gamma_W &=\SI{2.06}{GeV}\,, \\
m_H &= \SI{125}{GeV}\,, & \Gamma_H &=\SI{0.00407}{GeV}\,, \\
m_t &= \SI{175}{GeV}\,, & \Gamma_t &=\SI{1.5}{GeV}\,, \\
G_\mu &= 1.16639\times 10^{-5}\,{\rm GeV}^{-2}\,,&
\sin^2\theta_W &= 1-{M}^2_W/\tilde{M}^2_Z 
\,.
\end{align}
and the complex-mass scheme~\cite{Denner:1999kn} is used.  Apart from the
top-quark, all other quarks are assumed massless, and top-quark mass effects
are included in the running of $\alpha_s$.

\clearpage
\section{Scaling patterns in Jets and V+Jets production}
\label{sec:scaling}
When considering hadron collisions at very high energies, QCD jet
production processes are omnipresent. Even processes with very 
large multiplicity of (associated) jets exhibit sizeable rates. 
Accurate predictions for such final states pose a severe 
challenge for Monte-Carlo event generators and one might
have to resort to approximate methods. This section focuses on
one such approach, which is based on the scaling behaviour of 
QCD jet rates with respect to jet multiplicity. 

\begin{figure}[hbt]
  \centering
  \includegraphics[]{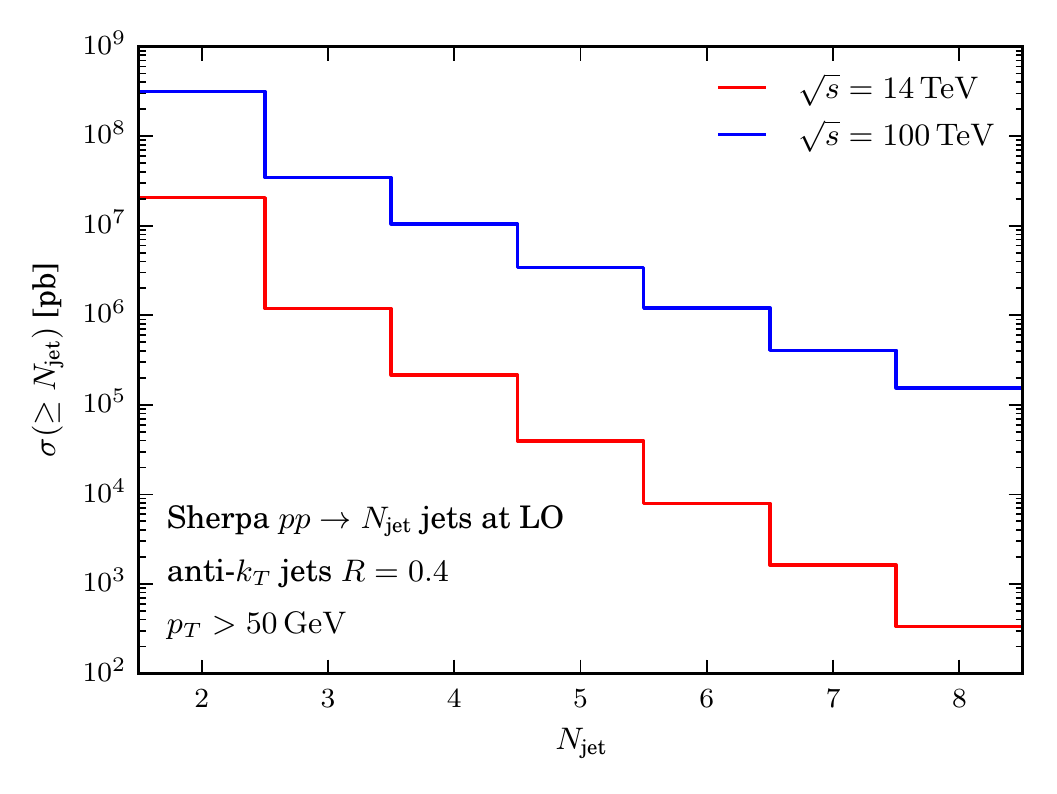}
  \caption{A comparison of LO QCD production rates
  of at least $N_\text{jet}$ anti-$k_T$ jets with $p_T>\SI{50}{GeV}$ and $R=0.4$ 
  in proton-proton collisions at $\sqrt{s}=\SI{14}{TeV}$ and 
  $\sqrt{s}=\SI{100}{TeV}$.}
  \label{figJetRatesLO14vs100}
\end{figure}

To give an impression of jet-production rates at the FCC, 
Fig.~\ref{figJetRatesLO14vs100} compares the leading-order
multijet cross sections at hadronic collision energies of 
$\sqrt{s}=\SI{14}{TeV}$ (LHC) and $\sqrt{s}=\SI{100}{TeV}$ (FCC).
For these estimates, anti-$k_T$ jets with $R=0.4$ and a 
minimal jet transverse momentum of $p_{T,\text{min}}=\SI{50}{GeV}$ 
are considered. While the two-jet
cross section increases by one order of magnitude, an 
increase of more than two orders of magnitude for the 
production of at least eight jets is to be expected. 
In absolute numbers, the total LO dijet rate at the FCC is
of order \SI{300}{\micro\barn} while the LO eight-jet cross
section for jets with transverse momentum above \SI{50}{GeV} 
amounts to \SI{150}{\nano\barn}.

\begin{figure}[hbtp]
  \centering
  \includegraphics[]{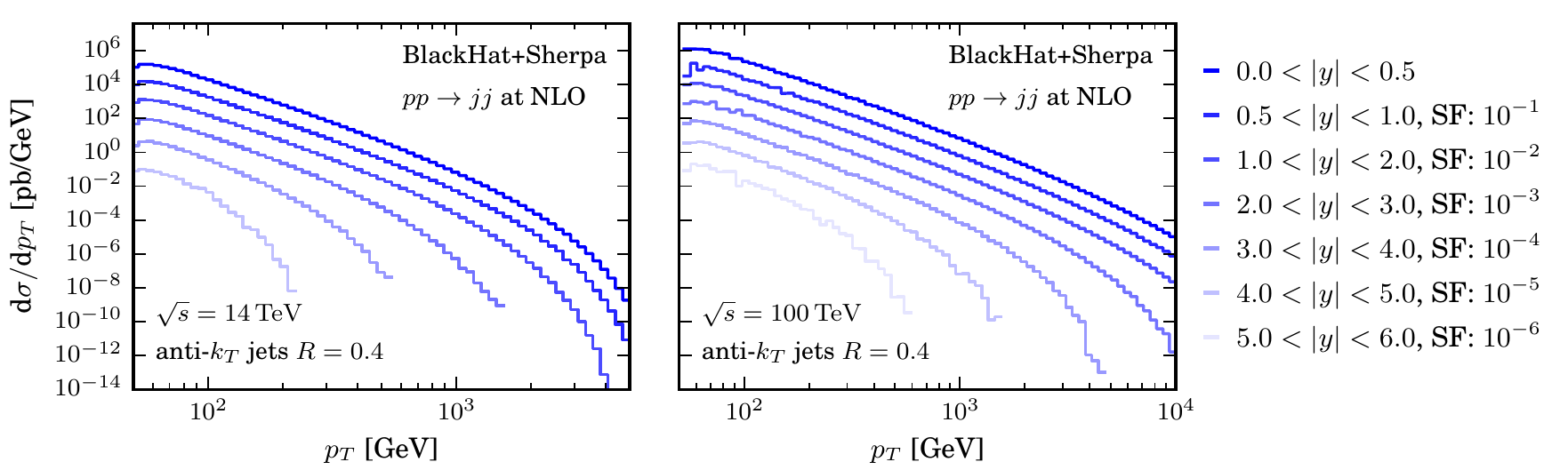}
  \caption{NLO QCD inclusive jet cross sections for LHC (left) and FCC (right)
    collision energies, differential in $p_T$ for different bins in jet rapidity
    $y$.  Note that for illustrative purpose results have been multiplied
    by variable scaling factors (SF), as indicated in the legend.}
  \label{figDiffJetRatesLO14vs100}
\end{figure}

In Fig.~\ref{figDiffJetRatesLO14vs100}, anti-$k_T$ jet rates 
at NLO QCD differential in jet transverse momentum and additionally 
binned in jet rapidity $y$ are presented. Results have been obtained with 
\Blackhat\!\!+\Sherpa~\cite{Berger:2008sj}, both the renormalisation and
factorisation scale have been set to $\mu_R=\mu_F=\frac12 H_T$. 
Comparing rates for $14$ and \SI{100}{TeV} centre-of-mass energy, 
an increase of about one order of magnitude for central jets 
with low and moderate $p_T$ is observed. This effect becomes
 more extreme at larger $p_T$ values, for example at 
$p_T=\SI{3.5}{TeV}$ the FCC rates are more than three orders of 
magnitude larger than at the LHC.
In fact, the FCC provides substantial jet rates even for very large rapidities:
\SI{200}{GeV} jets with $5<|y|<6$ come with rates about two orders of
magnitude larger than those for \SI{200}{GeV} jets in the more central $4<|y|<5$ bin at
the LHC. From these rate estimates, it can be concluded that one can expect
at least ten times more jets at FCC than at LHC.  This massive
enhancement will become even larger in high--$p_T$ and/or high--$|y|$ regions
or when demanding large jet multiplicities. Accordingly, the rapidity 
coverage of general-purpose detectors at the FCC should increase with 
respect to ATLAS or CMS. 

\begin{figure}[hbtp]
  \centering
  \includegraphics[width=.47\textwidth]{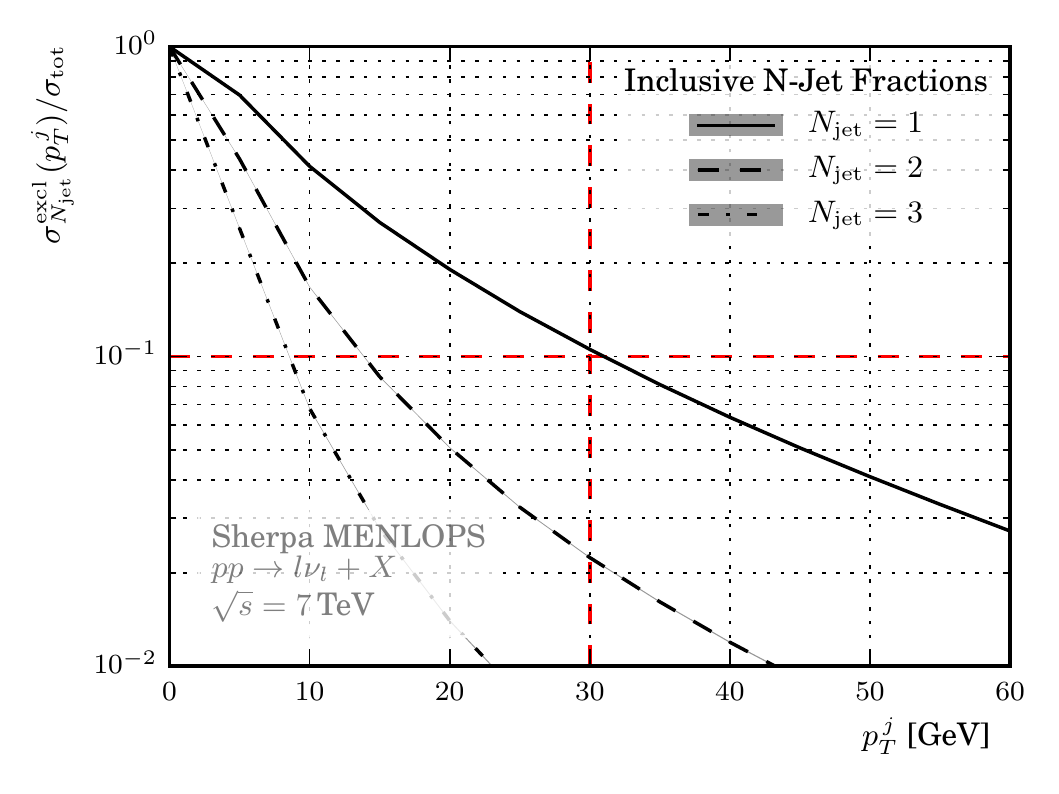}
  \hfill
  \includegraphics[width=.47\textwidth]{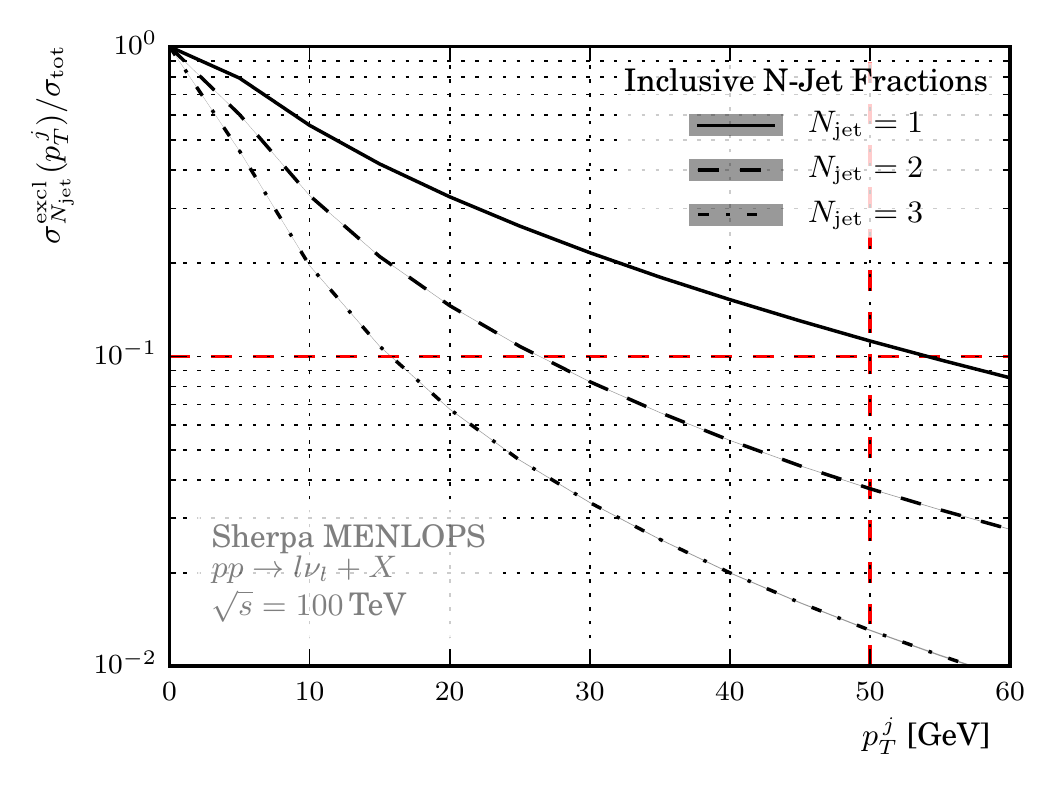}
  \caption{Normalised jet rates for different inclusive jet
    multiplicities differential in jet $p_T$ for $W$+jets production at
    the LHC and the FCC. The dashed red lines indicate the effect of
    cutting at $p_{T,\text{min}}^\text{jet}=\SI{30}{GeV}$, i.e.
    $\SI{50}{GeV}$.}
  \label{figPTComparison}
\end{figure}

So far, identical jet-selection criteria, and in particular $p_T$ thresholds,
have been considered, both for the LHC and the FCC machine.  However, moving
to a \SI{100}{TeV} collider will certainly entail adjustments in the jet-cut
choices.  At LHC in analyses looking for New Physics, typical jet $p_{T,\text{min}}$
cuts often are around 25--\SI{30}{GeV}.  To gain some insight into which range
of $p_{T,\text{min}}$ cuts would lead to a similar relative suppression 
of low $p_T$ jets at the FCC as using 25--\SI{30}{GeV} produces at the LHC, 
one could use a standard candle such as
$W$+jets production.  In Fig.~\ref{figPTComparison}, inclusive jet fractions
in $W+n$-jet production, differential in $p_{T,\text{min}}$, are presented for
the \SI{7}{TeV} LHC and the \SI{100}{TeV} FCC setup.  A jet transverse momentum
cut of $p_{T,\text{min}}=\SI{50}{GeV}$ at the FCC leads to a one-jet inclusive
fraction of order $10\%$, similar to the effect of a
$p_{T,\text{min}}=\SI{30}{GeV}$ cut at the \SI{7}{TeV} LHC.  Accordingly, in
what follows a $p_{T}$ threshold of $\SI{50}{GeV}$ will be considered unless
explicitly stated otherwise.

The QCD jet production rates to be anticipated at the FCC demand suitable
theoretical methods even for very large jet multiplicities.  While fixed-order
predictions for given jet processes are suitable to describe the 
corresponding jet-multiplicity bin, matrix-element parton-shower merging
techniques provide inclusive predictions, differential in the jet multiplicity,
with high jet multiplicities beyond the reach of fixed-order technology being
modelled through the parton shower.  Alternatively, there has recently been
progress in making (semi-)analytical predictions for jet rates at hadron
colliders that account for small jet radii and high jet
counts~\cite{Gerwick:2012fw,Gerwick:2013haa,Dasgupta:2014yra}.  With the
advent of such methods, the morphology of the entire jet-multiplicity
distribution can be studied.  Guided by phenomenological evidence, and
supported by both fixed-order calculations and parton-shower simulations,
certain jet-multiplicity scaling patterns can be
identified~\cite{Gerwick:2012hq} that find their analogue in the analytical
jet-rate predictions~\cite{Gerwick:2012fw,Gerwick:2013haa}.  As already
visible in Fig.~\ref{figJetRatesLO14vs100}, jet rates differential in the
number of jets exhibit a high degree of regularity. To study this feature
one considers the ratio $R_{(n+1)/n}$ of the exclusive $n+1$ over the $n$-jet
cross section, i.e.
\begin{equation}
  R_{(n+1)/n} \equiv \frac{\sigma_{n+1}^\text{excl}}{\sigma_{n}^\text{excl}} \,.
\end{equation}
The approximately equal step size (on a logarithmic scale) between subsequent
exclusive jet rates observed in Fig.~\ref{figJetRatesLO14vs100} translates
into a flat plateau for $R_{(n+1)/n}$, i.e.\ $R_{(n+1)/n} \sim \text{constant}$.
This implies a simple exponential form of the jet-rate distribution, also
known as the Staircase Pattern. A second possibility for the jet rate
distributions is the Poissonian Pattern.  Jet cross sections following 
a simple Poisson statistics result in $R_{(n+1)/n} \sim \bar{n}/(n+1)$, 
with the average number of jets given by $\bar{n}$.

Both these patterns have been observed in LHC 
data~\cite{Aad:2011tqa,Aad:2013ysa,Aad:2014qxa,Khachatryan:2014uva} 
and in Monte-Carlo studies~\cite{Englert:2011cg,Englert:2011pq,Bern:2014fea}. 
They can be understood as the limiting cases for the jet-emission probability: 
for $\alpha_s/\pi \log^2 Q/Q_0 \ll 1$ a Staircase-like pattern is
induced while for $\alpha_s/\pi \log^2 Q/Q_0 \gg 1$ a Poisson-Scaling pattern
is found~\cite{Gerwick:2012fw,Gerwick:2012hq,Gerwick:2014koa}.
Here $Q$ denotes the hard process scale and $Q_0$ is of the order 
of the jet-resolution scale, i.e. $Q_0\sim p_{T,\text{min}}$. The derivation
is based on the language of generating functionals for the jet rates.
The two distinct regimes correspond to additional parton emissions being
distributed either equally among all other partons or stemming predominantly
from a single hard parton line. The latter follows a simple Sudakov 
decay-like model which results in a Poisson distribution, as it is the
case for photon emissions from a hard electron line~\cite{Peskin:1995ev}.
The case of democratic emissions (mainly gluons from gluons), on the
other hand, is exclusive to field theories with a non-abelian group
structure as QCD.

In realistic measurements, jet patterns will be overlaid and cut off by other
effects, such as phase-space constraints. When the available energy for
further jet emission is depleted or jets already radiated cover a good
fraction of the available solid angle~\cite{Gerwick:2014koa}, then higher
multiplicities will quickly tend to zero. On the other hand, the first few
emissions carry away sizeable parts of the total energy available, such that 
the increase in the partonic momentum fractions at which any 
participating PDFs are evaluated is comparably large. This leads 
to somewhat steeper decrease of jet rates for the first few 
emissions and is known as the PDF suppression effect~\cite{Gerwick:2012hq}.
Furthermore, the described regimes represent limiting cases and one
can expect a transition from the Poisson scaling to the Staircase
scaling when the jet evolution begins to wash out any initial large 
scale hierarchy~\cite{Gerwick:2012fw}. When ultimately the energy of 
the partons in the jet evolution becomes of order $Q_0$, one expects 
such patterns to break down giving way to a faster reduction of 
higher multiplicities. 

In view of the enormous phase space available for producing additional jets
at the FCC collider, studies of the jet multiplicity distribution based on
scaling patterns will provide a sensitive handle to estimate and probe the
tails of the distribution, where otherwise one has to largely rely on
parton-shower simulations alone. Based on these predictions,
background subtractions for New Physics signatures resulting from 
decays of new heavy coloured particles yielding a distinct imprint 
on the multiplicity distribution might become 
feasible~\cite{Englert:2011cg,Hedri:2013pvl}.

To study in how far simple jet scaling patterns describe the jet-multiplicity 
distributions at FCC energies, fits of $R_{(n+1)/n}$ in 
Monte-Carlo predictions are considered. For that purpose, Monte-Carlo samples 
for pure jet production and vector-boson production are explored,
triggering scaling patterns using either democratic or hierarchical, i.e.
staggered, jet cuts. Here \emph{democratic} reflects the fact that all 
jet $p_{T,\text{min}}$ are of the same order, i.e. uniform, whereas 
\emph{hierarchical} refers to the scenario where the cut on the leading 
jet, $p_{T,\text{min}}^{\text{leading}}$, is significantly increased. 

\begin{table}[hbtp]
\begin{center}
\begin{ruledtabular}
\begin{tabular}{@{}lrrlll@{}}
label & $p_{T,\text{min}}^\text{leading}\;\text{[GeV]}$ & $p_{T,\text{min}}\;\text{[GeV]}$
& fit function & fit region & fit parameters
\\
\hline \noalign{\smallskip}
S1 (democratic)    & 100   & 50  & $f_\text{Staircase}$ & $3 \leq n \leq 5$ & $c=0.342,\;m=0.006$       \\
S2 (democratic)   & 200   & 100 & $f_\text{Staircase}$ & $3 \leq n \leq 5$ & $c=0.274,\;m=0.003$       \\
P1 (hierarchical)  & 500   & 50  & $f_\text{Poisson}$ & $1 \leq n \leq 5$ & $\bar{n}=2.21,\;c=0.16$ \\
P2 (hierarchical)  & 2000  & 50  & $f_\text{Poisson}$ & $1 \leq n \leq 5$ & $\bar{n}=2.64,\;c=0.25$ \\
\end{tabular}
\end{ruledtabular}
\caption{The jet-cut scenarios considered for pure jet production at FCC
  energies. Furthermore, the fit hypothesis used, cf.\
  Eqs.~(\ref{eq:fitfuncs1}) and (\ref{eq:fitfuncs2}), and the corresponding
  parameters are listed.}
\label{tabPureJetsScenarios}
\end{center}
\end{table}

\begin{figure}[hbtp]
  \centering
  \includegraphics[width=0.49\textwidth]{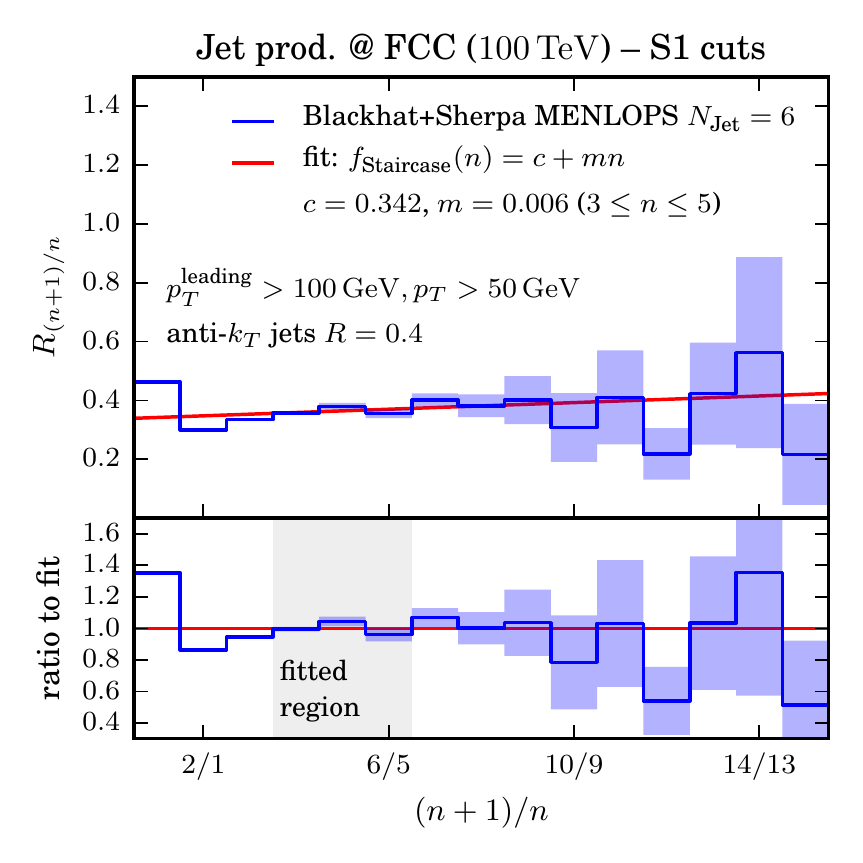}
  \includegraphics[width=0.49\textwidth]{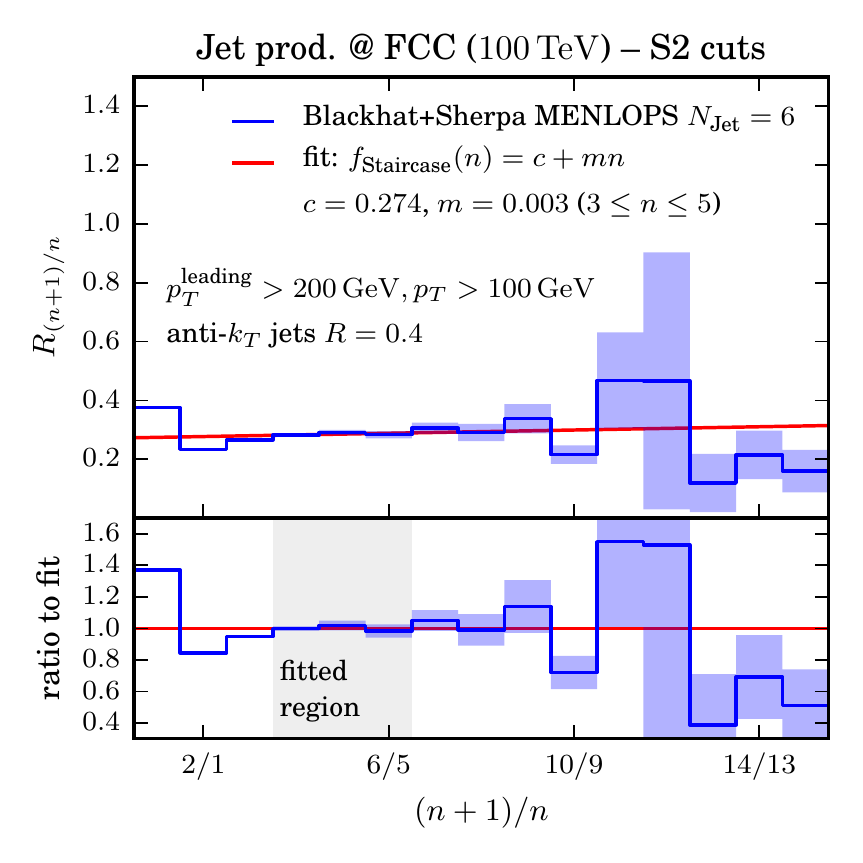}
  \\
  \includegraphics[width=0.49\textwidth]{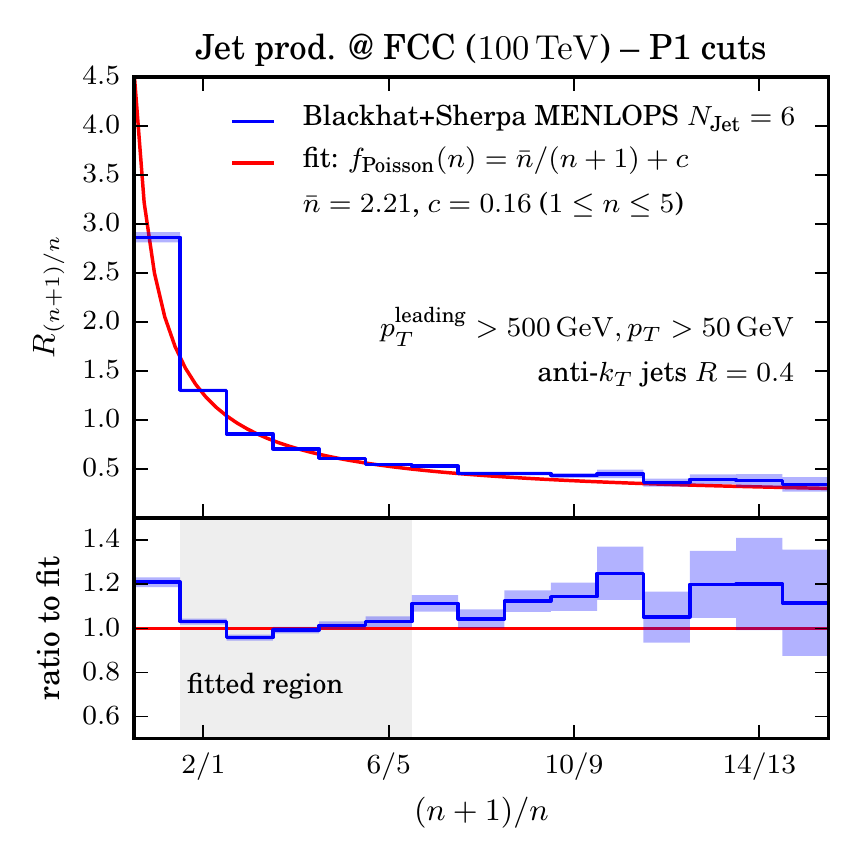}
  \includegraphics[width=0.49\textwidth]{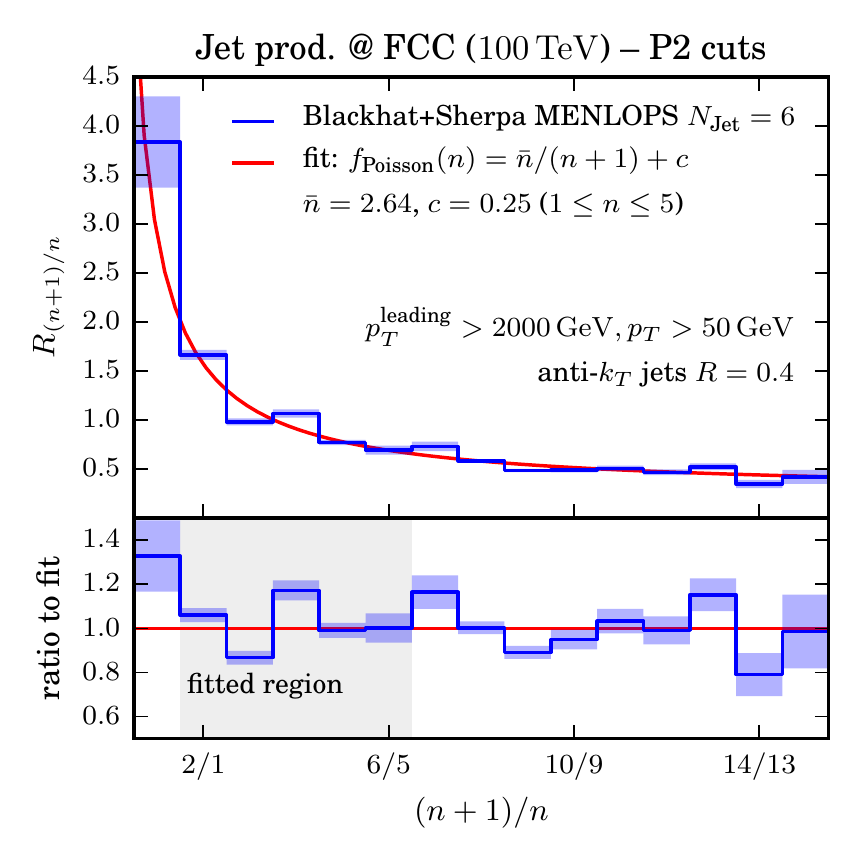}
  \caption{The exclusive jet multiplicity ratio $R_{(n+1)/n}$ in pure jet production at the FCC. 
  Results are presented for the four cut scenarios described in Tab.~\ref{tabPureJetsScenarios}
  with fits  for the Staircase and Poisson patterns,
  cf. Eqs.~(\ref{eq:fitfuncs1}), (\ref{eq:fitfuncs2}).}
  \label{figR}
\end{figure}

The cut scenarios considered for pure jet production are listed in 
Tab.~\ref{tabPureJetsScenarios}. In all case the $2\to 2$ core process
has been considered at \MCatNLO accuracy, furthermore LO matrix elements 
for final-state multiplicities up to six partons are included, all 
consistently merged with the parton shower. In Fig.~\ref{figR} the 
resulting $R_{(n+1)/n}$ distributions are presented for the four 
considered selections. Note, the index $n$ counts the number of 
jets radiated off the hard two-to-two core, i.e. $n=1$ corresponds
to the production of three final-state jets. 

As discussed in Ref.~\cite{Englert:2011pq}, jets assigned to the core process 
behave differently from jets emitted thereof, which is why they have to be
dismissed from pattern fits through the data. Furthermore, PDF effects leave
a non-universal imprint on the first few bins. Therefore, for the 
Staircase-like patterns found for the democratic cut scenarios, cf.\ the two 
upper panels of Fig.~\ref{figR}, the fits are based on the values from 
$R_{4/3}$ through $R_{6/5}$. For the hierarchical cut scenarios, PDF suppression
effects are less prominent, due to hard cuts on the leading jet that
induces a much higher scale $Q$ for the core process. Accordingly, the fits
for the Poisson-like patterns, cf.\ the two lower panels in Fig.~\ref{figR},
are based on $R_{2/1}$ up to $R_{6/5}$.  To quantify the quality of the fits,
terms linear in $n$ for the Staircase pattern and a constant term for the
Poisson pattern have been added to the ideal scaling hypotheses. The resulting
fit functions for the two scenarios read
\begin{align}
  f_\text{Staircase}(n) &= c + m \, n \,,\label{eq:fitfuncs1}\\
  f_\text{Poisson}(n)   &= \frac{\bar{n}}{n+1} + c\,.\label{eq:fitfuncs2}
\end{align}
All resulting fit parameters are listed in Tab.~\ref{tabPureJetsScenarios}.
For all cut scenarios the fit function and its extrapolation to higher jet
bins describe the simulated data very well. For the two democratic scenarios, 
the constant $c$ decreases from 0.35 to 0.29 when we increase the jet cuts, 
reflecting the fact that the {\em costs} for adding an additional jet gets
higher.

Poissonian emission patterns are obtained when hierarchical cuts are applied.
Although the constant offset, $c$, increases from $0.16$ to $0.25$ when 
enlarging the gap between the leading jet cut and the overall jet cut
$p_{T,{\rm min}}$, one can see by eye that the fit quality is better for the
larger cut gap, i.e.\ \SI{2000}{GeV} vs. \SI{50}{GeV}.  For the smaller cut
gap, i.e.\ \SI{500}{GeV} vs.\ \SI{50}{GeV}, the fit increasingly
underestimates $R_{(n+1)/n}$ for growing $n$, which might indicate a faster
transition to a more Staircase-like behaviour.  As expected, the average jet
multiplicity $\bar{n}$ found from the fit increases with a larger leading jet
cut (from $2.2$ to $2.6$). In particular the S2 and P2 cut scenarios are very
well modelled by the simple scaling-pattern hypotheses and allow for reliable
extrapolations where explicit calculations based and fixed order or even
parton-shower simulations become computationally unfeasible. 
\begin{figure}[hbtp]
  \centering
  \includegraphics[width=.49\textwidth]{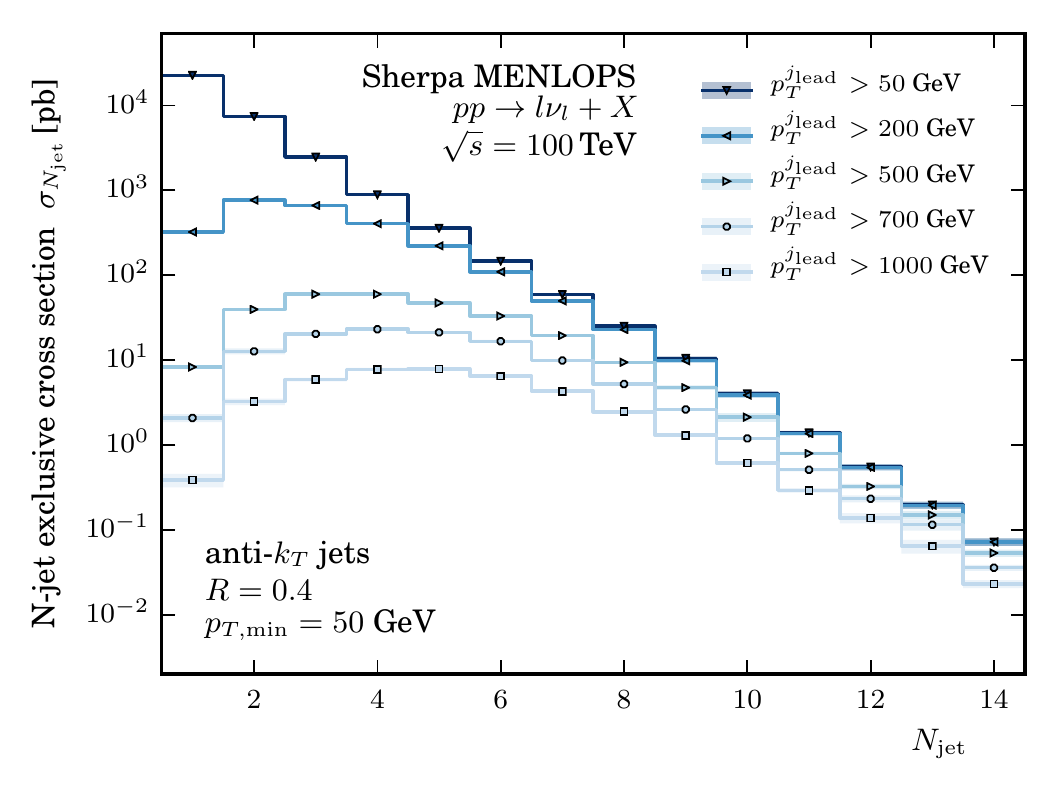}%
  \includegraphics[width=.49\textwidth]{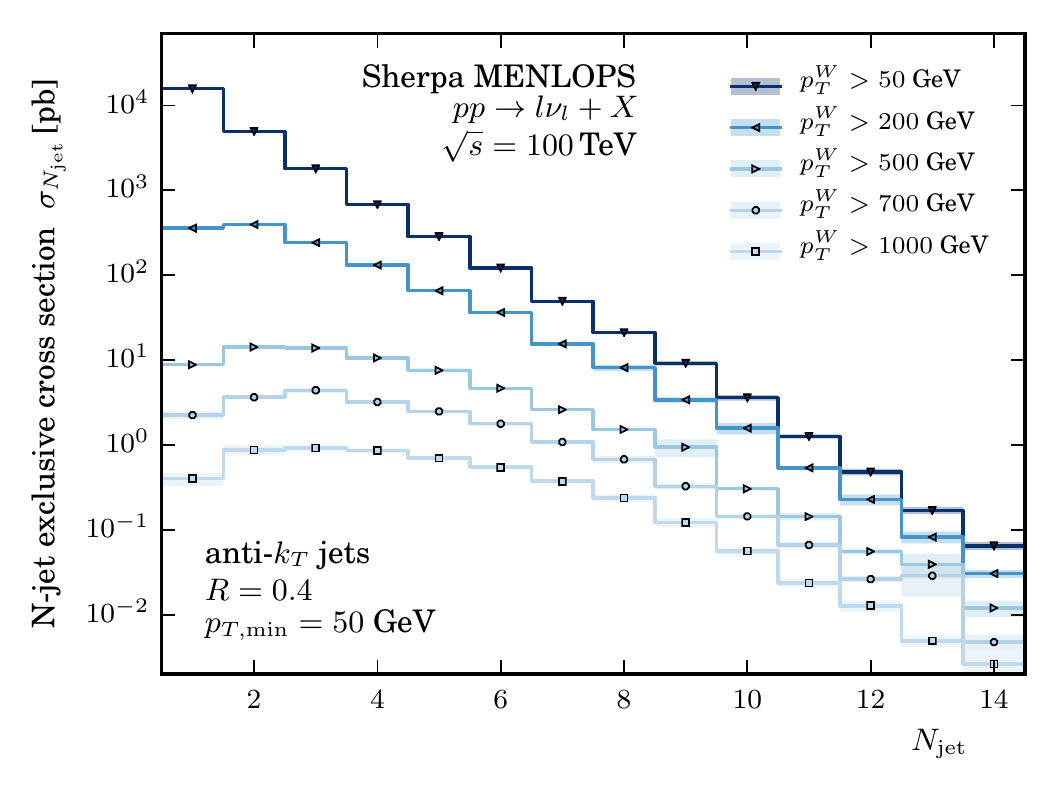}%
  \caption{\label{figWJetsMultiScalingCutChoice}
    Exclusive jet rates for $W$-boson production in association with $n$ jets 
    for different cuts on the leading jet (left panel) and the
    $W$-boson (right panel) transverse momentum, respectively. For all
    (subsequent) jets the universal cut of $p_T>50$ GeV is applied.}
  \end{figure}
  
\begin{figure}[hbtp]
  \centering
    \includegraphics[width=0.49\textwidth]{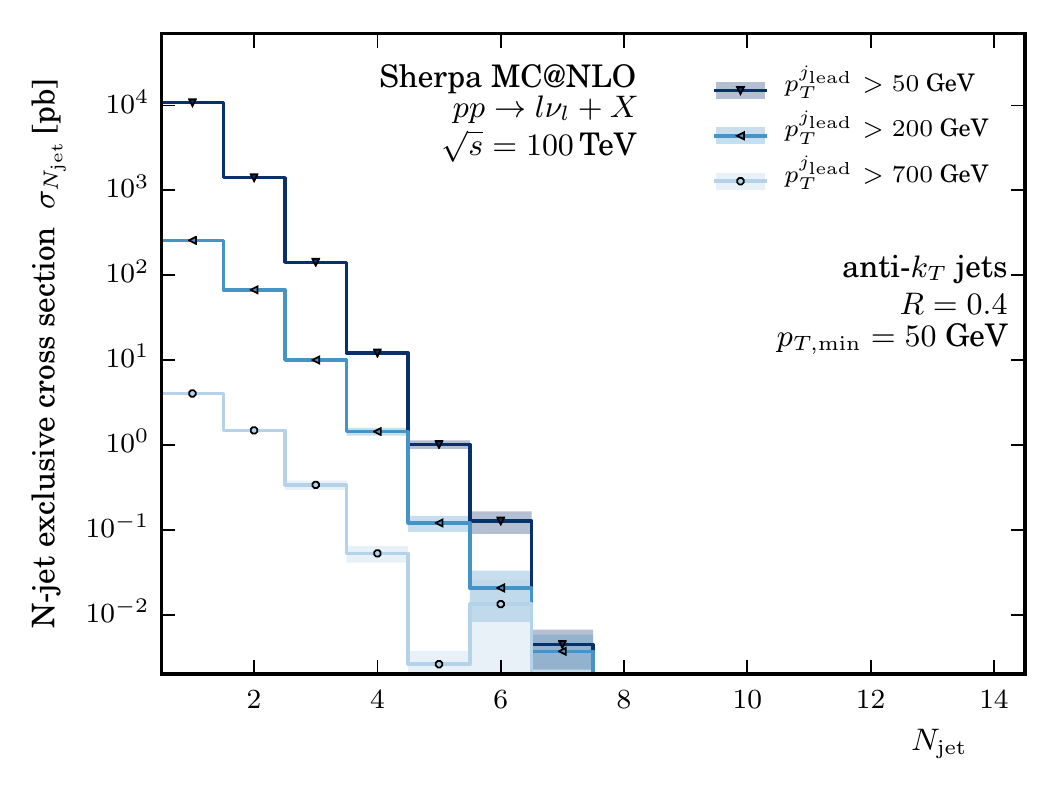}%
    \includegraphics[width=0.49\textwidth]{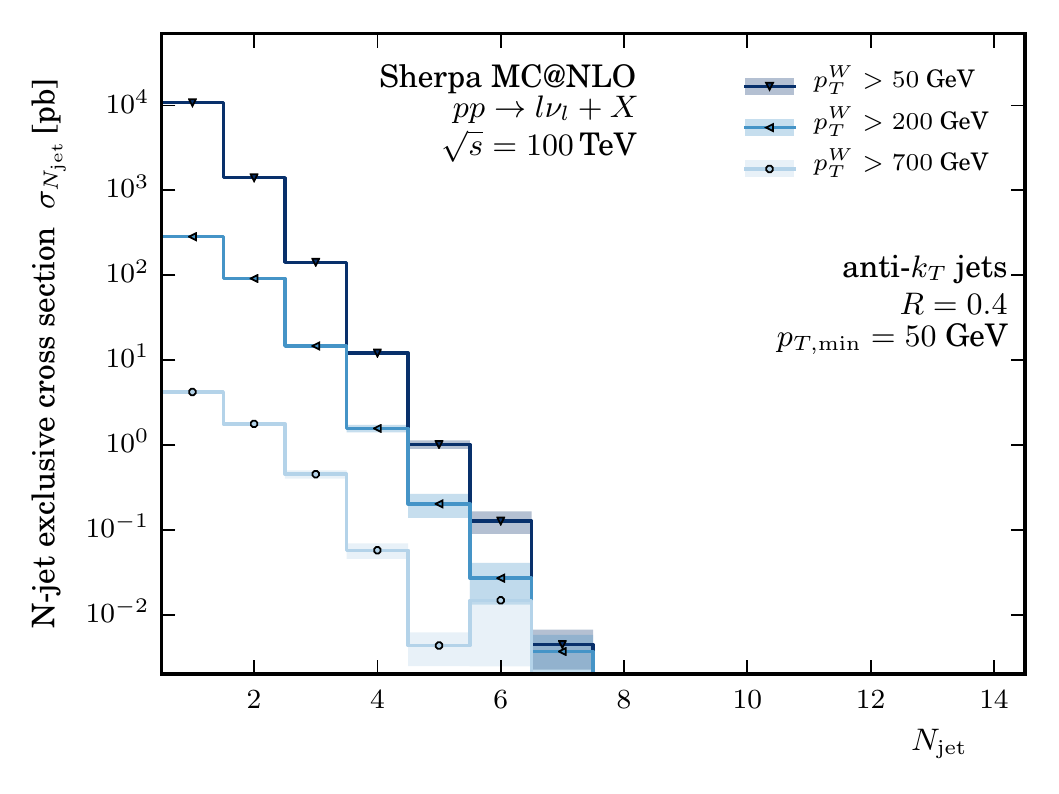}%
    \caption{The same as in Fig.~\ref{figWJetsMultiScalingCutChoice}, but %
      with multijet merging disabled.\label{figWJetsMultiScalingMCatNLO}}
\end{figure}
  
As explained above, jet-multiplicity scaling patterns are a generic feature
of associated jet-production processes. To illustrate this, vector-boson
production, and in particular $W$-boson production, in association with jets
will be considered in the following. Once again, samples based on an
\MCatNLO simulation of $pp\to W$ merged with additional LO matrix elements
for up to five jets dressed with parton showers have been produced.  In
Fig.~\ref{figWJetsMultiScalingCutChoice}, the predictions for exclusive jet
rates imposing a jet cut of $p_{T,{\rm min}}=\SI{50}{GeV}$ and variable cuts on
the leading jet (left panel) or on the $W$ boson (right panel) transverse
momentum are presented. The two cut schemes induce very similar shapes of the
multiplicity distributions but the overall rates are significantly smaller
when demanding the $W$-boson to have large transverse momentum. In fact, a
sizeable part of the $W+$jets cross section originates from hard jets
accompanied by a vector boson with comparatively low transverse
momentum~\cite{Rubin:2010xp}.

For comparison, Fig.~\ref{figWJetsMultiScalingMCatNLO} shows the same
event selections but for a pure \MCatNLO simulation of the inclusive
vector-boson production process, i.e.\ without any additional tree-level
matrix elements taken into account. Noticeably, with the lack of
higher-multiplicity matrix elements the rate estimates for the
high-multiplicity bins are orders of magnitude smaller than in the merged run.
From similar comparisons at LHC energies, it is apparent that the predictions
based on higher-multiplicity matrix elements are more reliable and describe
data much better, see for instance
Refs.~\cite{CMS2012:aa,Aad:2013ysa,Aad:2014qxa,Khachatryan:2014uva}.

\begin{figure}[hbtp]
  \centering
  \includegraphics[width=.6\textwidth]{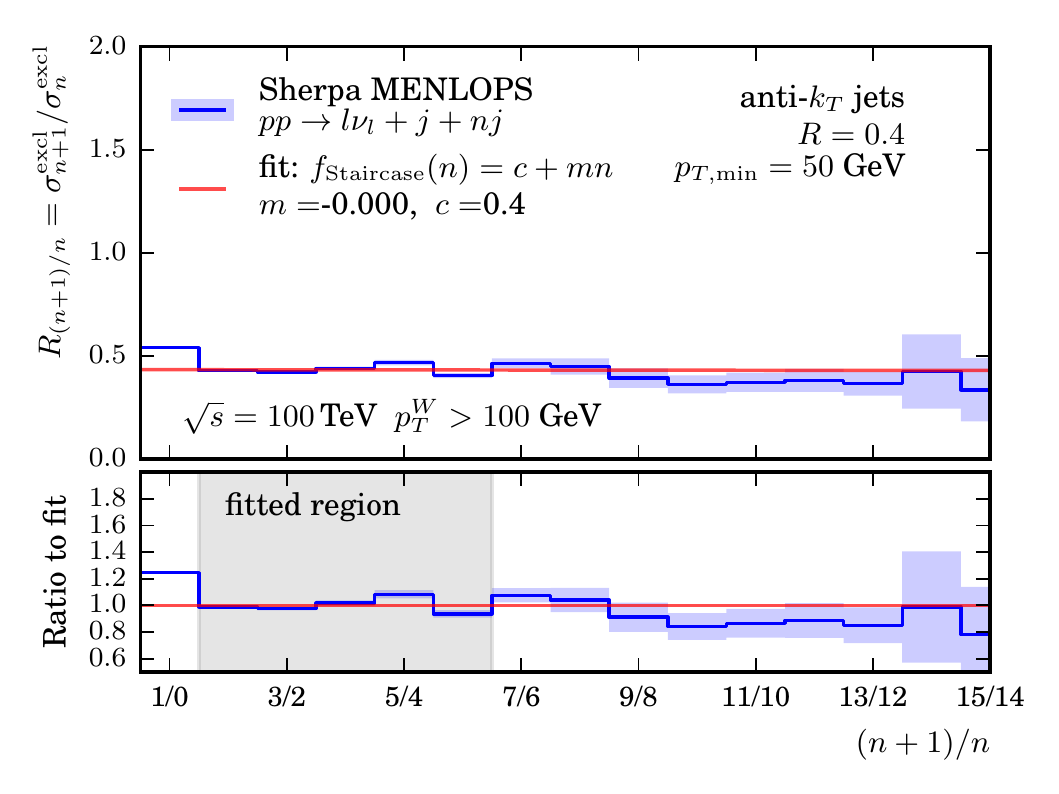}%
  \caption{Exclusive jet multiplicity ratios in $W$ production for a
    democratic jet selection, i.e. applying a universal jet cut of 
    $p_{T,\rm{min}}=\SI{50}{GeV}$ and requiring $p_{T,W}>\SI{100}{GeV}$.
    As the fit function the Staircase hypothesis given in
    Eq.~(\ref{eq:fitfuncs1}) has been used.}
  \label{w_jetratios_democratic}
\end{figure}

\begin{figure}[hbtp]
  \centering
  \includegraphics[width=.49\textwidth]{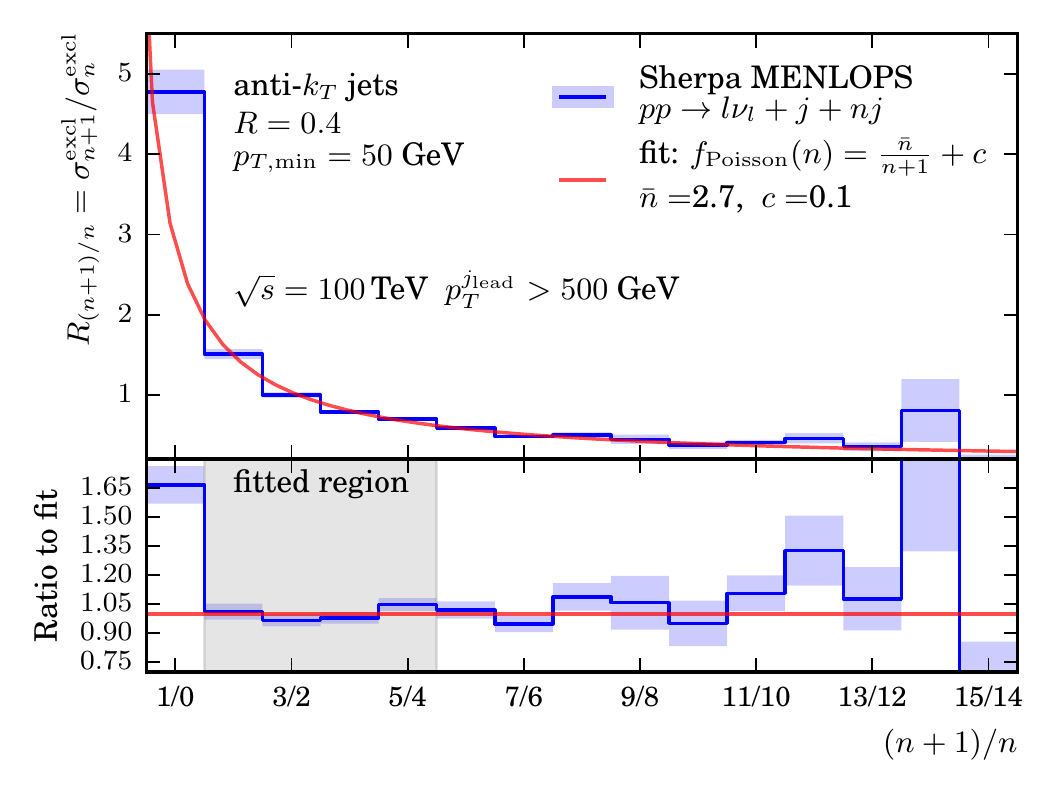}%
  \includegraphics[width=.49\textwidth]{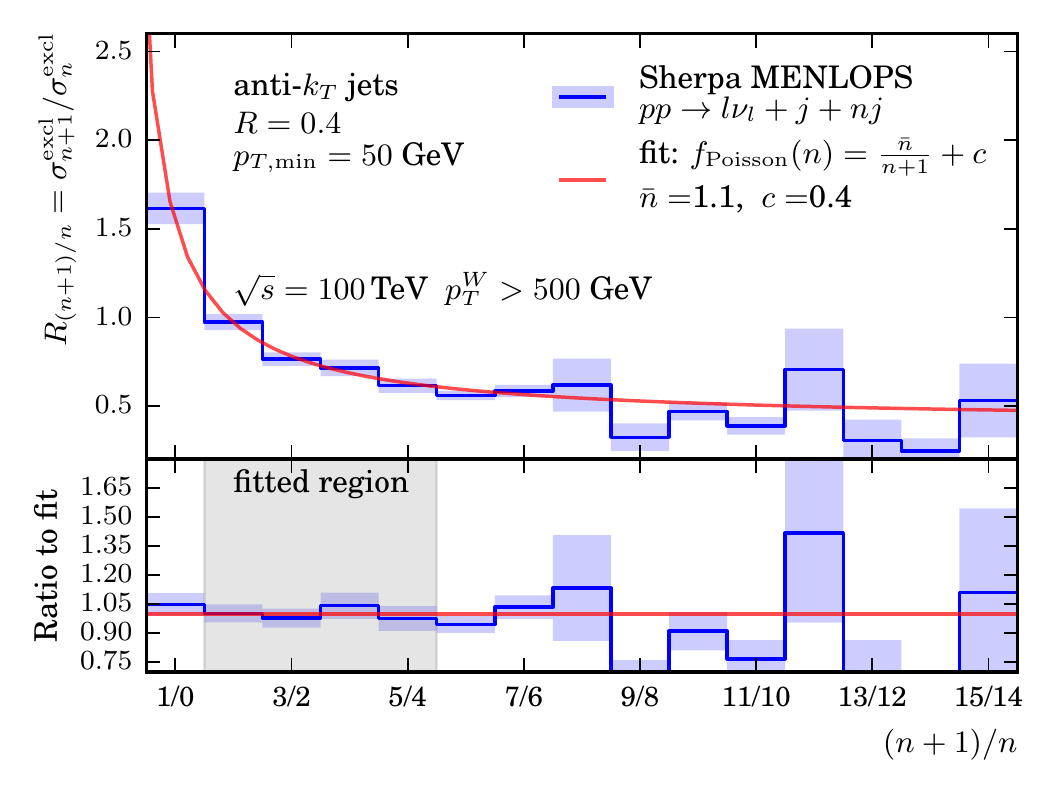}%
  \caption{Exclusive jet multiplicity ratios in $W$ production with
    hierarchical event selection cuts. In the left panel, a cut 
    on the leading jet of $p^{j_{\rm lead}}_T>\SI{500}{GeV}$ is applied, 
    whereas in the right panel, the cut on the $W$ transverse momentum 
    is $p^{W}_T>\SI{500}{GeV}$. As fit function the Poisson hypothesis 
    given in Eq.~(\ref{eq:fitfuncs2}) has been used.}
  \label{w_jetratios_hierarchical}
\end{figure}

In Figs.~\ref{w_jetratios_democratic} and \ref{w_jetratios_hierarchical},  
the exclusive jet multiplicity ratios $R_{(n+1)/n}$ for the multijet-merged
sample described above are plotted alongside with fits following the
functional forms given in Eqs.~\eqref{eq:fitfuncs1},
\eqref{eq:fitfuncs2}.  In this context, the jet multiplicity, $n$, counts
the number of jets in addition to the core process
$pp\rightarrow l\nu_l+j$, i.e.\ $W$~production in association with at least
one jet. In Fig.~\ref{w_jetratios_democratic} results for the democratic
selection scenario, i.e. a universal jet cut of $p_{T,\rm{min}}=\SI{50}{GeV}$,
requiring $p_{T,W}>\SI{100}{GeV}$, are presented. A fit of the Staircase
hypothesis in the range $1\leq n\leq 5$ results in a vanishing parameter $m$.
This presents an ideal Staircase scaling, with a constant ratio of $c=0.4$.
The extrapolation of this scaling function to higher values of $n$ is in
very good agreement with the Monte-Carlo simulation results.  In 
Fig.~\ref{w_jetratios_hierarchical}, the corresponding results for 
hierarchical selection criteria are presented. Two cut scenarios have been
considered, namely $p^{j_{\rm lead}}_{T}>\SI{500}{GeV}$ and
$p^{W}_{T}>\SI{500}{GeV}$ while $p_{T,\rm{min}}=\SI{50}{GeV}$ is still required. 
The results for the fits of the Poisson hypothesis in the range
$1\leq n\leq 4$ illustrate the significantly larger average
jet number $\bar{n}=2.7$ in the first case vs. $\bar{n}=1.1$ in
the latter case. The constant offset parameters $c$ are determined as
$c=0.1$ and $c=0.4$, respectively. The extrapolations of both fits
yield a good description of the simulated data up to very high jet counts.

\begin{figure}[hbtp]
  \centering
  \includegraphics{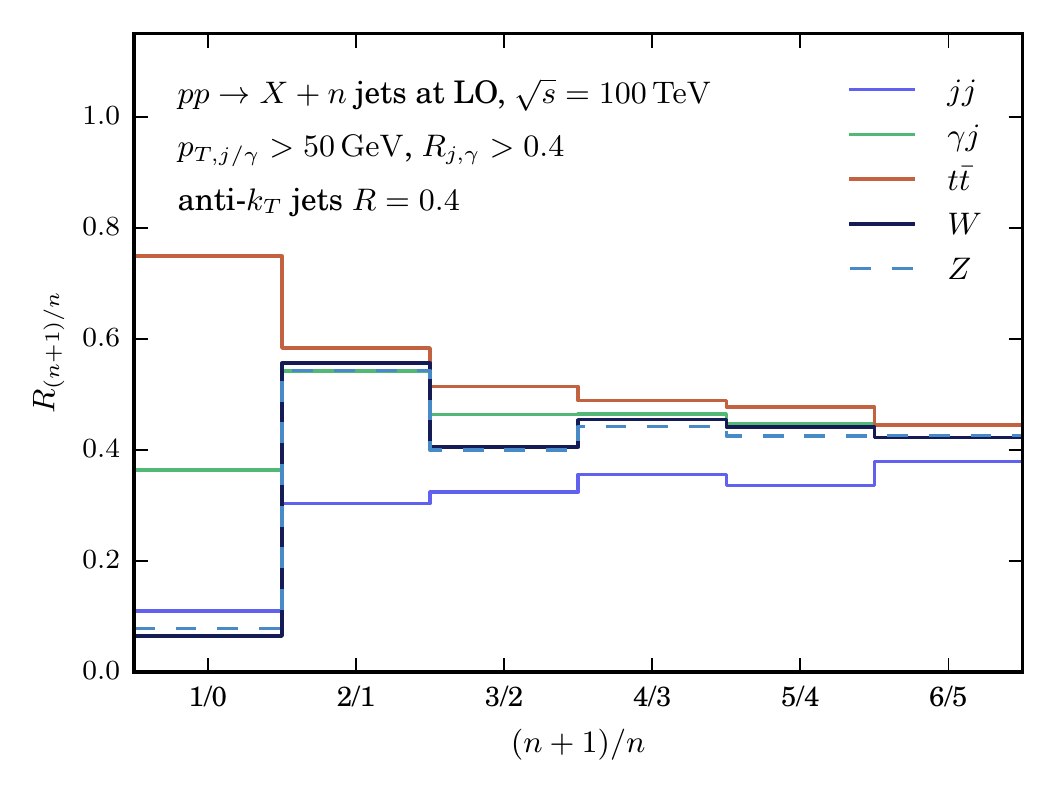}
  \caption{The jet multiplicity ratio $R_{(n+1)/n}$ for several processes
  calculated at LO for each final-state multiplicity. Note, the index 
  $n$ counts jets associated to the core process listed in the legend.
   }
  \label{figLOJetsMultiScaling}
\end{figure}

To illustrate the universality of jet scaling patterns,
Fig.~\ref{figLOJetsMultiScaling} compiles the exclusive jet multiplicity
ratios for a variety of processes, including pure jets, $\gamma+$jets,
$t\bar{t}+$jets and $W/Z+$jets.  The predictions are based on dedicated $n-$jet 
tree-level matrix element calculations, without invoking parton showers.
Democratic jet selection cuts are applied, requiring $p_{T,j}>\SI{50}{GeV}$ in
all processes. In addition, the photon production processes are regulated by
the selection criteria $p_{T,\gamma}>\SI{50}{GeV}$ and $R_{j,\gamma} > 0.4$, with
$R_{j,\gamma}$ the $\eta-\phi$ distance between all jets and the photon.

There are a few remarkable aspects to note here. Apparently, for the pure
jets and $W+$jets processes these LO rate estimates nicely reproduce the
Staircase scaling parameters found in the matrix-element plus parton-shower
samples for the analogous jet-selection cuts, c.f.\ Fig.~\ref{figR} (upper
left panel) and Fig.~\ref{w_jetratios_democratic}.  This is supported by
the fact that for exact Staircase scaling the cross section ratios for
subsequent jet multiplicities are identical for exclusive and inclusive cross 
sections~\cite{Gerwick:2012hq}:
\begin{equation}
\frac{\sigma_{n+1}^\text{excl}}{\sigma_{n}^\text{excl}}= \frac{\sigma_{n+1}^\text{incl}}{\sigma_{n}^\text{incl}}=R={\rm const.}
\end{equation}
Also note that the ratios of the three vector-boson production processes,
$W/Z/\gamma+$jets, are basically the same, illustrating the fact that the
actual gauge-boson mass does not yield a big imprint on the jet-production
probabilities. The production of a pair of top-quarks, however, induces a
large upper scale for subsequent jet emission. Correspondingly, the jet rates
for the first few emissions are sizeable, resulting in ratios $R_{(n+1)/n}>0.5$, 
indicating that a pure leading-order approximation is inappropriate.

To summarise this section: it is possible to fit jet
multiplicities $n$ reasonably well up to values of $n=15$ or even higher,
using results for much lower $n$.  The underlying fits are based on the theoretical hypothesis
of simple scaling patterns, namely Staircase and Poissonian scaling.  These
extrapolations allow meaningful predictions for very high jet-multiplicity
bins that will be populated by a variety of production processes at FCC
energies.  The methods discussed enable the use of techniques that discriminate 
New Physics signals and QCD backgrounds based on the shape of the 
jet-multiplicity distribution.


\clearpage
\section{Jet substructure and boosted topologies at high transverse momentum}
\label{sec:substructure}
The enormous collision energies at a \SI{100}{TeV} hadron-collider machine 
not only allow for the production of final states with a large number of
well isolated QCD jets, but furthermore enable the creation of heavy 
resonances that subsequently decay.  Even for massive particles, the 
fraction of events where these states are produced with large transverse 
boosts can become quite sizeable.  This becomes even more apparent when these
particles originate from the decay of heavy new particles in the multi-TeV
mass range.  Identifying such production processes by reconstructing the decay
products, in particular for hadronic decays, provides a severe challenge for
the detector designs. The separation of very collimated event substructures
requires a fine granularity of the calorimeters to be supplemented by the use
of particle-tracking information, known as particle-flow
techniques~\cite{CMS:2009nxa,Thomson:2009rp}. 
Certainly, the methods currently used at the LHC for tagging such processes, 
e.g.\ for boosted hadronic top-quarks or Higgs bosons, will need to be 
revised and probably overhauled~\cite{Schaetzel:2013vka,Larkoski:2014bia,
        Cohen:2014hxa,Larkoski:2015yqa}. 

The usage of large-area QCD jets, ``fat-jets'', which are assumed to contain
the hadronic decay products of the produced resonance and the majority of the
associated QCD radiation, is prototypical for substructure analyses.  At the
LHC, typical radii for such fat-jets are of the order of $R\approx 1$, but it
is clear that larger boosts -- larger transverse momenta -- will necessitate
smaller radii, usually of the order of $R\approx 2M/p_T$, where $M$ is the mass
of the heavy particle.  Assuming a top quark with a transverse momentum of
around \SI{3.5}{TeV}, originating from a hypothetical \SI{7}{TeV} resonance,
the resulting fat-jet will have a radius of around $R\approx 0.1$ only.  This clearly
poses a considerable challenge for the granularity of future detectors.
However, assuming suitable fat-jets have been identified, specialised tagging
methods are used, which analyse their substructure.   This is achieved through,
for example, re-clustering the large-jet constituents into smaller subjets,
or in terms of jet-shape like measures.  For reviews of the currently
available techniques see 
Refs.~\cite{Abdesselam:2010pt,Plehn:2011tg,Altheimer:2012mn,Altheimer:2013yza}.
Vital for all these approaches is a good theoretical understanding of both
the backgrounds from pure QCD jets and the radiation pattern of the heavy
resonance and its decay products. The complexity of the tagging methods used
often allows for a comparison of the response from different Monte-Carlo
generators only. However, there is a lot of activity to develop predictive
analytical techniques, see 
for instance Refs.~\cite{Dasgupta:2013ihk,Dasgupta:2013via,Larkoski:2014wba,
        Dasgupta:2015lxh,Larkoski:2015kga,Frye:2016aiz}.

In the following the focus will be on some rather coarse feature of large-area 
QCD jets at high transverse momentum, namely the mean number of small-$R$
subjets $\langle n_{\rm subjets}\rangle$ found inside fat-jets. Results will be
finally compared to the corresponding observable for highly-boosted hadronic
decays of top quarks and $W$-bosons. The number of subjets found inside a
larger jet is expected to carry information on the QCD colour charge of the jet
initiating particle. Broadly speaking, at lowest order one expects the scaling
behaviour ${\langle n_{\rm subjets}\rangle \propto C_A}$ for colour octets and
${\langle n_{\rm subjets}\rangle \propto C_F}$ for colour triplets.  Based on such
considerations one can attempt to discriminate gluon from quark
jets~\cite{Gallicchio:2011xq,Bhattacherjee:2015psa}, i.e. assign 
a corresponding likelihood based on the jet-internal QCD activity.  For hadronic 
decays of colour singlets, a reduced and more collimated QCD radiation can be 
expected, resulting in a smaller number of subjets to be found.  Considering the 
physics case of highly-boosted hadronic decays, rather small radii
$R_{\rm subjet}$ need to be considered.  

\begin{figure}[h!]
\includegraphics[width=0.66\textwidth]{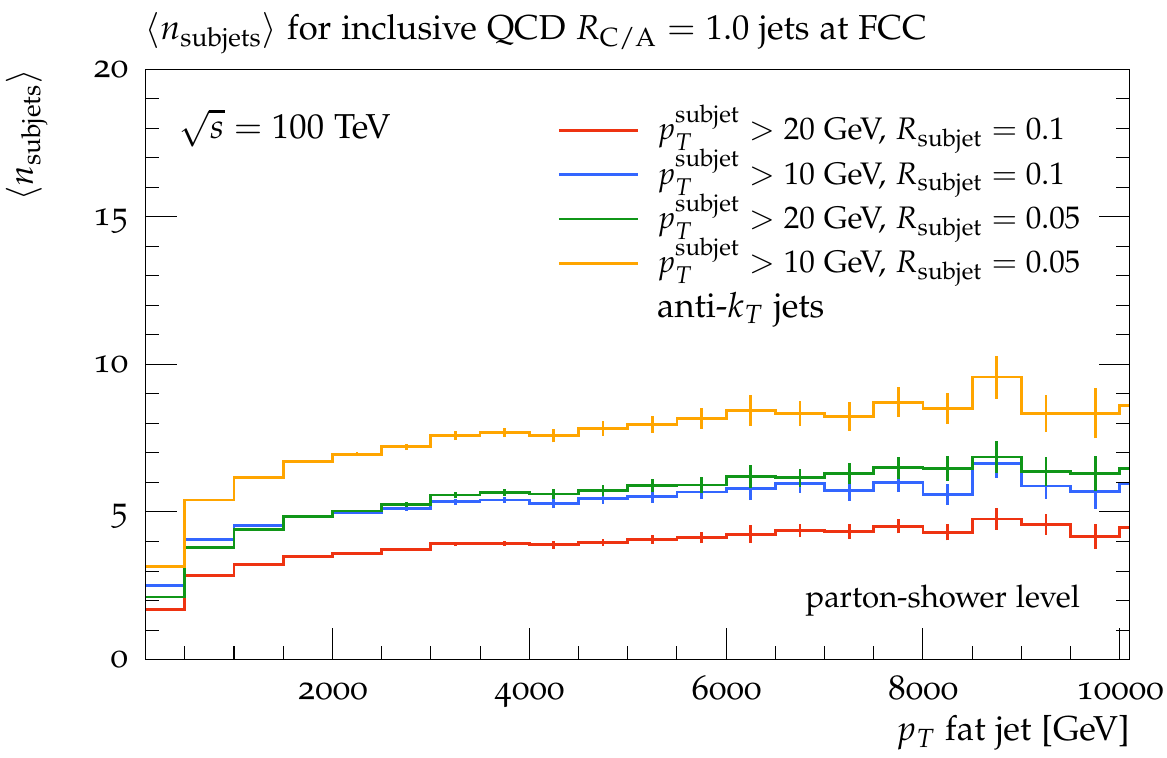}
\caption{\label{FIG:avN_QCD}Average number of subjets inside 
Cambridge--Aachen jets of $R=1.0$ in inclusive QCD-jet production. 
Subjets are reconstructed using the anti-$k_T$ jet finder with 
$R_{\rm subjet} = 0.05,\,0.1$ and $p^{\rm subjet}_{T} > 10,\,20\,\rm{GeV}$. 
Results are presented at parton level.}
\end{figure}

To set the stage, Fig.~\ref{FIG:avN_QCD} compiles the expectation for the
average number of anti-$k_T$ subjets found inside large-area Cambridge--Aachen
jets of size $R_{C/A}=1.0$ \cite{Cacciari:2011ma} as a function of the fat-jet
transverse momentum.  This potentially allows contact to be made with LHC results
in the future.  Results are obtained from a \Sherpa dijet simulation, invoking
parton showers but neglecting any non-perturbative corrections, like
parton-to-hadron fragmentation and the underlying event.  While the results
shown here were obtained from the parton shower based on Catani--Seymour
dipoles~\cite{Schumann:2007mg}, they have carefully been checked and confirmed
using the independent \DIRE shower implementation~\cite{Hoche:2015sya} in 
\Sherpa. 

In all results, two benchmark values for $R_{\rm subjet}$ are considered,
$R_{\rm subjet}=0.05$ and $0.1$. Furthermore, two threshold values for the subjet
transverse momentum are used, namely $p^{\rm subjet}_T>20,\,\SI{10}{GeV}$.
Clearly, $\langle n_{\rm subjets}\rangle$ grows with smaller $R_{\rm subjet}$ and
$p^{\rm subjet}_{T}$ cut. For the mixture of quark and gluon jets given by the
LO matrix elements in this calculational setup, a mean number of
subjets of $\langle n_{\rm subjets}\rangle\approx 5$ for
$p^{\rm fat}_T=\SI{3.5}{TeV}$, $p^{\rm subjet}_T>\SI{10}{GeV}$ and $R_{\rm subjet}=0.1$
is found. In the following, the LO matrix elements for quark and
gluon production will be considered separately, in order to contrast them 
individually.  However, for all considered parameter choices the slope of 
the $\langle n_{\rm subjets}\rangle$ distributions levels off for large values 
of the fat-jet $p_T$, corresponding to very collimated jet-energy profiles.  
In this regime of large $p_T$, the
actual jet inside the fat-jet area becomes to be of a size comparable to the
subjet size, and it becomes increasingly harder to push more subjets into
the jet.

Using very small $R_{\rm subjet}$ and $p^{\rm subjet}_T$ is not only an experimental
challenge for reconstruction algorithms but also induces large logarithms
that need to be resummed in order to obtain a reliable prediction. Both of
the jet algorithms considered here fall into the family of generalised
longitudinally-invariant $k_t$ algorithms.  They all rely on a distance
measure between all pairs of particles $i$ and $j$ given by
\begin{equation}
  d_{ij}=\min\left(p^{2p}_{T,i},p^{2p}_{T,j}\right)\frac{(\Delta R_{ij})^2}{R^2}
\end{equation}
and a separation between all particles and the beam direction,
\begin{equation}
d_{iB}= p^{2p}_{T,i}\,.
\end{equation}
With $p_{T,i}$ the transverse momentum of particle $i$ and 
$(\Delta R_{ij})^2=(y_i-y_j)^2+(\phi_i-\phi_j)^2$, where $y_i$ denotes the
rapidity and $\phi_i$ the azimuth angle of the particle $i$.  The parameter
$p$ determines the actual jet algorithm. The choices $p= 1,0,-1$ correspond
to the $k_T$ \cite{Catani:1993hr}, Cambridge--Aachen \cite{Dokshitzer:1997in}
and anti-$k_T$ \cite{Cacciari:2008gp} jet finder, respectively. 
$R$~corresponds to the jet-radius parameter already mentioned above.  For this
class of jet algorithms there are predictions resummed for small $R$ to
all-orders of $(\alpha_s\log R^2)$ \cite{Dasgupta:2014yra,Dasgupta:2016bnd},
and for small $R$ and small transverse-momentum threshold $p_{T,\rm min}$ of 
$(\alpha_s\log R^2\log(p_{T}/p_{T,\rm min}))$ to double and next-to-double
logarithmic approximation~\cite{Gerwick:2012fw,Bhattacherjee:2015psa}. 
In particular, Ref.~\cite{Gerwick:2012fw} derived resummed predictions for 
jet rates and the mean number of jets to double-logarithmic (DLA) 
and next-to-double-logarithmic approximation (NDLA), accounting 
for effects of the running of the strong coupling. It should be noted that 
at this level of accuracy the results are independent of the parameter 
$p$ that distinguishes the jet algorithms. 

\begin{figure}[bh!]
\includegraphics[width=0.49\textwidth]{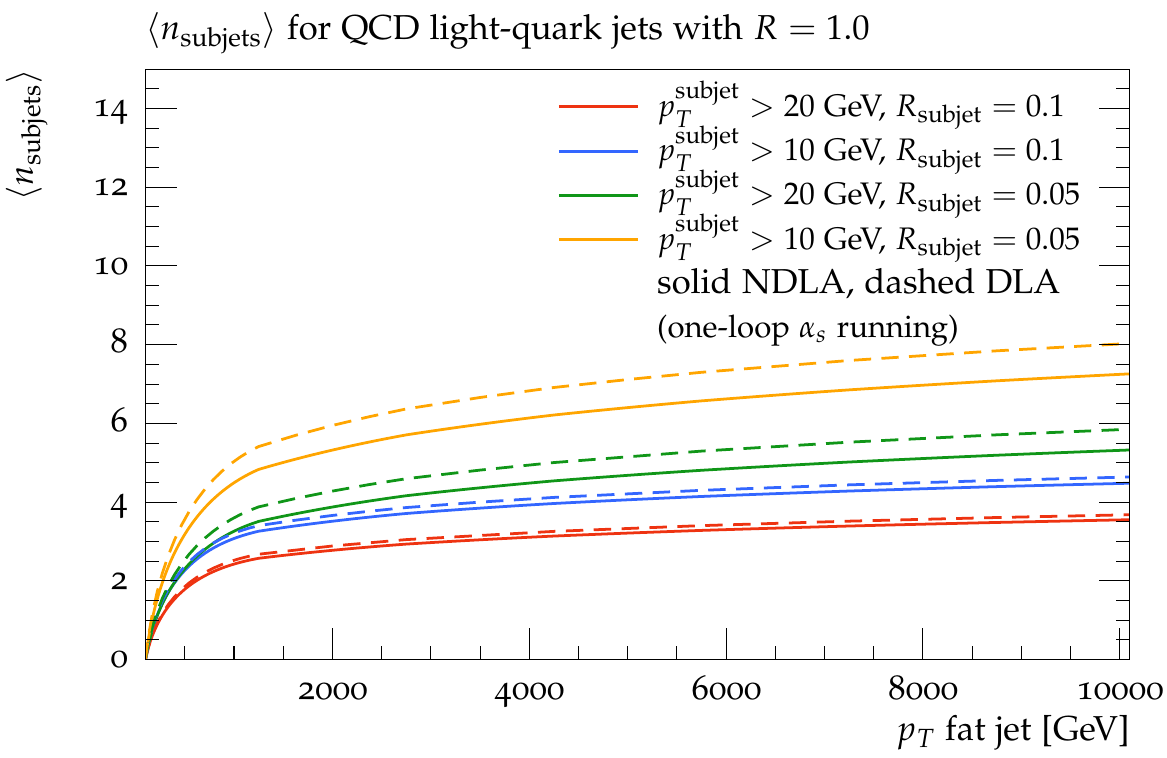}
\includegraphics[width=0.49\textwidth]{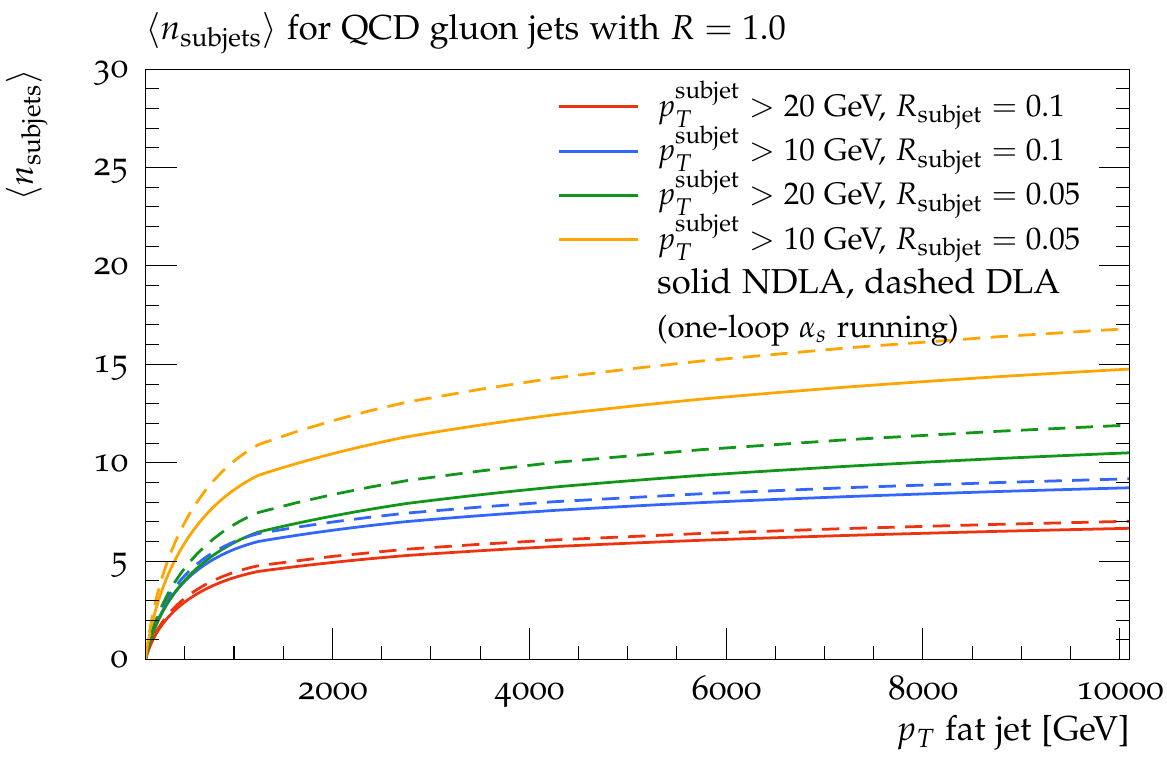}
\caption{\label{FIG:avN_QG_pure_NDLA}NDLA (solid) and DLA (dashed) 
predictions for the mean number of subjets inside $R=1.0$ light-quark 
(left panel) and gluon (right panel) initiated jets for different 
choices of $p^{\rm subjet}_T$ and $R_{\rm subjet}$.}
\end{figure}

\clearpage
In Fig.~\ref{FIG:avN_QG_pure_NDLA}, resummed predictions to DLA and NDLA 
accuracy including the effect of the running of $\alpha_s$ to one-loop order
are presented for $\langle n_{\rm subjets}\rangle$ for both light-quark and gluon
initiated jets of size $R=1.0$.  It can be observed that for all combinations
of $R_{\rm subjet}$ and $p^{\rm subjet}_{T,{\rm min}}$ gluons induce a larger mean
number of subjets than quarks, as naively expected from the colour charges.
The NDLA corrections are most sizeable for $R_{\rm subjet}=0.05$, where they
reduce the DLA prediction significantly. 

\begin{figure}[ht!]
\includegraphics[width=0.49\textwidth]{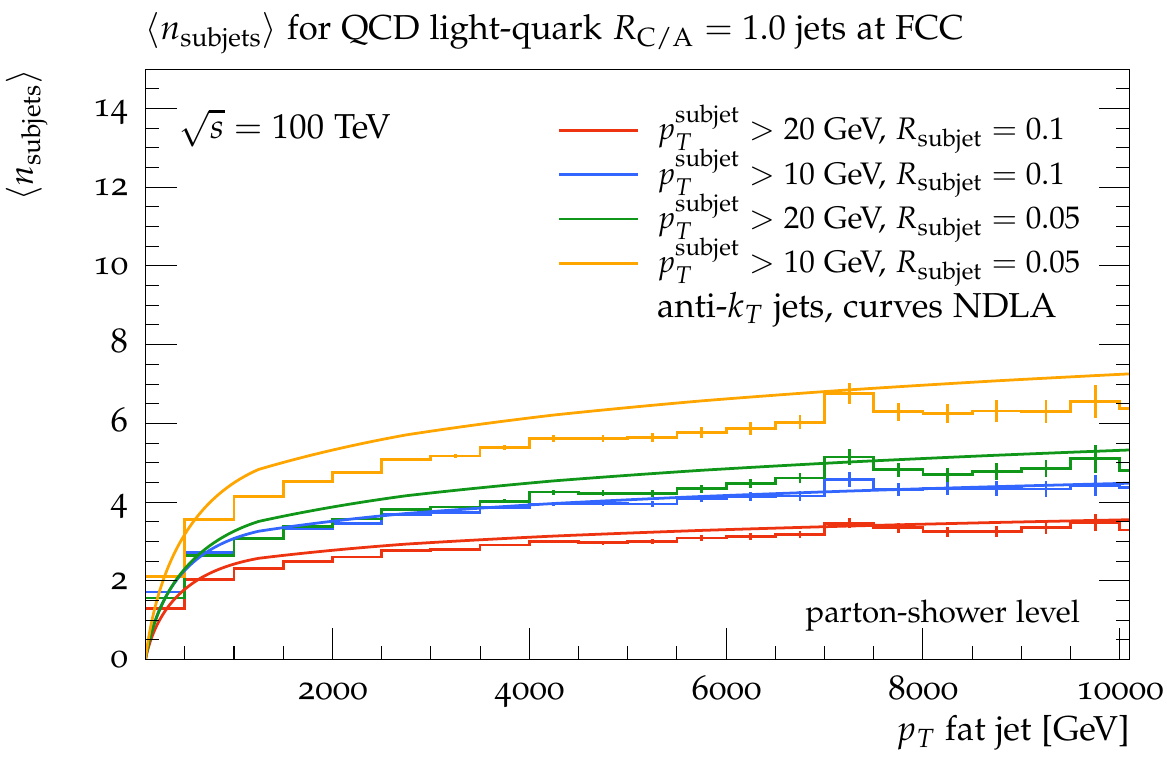}
\includegraphics[width=0.49\textwidth]{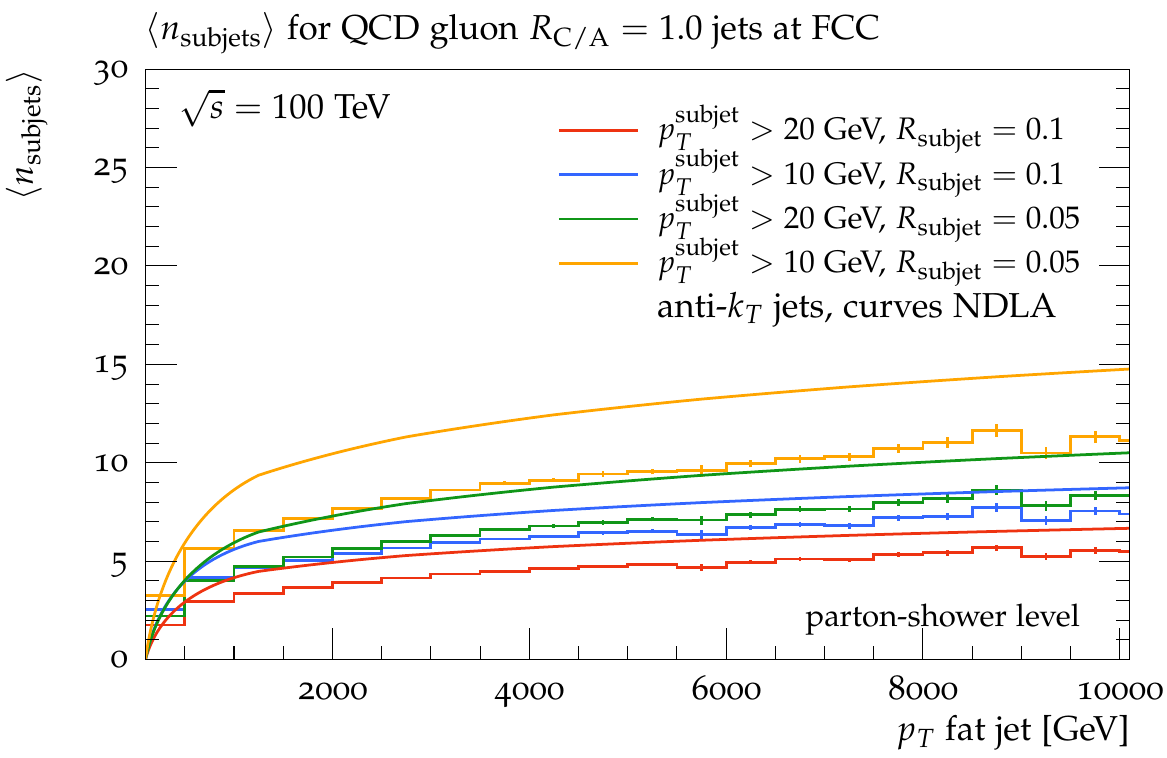}
\caption{\label{FIG:avN_QG_Sherpa_vs_NDLA}Average number of subjets 
inside $R=1.0$ light-quark (left panel) and gluon (right panel) initiated 
jets. A~\Sherpa parton-shower simulation (histograms) is compared to a
corresponding NDLA resummed predictions (solid curves) for different choices of
$p^{\rm subjet}_T$ and $R_{\rm subjet}$.}
\end{figure}

In Fig.~\ref{FIG:avN_QG_Sherpa_vs_NDLA} the comparison of the
$\langle n_{\rm subjets}\rangle$ distribution for a \Sherpa parton-shower 
simulation and the corresponding NDLA prediction for quark- and gluon-initiated
jets is presented.  For the shower simulation, the processes $pp\to q\bar{q}$
and $pp\to gg$ have been considered, respectively. Given the large jet 
transverse momenta investigated here, initial-state parton-shower effects 
are rather suppressed and a comparison to the pure final-state evolution 
hypothesis of the resummed calculation is applicable\footnote{This hypothesis 
has explicitly been checked and confirmed by switching off initial-state 
splittings in the \Sherpa parton shower.}. For the case of quark-initiated 
jets, the resummed predictions agree well with the parton-shower results, and 
the dependence on the fat-jet transverse momentum is very well reproduced. 
For $R_{\rm subjet}=0.05$ the resummation overshoots the shower prediction 
by about $10\%$.  When comparing the results for gluon jets, somewhat larger
deviations are observed. Once again the parton shower nicely reproduces the
shape of the resummed prediction. However, the NDLA results overshoot the
Monte-Carlo simulation by about $15-20\%$. It has been observed before that
resummed predictions for gluon jets tend to produce larger deviations from
shower generators~\cite{Bhattacherjee:2015psa} and that the latter predict
somewhat lower rates, in particular when considering small jet radii.  Since in
general gluons radiate more than quarks, they are thus more sensitive to
missing higher-order terms.  For $R_{\rm subjet}$ values as small as $0.1$ or
even $0.05$ the analytic resummation of terms like
$(\alpha_s\log(1/R^2_{\rm subjet}))^n$ to all orders~\cite{Dasgupta:2014yra} or
jet-clustering logarithms as discussed in \cite{Delenda:2006nf} might need to
be considered.  Overall, one can conclude that parton-shower predictions for
the mean-number of subjets in large-area jets give reliable results that are
in good agreement with analytical estimates from resummed calculations.  This
allows realistic perturbative predictions to be made for this observable even
for very large jet transverse momenta, small subjet radii and rather small
subjet $p_T$ thresholds. Certainly, for a dedicated comparison against data,
non-perturbative corrections from hadronisation and the underlying event
need to be included.  However, these are largely independent of the flavour
of the particle that seeds the jet evolution and thus will not critically
change the above picture.  Instead, apart from slightly washing out some
of the differences between quark and gluon jets, only a modest offset in the
mean number subjets is expected. 

\begin{figure}[ht!]
 \includegraphics[width=0.66\textwidth]{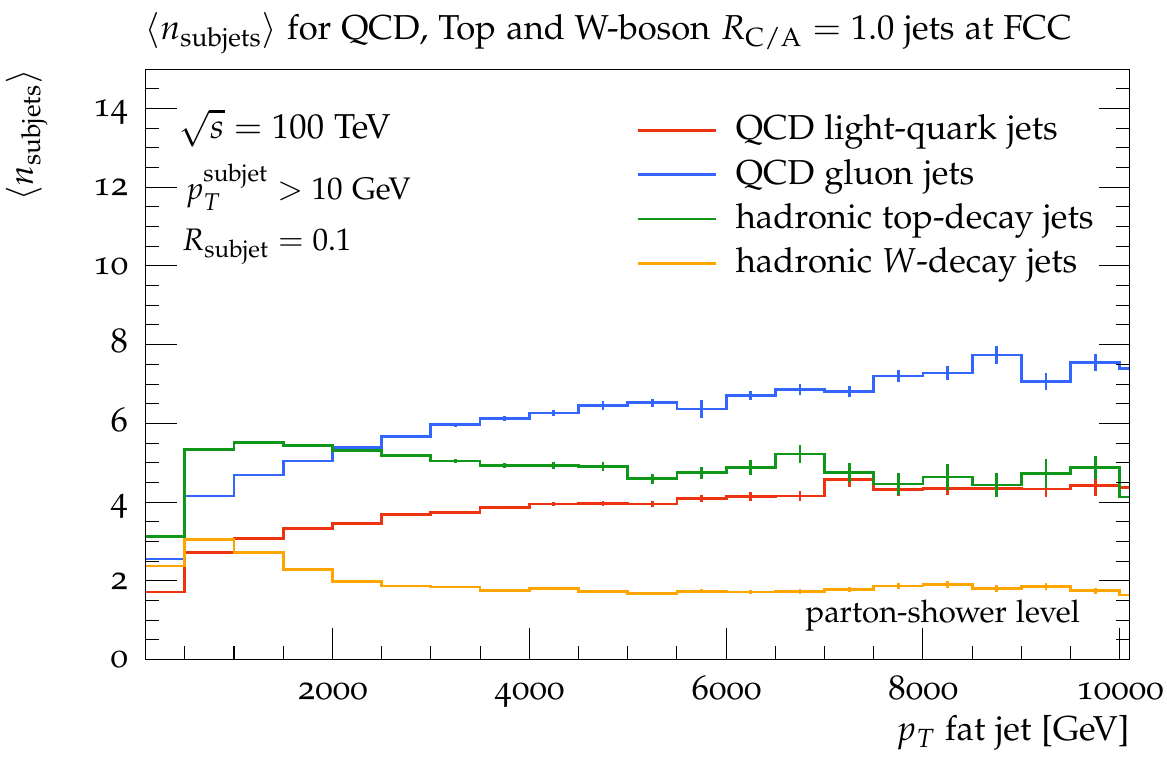}
 \caption{\label{FIG:avN_JTW}Average number of subjets inside 
Cambridge--Aachen jets of $R=1.0$ originating from hadronic 
top-quark and $W$-boson decays, and QCD quark-jet production. 
Subjets are reconstructed using the anti-$k_T$ jet finder with
$R_{\rm subjet} = 0.1$ and $p^{\rm subjet}_{T} > \SI{10}{GeV}$. Results
are presented at parton level.}
\end{figure}

The observable at hand, $\langle n_{\rm subjets}\rangle$ as a function of the
transverse momentum of a large-area jet, will now be considered as a
discriminator for QCD jets and hadronic decays of heavy particles. In 
Fig.~\ref{FIG:avN_JTW}, a comparison for the mean number of subjets found inside
Cambridge--Aachen jets of $R=1.0$ containing the hadronic decay products of 
top-quarks, $W$-bosons and light-quark QCD jets is presented. In the analysis 
of the top-quark and $W$-boson decays, the reconstructed fat-jet that is
closest to the direction of the actual resonance is being analysed. The
quark-jet distribution is obtained from the analysis of $pp\to q\bar{q}$
events.  For $R_{\rm subjet}$ and $p^{\rm subjet}_T$ the values $0.1$ and
$\SI{10}{GeV}$ are considered, respectively. 

Most notably, jets containing the decay jets of boosted $W\to q\bar{q}'$ decays
feature a rather small number of subjets. This is related to the colour-singlet 
nature of the $W$-boson.  Its decay jets are very collimated at high
transverse momentum, with no colour--connection to the rest of the event,
that characterise the quark or gluon jets.  This results in a rather constant
expectation of just two subjets for $p^{\rm fat}_T>\SI{2}{TeV}$. At 
$p^{\rm fat}_T\approx\SI{1}{TeV}$ three subjets are resolved on average, 
corresponding to the emission of one additional jet from the two decay 
partons. This prominent feature makes it easily possible to distinguish 
hadronic $W$-boson decays, or similarly Higgs-boson decays, from QCD jets. 

\begin{figure}[ht!]
 \includegraphics[width=0.49\textwidth]{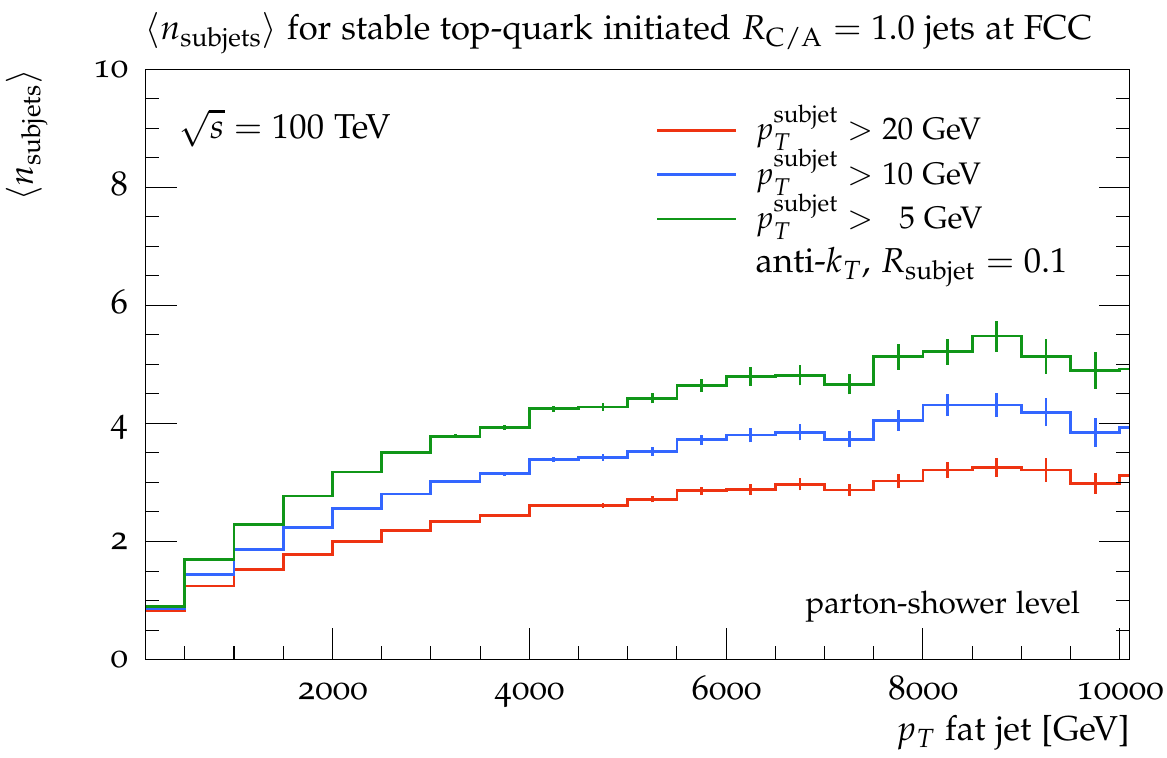}
 \includegraphics[width=0.49\textwidth]{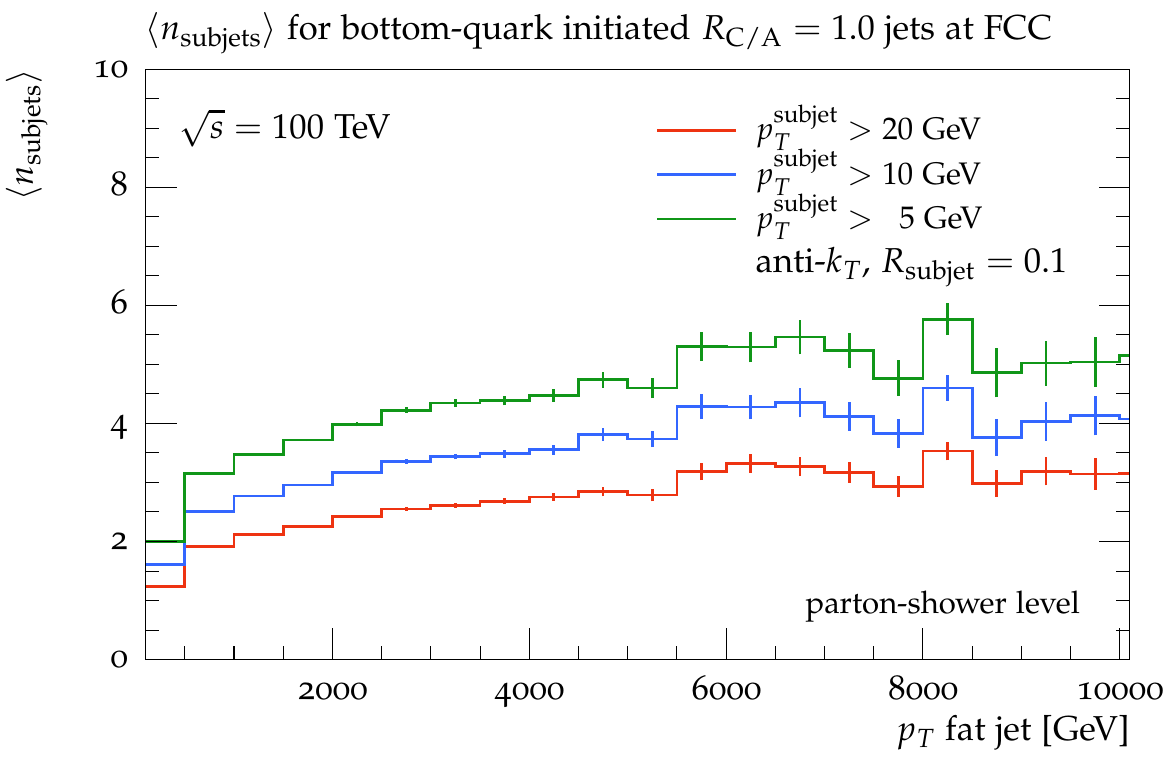}
 \caption{\label{FIG:avN_TopBottom} Average number of subjets inside 
Cambridge--Aachen jets of $R=1.0$ originating from stable 
top-quark and bottom-quark production. Subjets are reconstructed 
using the anti-$k_T$ jet finder with $R_{\rm subjet} = 0.1$ and 
$p^{\rm subjet}_{T} > 5\,, 10,\,20\,\rm{GeV}$. Results are presented 
at parton level.}
\end{figure}

The identification of top-quark decays based on $\langle n_{\rm subjets}\rangle$
seems much harder. The distribution peaks around
$p^{\rm fat}_T\approx\SI{1}{TeV}$ with a value of
$\langle n_{\rm subjets}\rangle\approx 5.5$.  This is significantly higher than
what is observed for light-quark jets and even for gluon jets, and it is due
to the hadronic decays assumed for the tops, i.e. $t\to bW^+\to bq\bar{q}'$, 
which yield three jets -- two more than the original quark.  With increasing 
transverse momentum the top-jet
distribution approaches the light-quark result, reflecting the fact that
beyond $p^{\rm fat}_T\approx\SI{4}{TeV}$ the decay products are extremely
collimated and basically radiate with their combined colour charge $C_F$ as
light-quark jets do. To illustrate this fact Fig.~\ref{FIG:avN_TopBottom}
compiles the $\langle n_{\rm subjets}\rangle$ distribution for stable (undecayed)
top quarks and bottom quarks. Three values of $p^{\rm subjet}_T$ are considered,
$5,\,10$ and $\SI{20}{GeV}$ while $R_{\rm subjet}$ is fixed to $0.1$. Mass
effects, namely the shielding of collinear singularities, yield a suppression
of radiation off top quarks up to $p_T$ values of $\SI{4}{TeV}$.  The
radiation off bottom-quarks is at high transverse momenta as considered here
compatible with the light-quark distributions presented in 
Fig.~\ref{FIG:avN_QG_Sherpa_vs_NDLA}. 

It can be concluded that at FCC collisions energies the identification of 
very boosted hadronic decays becomes extremely challenging. The observable 
presented here, i.e. $\langle n_{\rm subjets}\rangle$ of large-area jets, provides 
sensitivity to the QCD colour charge of the jet-initiating particle, either a 
QCD parton or a heavy resonance. For QCD quark and gluon jets the results
obtained from parton-shower simulations are in good agreement with predictions
from all-orders resummation calculations at NDLA accuracy. 

\section{Loop-induced processes at 100 TeV}
\label{sec:loop}
\subsection{Finite top mass effects in gluon fusion Higgs production}

At a \SI{100}{\tera\electronvolt} hadron-collider machine, the dominant 
production mechanism for a Standard-Model Higgs boson proceeds through 
top-quark loops that mediate an interaction between gluons and the Higgs 
boson. The increased calculational complexity due to the presence of this
loop, already at leading order, is commonly reduced by using 
the Higgs Effective Field Theory (HEFT) approximation. In this approach, 
the top-quark loop mediated couplings between gluons and the Higgs boson 
are approximated in terms of direct tree-level couplings that can be
derived from the effective Lagrangian
\begin{align}
  \mathcal{L}_\mathrm{HEFT} &=
  \frac{\alpha_s}{12\pi}G_{\mu\nu}^aG^{\mu\nu}_a\ln\left(1+\frac{H}{v}\right)\\
  &=\frac{\alpha_s}{12\pi}G_{\mu\nu}^aG^{\mu\nu}_a\left[\left(\frac{H}{v}\right)
    -\frac{1}{2}\left(\frac{H}{v}\right)^2 +
    \frac{1}{3}\left(\frac{H}{v}\right)^3
    - \ldots\right]\, ,\label{eq:heft_lag}
\end{align}
with the gluon field-strength tensor $G_{\mu\nu}$ and the Higgs
field~$H$. This Lagrangian gives rise to tree-level couplings that can be
understood as the infinite top-mass limit, i.e. $m_t \to \infty$, of
the respective loop-induced SM couplings between gluons and the Higgs
boson.

For the calculation of the inclusive total cross section at LHC energies, 
the application of the HEFT approximation is well motivated since finite
top-quark mass corrections turn out to be moderate~\cite{Marzani:2008az,
  Pak:2009bx,Pak:2009dg,Harlander:2009my,Harlander:2009mq,Harlander:2009bw}. 
This applies both to inclusive Higgs production and to Higgs production 
in association with one or two jets~\cite{Buschmann:2014sia,Frederix:2016cnl}. 
However, the tail of the Higgs-boson transverse momentum distributions is 
only poorly modelled in HEFT~\cite{Baur199038,Ellis1988221}, with the
infinite top-mass approximation overshooting the $m_t$-exact result. The
purpose of this section is to quantify such effects at a
\SI{100}{\tera\electronvolt} hadron collider. 

The results presented in this section are obtained from leading-order
calculations of inclusive Higgs production processes in association with one,
two, and three jets, including top-mass effects, based on one-loop matrix
elements from~\OpenLoops~\cite{Cascioli:2011va}.  For the evaluation of
scalar and tensor integrals the~\Collier library~\cite{Denner:2016kdg,
  Denner:2014gla} and \CutTools~\cite{Ossola:2007ax} is being employed. A
full NLO QCD calculation in the HEFT framework, but cluding top-mass effects
in the approximation of~\cite{Buschmann:2014sia} is feasible with \Sherpa
but not necessary in order capture the characteristics of the mass
corrections, as they have proven to factorize to a good approximation
from the NLO QCD corrections~\cite{Harlander:2012hf}. For the purpose of
studying the top-mass effects in the hard scattering process, parton-shower
effects are not considered here.

\begin{figure}[hbt]
  \centering
  \includegraphics[width=0.48\textwidth]{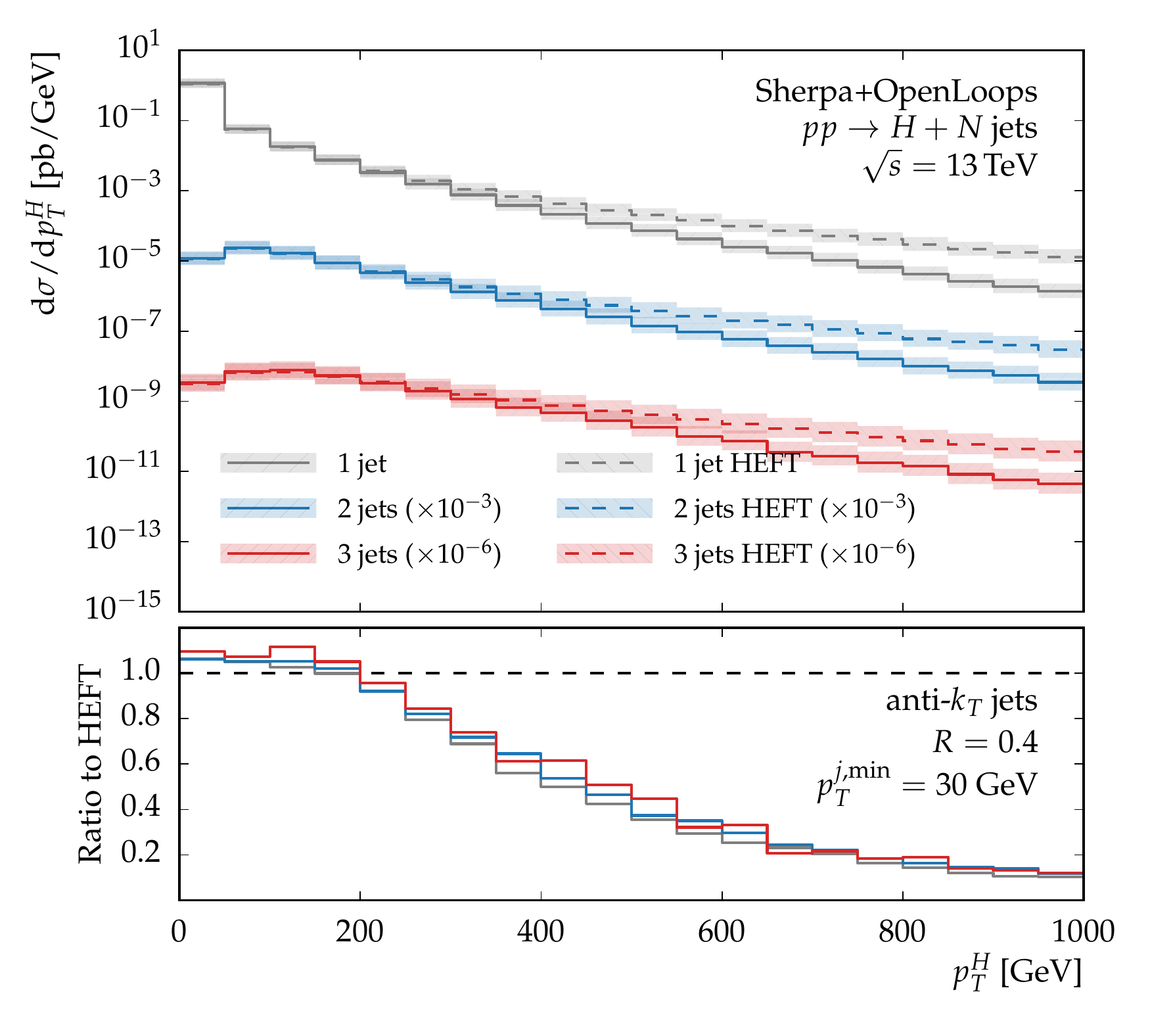}
  \hfill
  \includegraphics[width=0.48\textwidth]{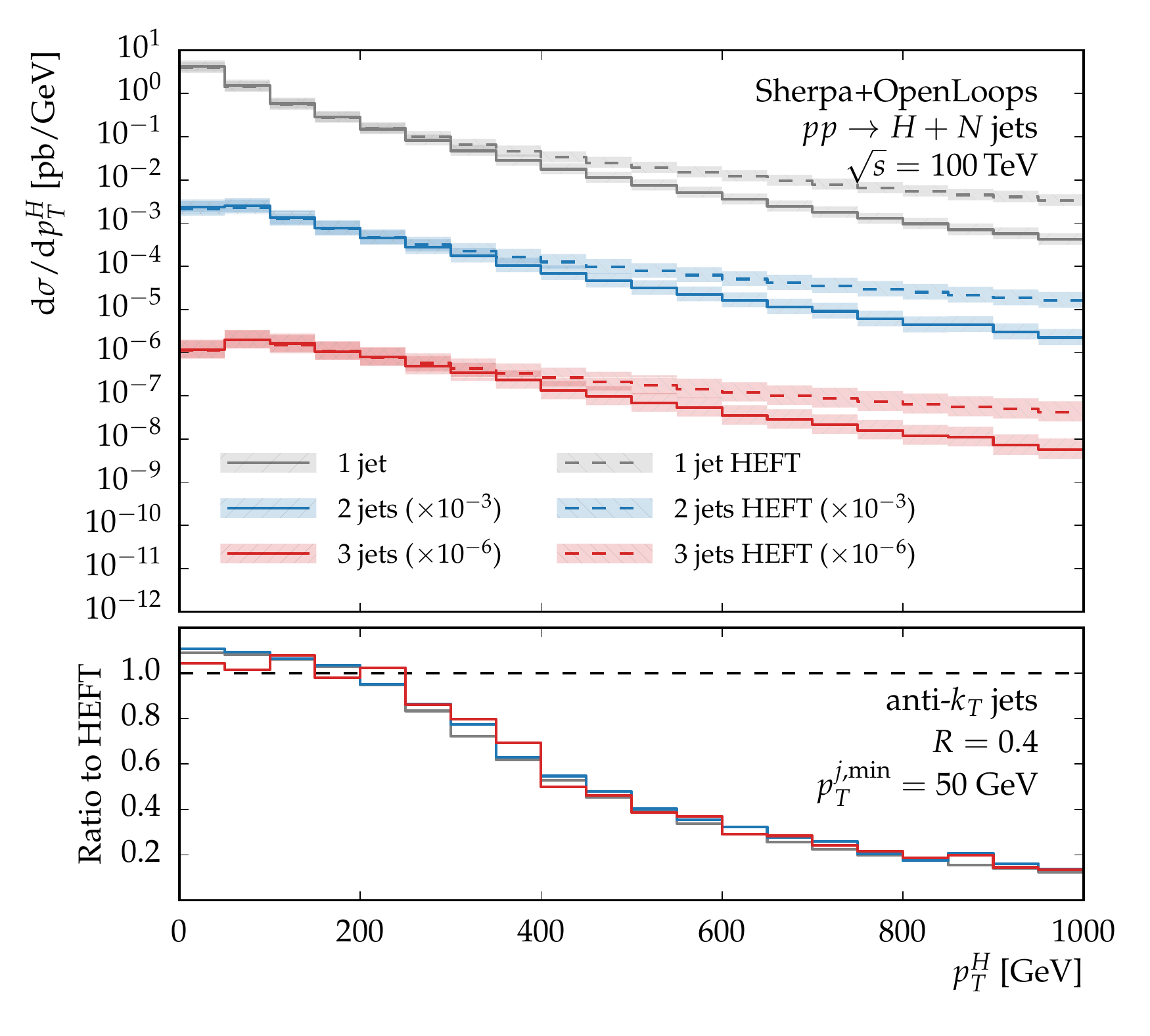}
  \caption{Higgs transverse-momentum distributions in gluon-fusion
    Higgs production in association with up to three anti-$k_T$ jets
    with $R=0.4$ and $p_{T,j}>\SI{50}{GeV}$. The distributions are
    obtained from leading-order calculations of the respective
    process. Results for the infinite top mass (HEFT) and the
    $m_t$-exact calculation are shown. The lower panels show the ratio
    of the full $m_t$-exact result and the HEFT
    approximation.}\label{fig:fo_ratios}
\end{figure}

In Fig.~\ref{fig:fo_ratios} the transverse-momentum distribution of
Higgs production at \SI{13}{\tera\electronvolt} and at
\SI{100}{\tera\electronvolt} with up to three jets is presented. From
the ratio plots in the lower panels of Fig.~\ref{fig:fo_ratios} it
is evident that the relative size of the finite top-mass corrections
to the $p_T$ distributions exhibit the same universal suppression
pattern for all jet multiplicities. This is in accordance with similar
findings in~\cite{Buschmann:2014twa,Buschmann:2014sia} where the
production of up to two associated jets at LHC energies has been
studied. When comparing the results for collider energies of
\SI{13}{\tera\electronvolt} and \SI{100}{\tera\electronvolt}, the
similarity of the relative size of the corrections is quite remarkable.
Considering the increase in partonic energy that is available at
\SI{100}{\tera\electronvolt}, one could have expected that for a given
value of Higgs $p_T$, the mean partonic centre-of-mass energy would be
higher, thus giving rise to larger discrepancies between the HEFT
calculation and the $m_t$-exact calculation. This is, however, not the
case, even for three-jet final states, where the jets can in principle
carry away large amounts of additional energy.

\begin{figure}[hbt]
  \centering
  \includegraphics[width=0.48\textwidth]{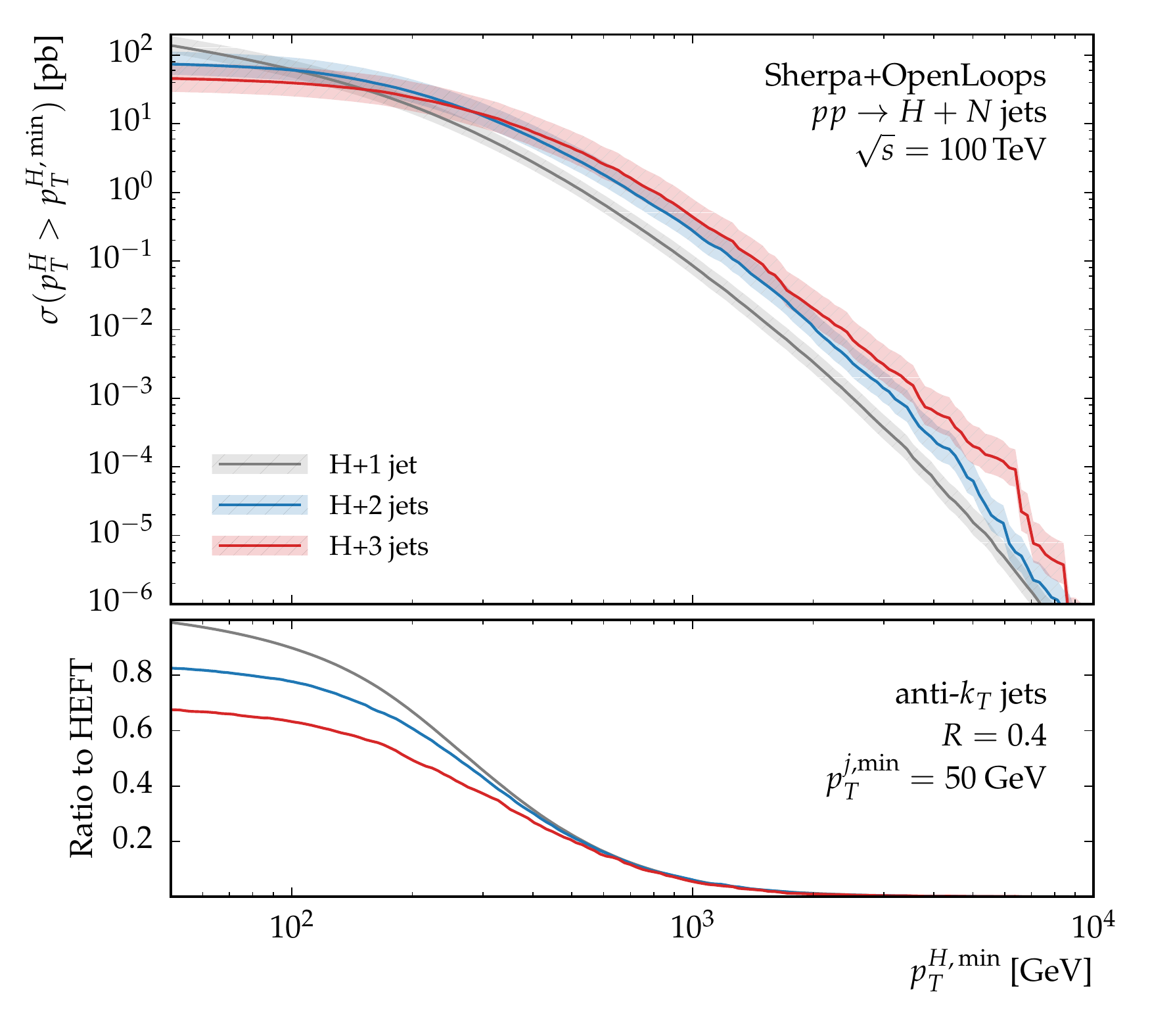}
  \hfill
  \includegraphics[width=0.48\textwidth]{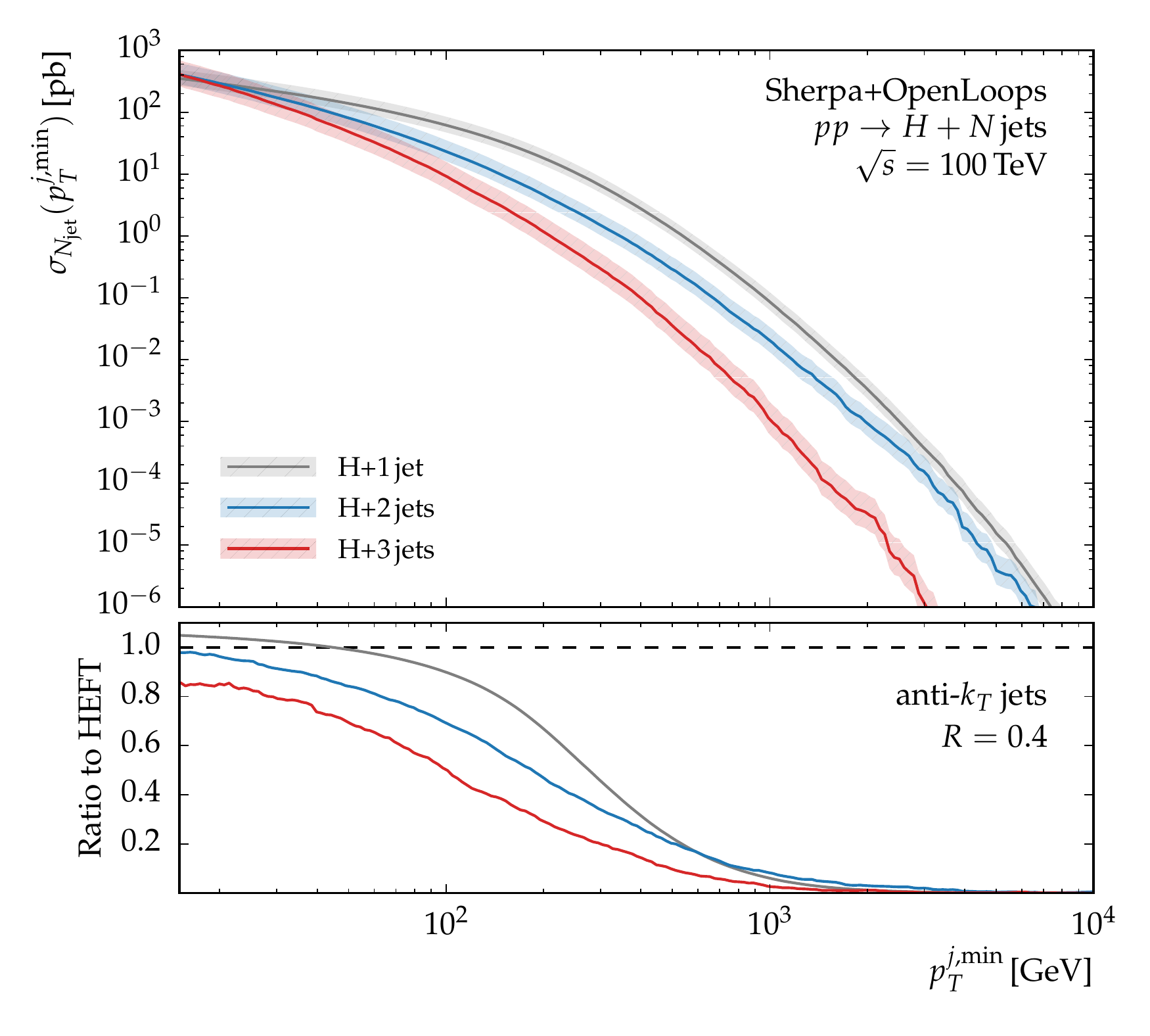}
  \caption{Fixed-order cross sections for Higgs production in
    association with up to three jets as a function of a minimum Higgs
    $p_T$ cut (left panel) and as a function of the minimum jet $p_T$
    (right panel) with finite top-mass effects taken into account. The
    lower panels show the ratios to the respective HEFT predictions
    and quantify the finite top-mass corrections.
  }\label{fig:ggh_cumulative}
\end{figure}

This effect can indeed be observed when considering cumulative
distributions, as exemplified in Fig.~\ref{fig:ggh_cumulative}. The
total cross section receives large finite top mass corrections at
\SI{100}{\tera\electronvolt} when applying a minimum 
transverse-momentum cut on the Higgs. This is due to the fact that 
the tail of the Higgs transverse-momentum spectrum exceeds further 
into the high energy regime, where finite top-mass effects are large. 
The effect is
already substantial for a cut of $p_T^H>\SI{50}{\giga\electronvolt}$,
where the HEFT result overshoots the $m_t$-exact result by more than
\SI{50}{\percent} in the three-jet case. The same effect can be
observed in the $n$-jet inclusive cross sections that are displayed in
the right panel of Fig.~\ref{fig:ggh_cumulative} as a function of
the minimum jet transverse momentum. Even for moderate cuts on the jet
transverse momenta around
$p_T^\mathrm{jet}=\SI{50}{\giga\electronvolt}$, large corrections of
the order of \SI{-30}{\percent} are observed, indicating a poor
description even of very inclusive observables in three-jet final
states. Tab. \ref{TAB:Hxsecs} illustrates this further with the
leading-order cross sections for all three jet multiplicities at
$p_T^{j,\mathrm{min}}=\SI{50}{\giga\electronvolt}$. Higher jet
multiplicities are particularly relevant for the description of the
Higgs $p_T$ spectrum above \SI{200}{\giga\electronvolt}. As
demonstrated in the left panel of Fig.~\ref{fig:ggh_cumulative}, the
three-jet contributions exceed both the one- and the two-jet
contributions in magnitude in this boosted regime. This feature of the
transverse-momentum distribution is however not specific to FCC
energies but can be observed at the LHC as well
\cite{Greiner:2015jha}.

\begin{table}[tbh!]
  \begin{tabular}{l||c|c|c|c}
  \hline
  \hline
    & $H$ & $H\,j$ & $H\,j^2$ & $H\,j^3$ \\
  \hline
  SM   & \SI{229}{\pico\barn} & \SI{137}{\pico\barn} &  \SI{79}{\pico\barn} &  \SI{47}{\pico\barn} \\
  HEFT & \SI{215}{\pico\barn} & \SI{138}{\pico\barn} & \SI{94}{\pico\barn}  &  \SI{68}{\pico\barn} \\
  \hline
  \hline
  \end{tabular}
  \caption{\label{TAB:Hxsecs}
    Leading-order inclusive production cross sections at the FCC for 
    $H$-boson production in association with $n=0,1,2,3$
    jets based on the $m_t$-exact calculation (SM) and the 
    HEFT approximation. Jets are reconstructed using the anti-$k_T$ algorithm 
    with $R=0.4$ and $p_T^{j,\mathrm{min}}=\SI{50}{\giga\electronvolt}$.
  }
\end{table}

It can be concluded that in contrast to the situation at the LHC,
finite top-mass effects are sizable at \SI{100}{\tera\electronvolt} even
when considering inclusive jet cross sections with moderate cuts on
the jet transverse momentum. It should furthermore be noted, that at a
future \SI{100}{\tera\electronvolt} hadron collider the event rates
for Higgs-boson production in association with
\si{\tera\electronvolt}-scale jets easily exceed several
femtobarns. In this kinematic regime, the HEFT approximation
completely fails and finite top-mass effects must be taken into
account in order to obtain meaningful predictions.

\begin{figure}
\centering
\includegraphics[width=.5\textwidth]{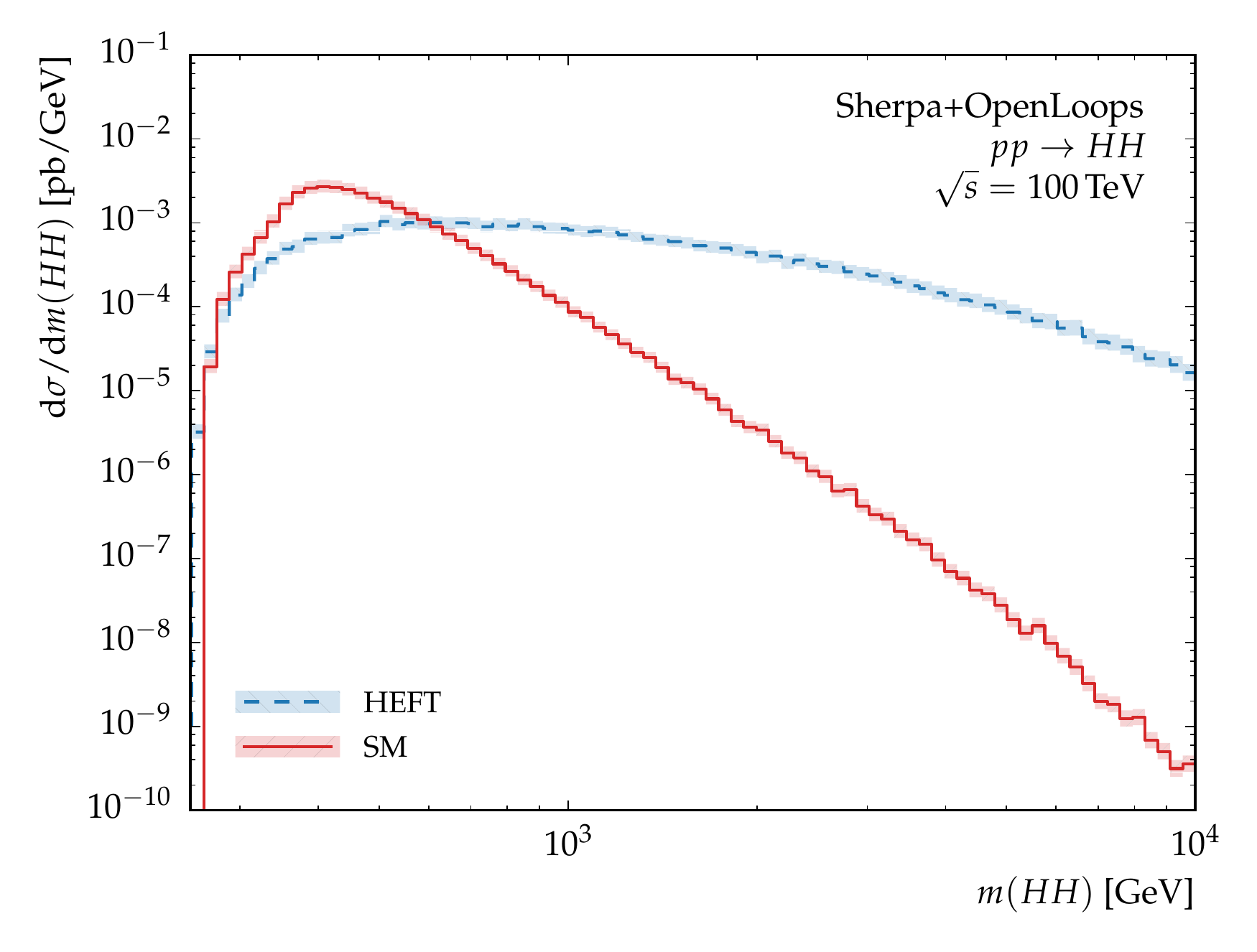}
\caption{Di-Higgs invariant-mass distributions as calculated in the
  infinite top-mass limit (HEFT) and with the exact top-mass
  dependence taken into account (SM) at leading order.}
\label{fig:hh_ratios_fcc}
\end{figure}

For completeness, we also compare the infinite top mass approximation to the
di-Higgs production with the $m_t$-exact SM result. It is well-established
that for this process even the fully-inclusive cross section at the LHC is
poorly approximated in the HEFT approximation~\cite{Baur:2002rb}. We
demonstrate this for FCC energies in Fig.~\ref{fig:hh_ratios_fcc}, where
we show the di-Higgs invariant mass spectrum. Even near threshold, the HEFT
approximation fails to even remotely reproduce the shape of the invariant
mass distribution. The same holds for the inclusive production cross sections
quoted in table \ref{tab:hh_xs}.

\begin{table}[tbh!]
  \begin{tabular}{l||c|c}
    \hline
    \hline
    & SM & HEFT \\
    \hline
    $\sigma(gg\to HH)$ &  $\SI{0.74}{pb}$ & $\SI{2.1}{pb}$\\
    \hline
    \hline
  \end{tabular}
  \caption{\label{tab:hh_xs}
    Leading-order inclusive production cross sections for Higgs pair 
    production at the FCC based on the $m_t$-exact calculation (SM) and the 
    HEFT approximation. 
  }
\end{table}

\subsection{Loop-induced di-boson processes}

In contrast to single- and di-Higgs production, where loop-induced
contributions are dominant, the production of other diboson final
states is typically dominated by quark-induced partonic channels that
proceed through tree-level diagrams at leading order. Loop-induced
contributions with gluons in the initial state nonetheless often
exist and contribute to the production. Despite being part of an 
NNLO QCD correction to the
tree-induced process, their relative size for fairly inclusive LHC
cross sections energies can range from a few percent, in case of the
$pp\to WW\to l\bar\nu_l\bar\l^\prime\nu_{l^\prime}$ process~\cite{Binoth:2006mf},
to around \SI{20}{\percent} for $pp\to ZZ\to 4l^\pm$~\cite{Zecher:1994kb},
thus exceeding the naive expectation of $\alpha_s^2\approx \SI{1}{\percent}$.
This is in part due to the large gluon luminosity in the Bjorken-$x$ range
probed by the process under consideration. At a \SI{100}{\tera\electronvolt}
collider, we expect this effect to be enhanced, since, for a final
state of given mass, the $x$ range probed will be shifted towards
smaller values where the gluon PDFs dominate. Furthermore, corrections
to Higgs processes which are mediated by heavy quark loops can feature
additional threshold effects that increase their relative size even
further~\cite{Goncalves:2015mfa}. As in the previous
section, we employ the \OpenLoops one-loop matrix element provider
along with \Collier and \CutTools and calculate all loop-induced
processes at leading order. We set the renormalization, factorization
and parton shower starting scales to $\sqrt{\hat{s}}/2$ for all
processes and take into account parton shower effects. 

Representative for processes whose gluon-induced components proceed
predominantly through loops of light-quark flavours, we consider
$W$-pair production, $Z$-pair production and photon-associated $Z$-boson 
production. We use matrix elements for the full leptonic final
states with all interference and off-shell effects taken into account.
For $W$-pair production, we consider the $e^-\bar\nu_e \mu^+\nu_\mu$
final state, while for the $ZZ$ process we generate
$pp\to e^-e^+ \mu^-\mu^+$. For photon-associated $Z$-production, we
consider the $e^+e^-\gamma$ final state. All leptons are required to
pass a minimum transverse momentum cut of \SI{50}{\giga\electronvolt}.
The photon is required to pass the same transverse-momentum cut. In
addition, we require the photon to be well separated from the leptons
with $\Delta R(\gamma,e^\pm)>\num{0.4}$. Invariant-mass cuts on lepton
pairs are applied in such a way as to ensure that the intermediate
vector bosons are near their mass shell with
$|m_V-m(ll)|<\SI{15}{\giga\electronvolt}$. The quark-induced processes
that proceed through tree-level diagrams at leading order are
calculated at NLO QCD accuracy with the exception of the
$e^+e^-\gamma$ process, which is calculated only at leading order.

\begin{figure}
\centering
\includegraphics[width=.48\textwidth]{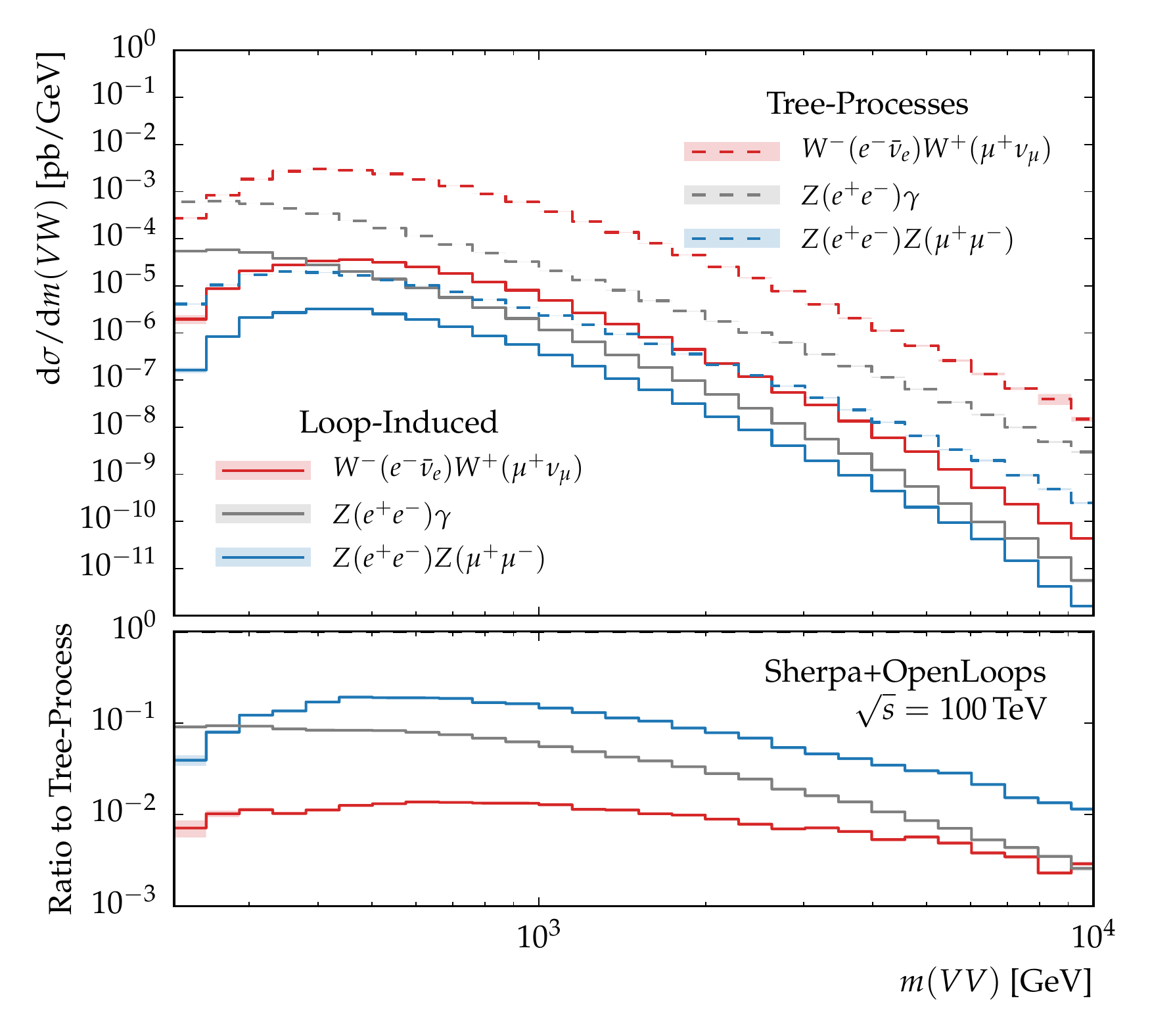}
\hfill
\includegraphics[width=.48\textwidth]{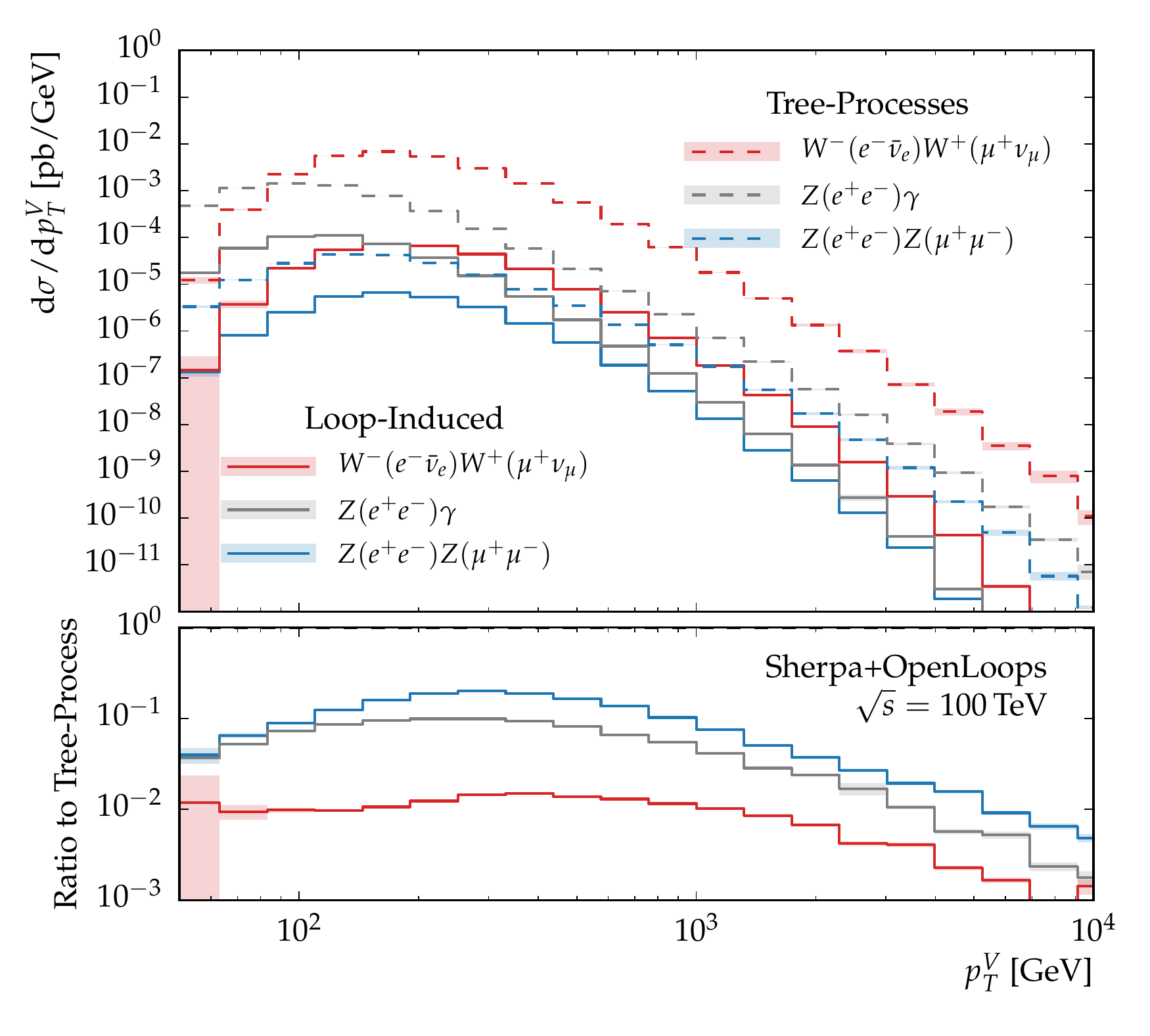}
\caption{Left panel: Invariant mass of the vector-boson pairs in
  $W^+W^-$ and $ZZ$ production. Right panel: Transverse-momentum
  distribution of the $W^-$ and the $Z$ decaying to electrons. The
  solid curves show loop-induced contributions while the dashed curves
  represent tree-like contributions. The lower panel shows the ratio
  of the loop-induced contributions to the tree-level contributions.}
\label{fig:l2_mass_pt}
\end{figure}

In Fig.~\ref{fig:l2_mass_pt} we show the diboson invariant-mass
distributions and the single-boson transverse-momentum spectra for the
three aforementioned processes. Individual curves for the
gluon-induced one-loop processes and for the quark-induced tree-level
processes illustrate the relative size of the loop-induced components.
With the set of cuts applied, we observe that the relative size of the
loop-induced contributions remains moderate. For $W$-pair production,
the corrections remain at the percent level, while the $pp\to ZZ$
process receives corrections of the order of \SI{15}{\percent} from
gluon-induced channels. In the $Z\gamma$ process, these channels still
contribute a substantial, yet somewhat smaller correction as in the
case of $ZZ$ production. Due to the different structure of
contributing Feynman diagrams and because of the differences in the
underlying PDFs, there are however significant shape differences when
comparing loop-induced processes with their respective tree-induced
counterparts.

\begin{figure}
\centering
\includegraphics[width=.48\textwidth]{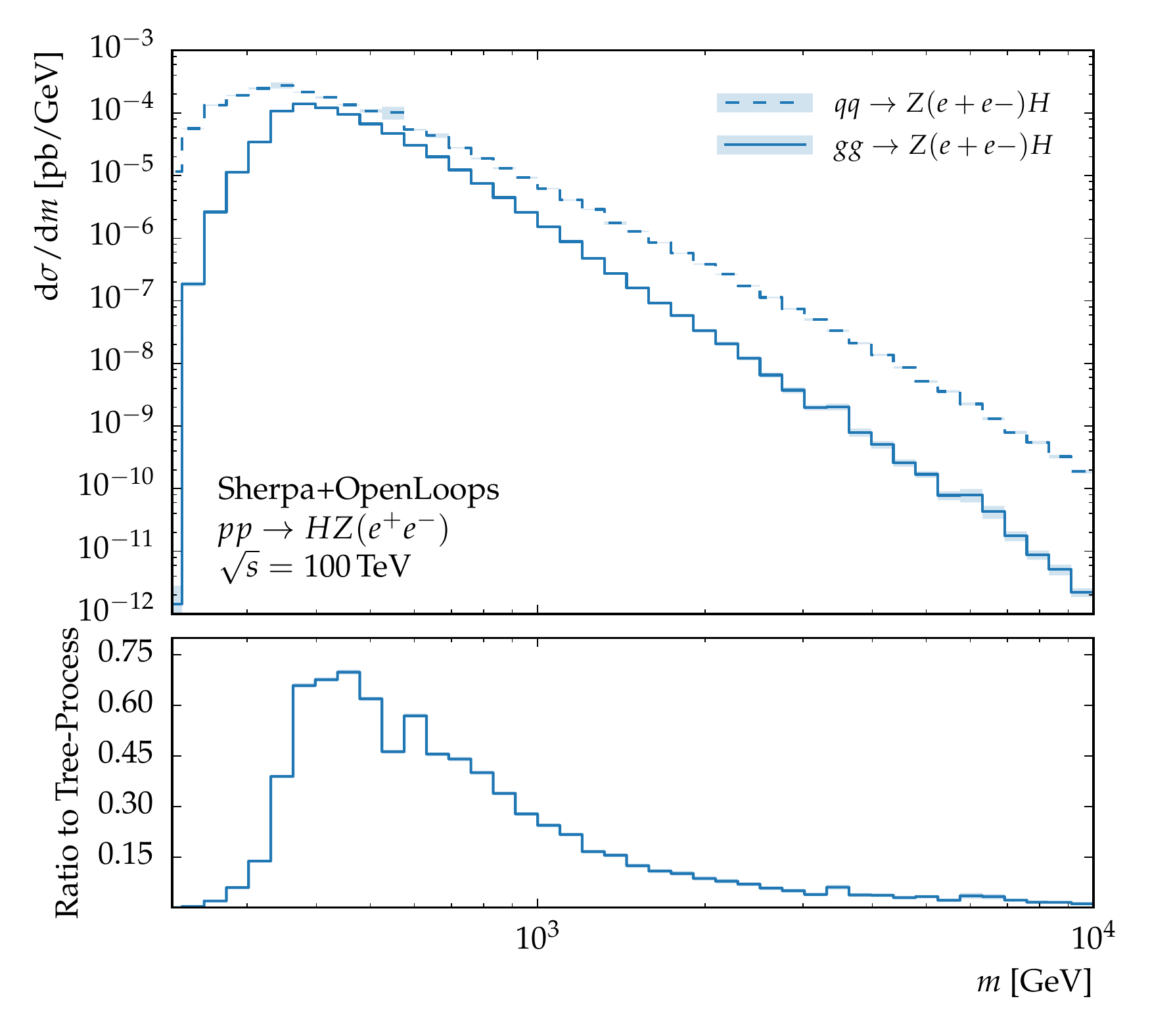}
\hfill
\includegraphics[width=.48\textwidth]{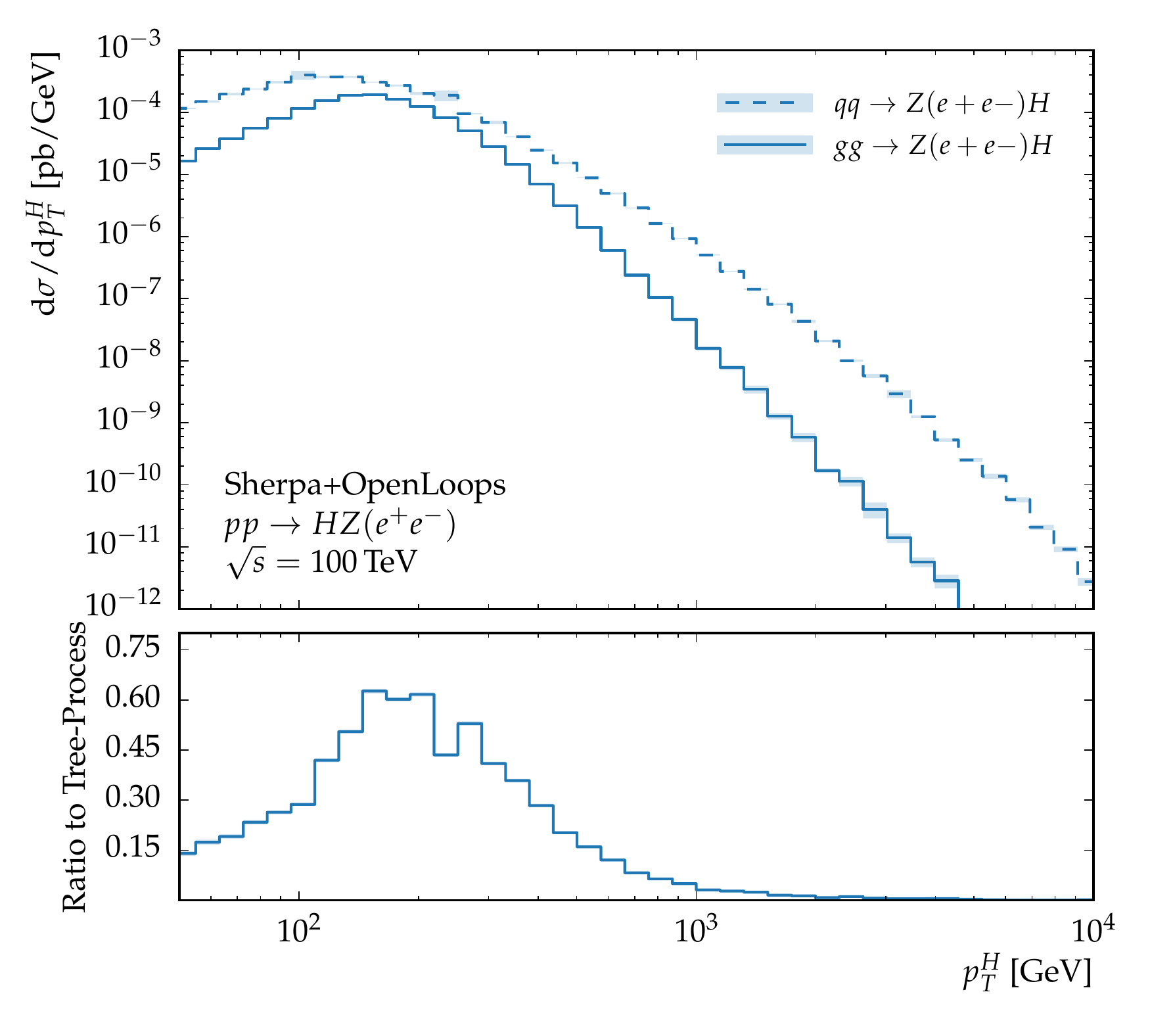}
\caption{Diboson invariant mass (left) and Higgs transverse momentum
  distribution (right) $Z$-associated Higgs production. We show both
  the gluon-induced contributions (solid) and the quark-induced
  tree-like contributions as well as the ratio of the two (lower
  panels).}
\label{fig:l2_t_mass_pt}
\end{figure}

In addition to the processes above we also consider $Z$-associated
Higgs-boson production. In contrast to the previously mentioned
processes, the gluon-induced component of the $ZH$ process is mediated by
loops of heavy quarks. This is manifest in the presence of a top-pair
threshold which is clearly visible in the $ZH$ invariant mass
distribution shown in the left panel of Fig.~\ref{fig:l2_t_mass_pt}.
Differentially, the gluon-induced corrections are strongly enhanced
with respect to the tree-like contributions for invariant-mass values
around $2m_t$. The magnitudes of both contributions are in fact of the
same order in this region. This effect can be observed already at LHC
energies to some extent \cite{Goncalves:2015mfa}. With the
loop-induced contributions being comparable to the tree-like
contributions in terms of magnitude, it is, however, much more
pronounced at FCC energies.

It should be stressed here that the results for loop-induced processes
presented so far are only leading-order accurate. NLO corrections to
the corresponding partonic channels are expected to be very large as
in the case of single-Higgs production in gluon fusion. In most cases,
NLO corrections have not been calculated for loop-induced processes
due to the considerable complexity of the required two-loop
amplitudes. Critical features of NLO real corrections to such
processes can, however, be captured by means of leading-order multijet
merging with \Sherpa \cite{Cascioli:2013gfa,Goncalves:2015mfa}. For
this purpose, QCD real-emission matrix elements are combined with the
core process and the parton shower in such a way as to correct hard
parton-shower emissions with appropriate matrix elements. For these
hard emissions, the fixed-order matrix-element accurracy is thereby
ensured while maining the logarithmic accurracy of the parton shower
as well. Given the increased energy available to produce additional
hard QCD radiation (cf. section \ref{sec:scaling}), an accurate
description of the corresponding event topologies is of utmost
importance.

\begin{figure}
\centering
\includegraphics[width=.48\textwidth]{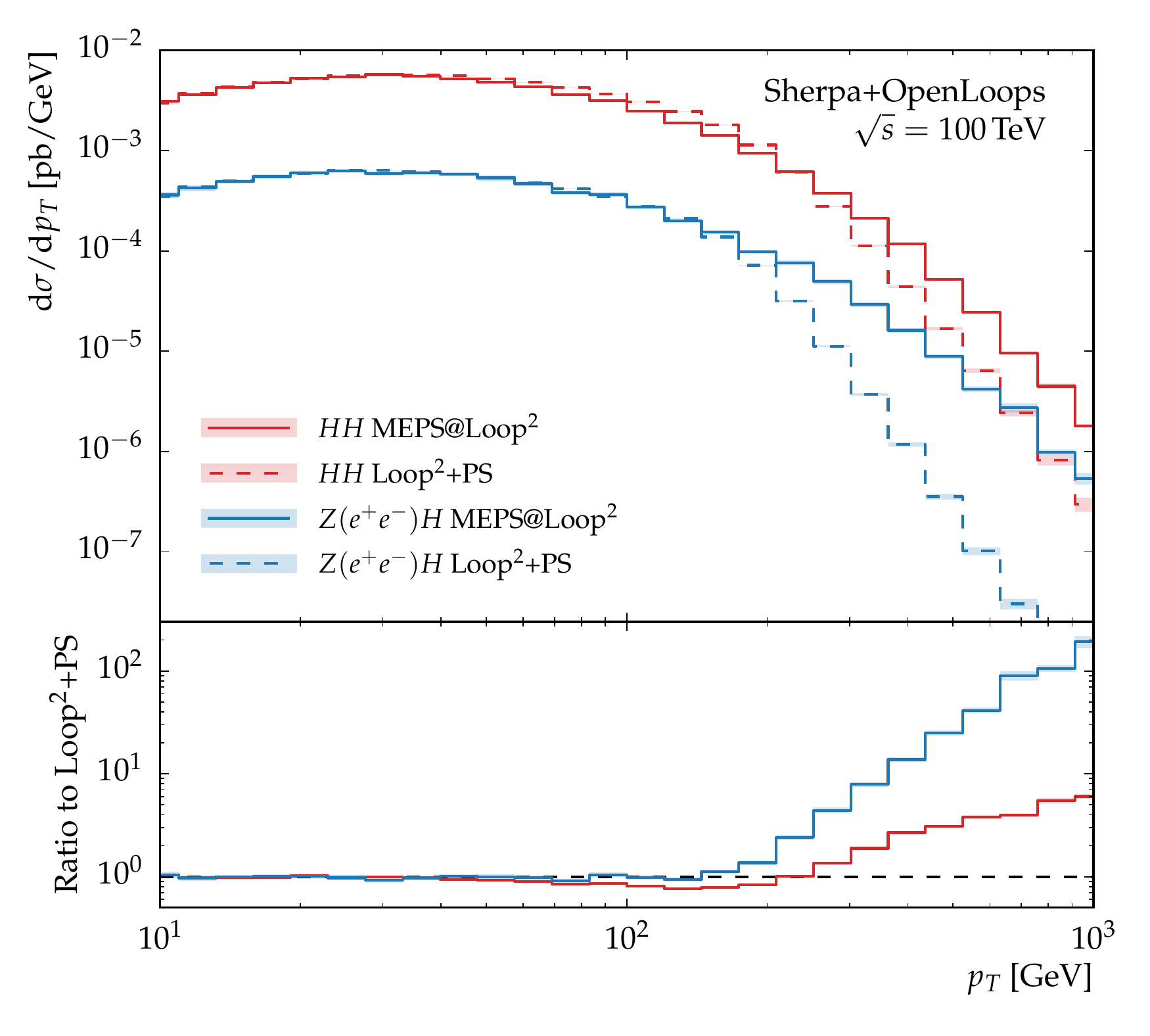}
\hfill
\includegraphics[width=.48\textwidth]{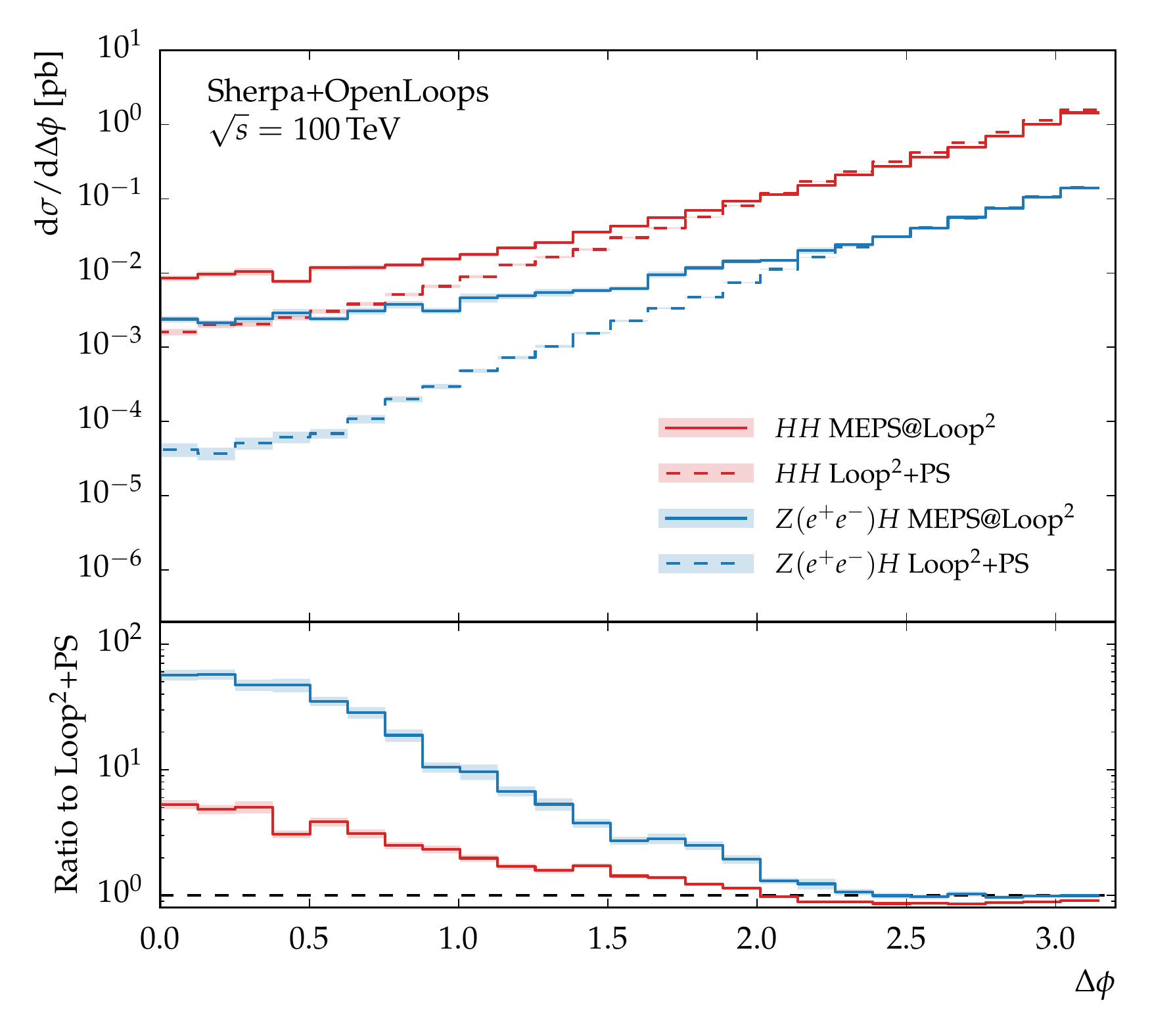}
\caption{Invariant mass of the diboson final state (left) and
  azimuthal separation of the diboson system (right) in Higgs-pair
  production and loop-induced $Z$-associated Higgs production. We show
  results obtained from a merged calculation (solid) and results as
  obtained from a pure parton-shower simulation (dashed) as well as
  the ratio of the two (lower panels).}\label{fig:l2_merging}
\end{figure}

The effects of incorporating higher-multiplicity matrix elements are
particularly notable when considering observables that are generated
entirely by QCD corrections to the leading-order process. As examples,
we consider loop-induced Higgs-pair production and $Z$-associated
Higgs production and calculate the transverse-momentum distribution of
the $HH$ system and the $ZH$ system, respectively.
Fig.~\ref{fig:l2_merging} shows the corresponding results as obtained
from a simulation based on the leading-order matrix element combined
with a parton shower (Loop$^2$+PS) in comparison to a merged
calculation (MEPS@Loop$^2$). In the merged calculation, we take matrix
elements with up to one extra jet in the final state into account. At
leading order, the transverse momentum of the diboson system is zero,
as required by momentum conservation. In the Loop$^2$+PS simulation,
the spectrum is therefore generated purely by the parton shower. For
kinematic configurations that contribute in the tail of this
distribution, the soft/collinear approximation inherent to a parton
shower breaks down, and the corrections one observes when utilising
multijet-merging techniques are correspondingly large, as quantified
in Fig.~\ref{fig:l2_merging}. Also shown in this figure is the
azimuthal separation $\Delta\phi$ of the diboson pair. Similarly to
the transverse-momentum spectrum of the diboson system, a nontrivial
distribution is generated entirely though QCD real-emission
corrections. Any configuration with $\Delta\phi\neq \pi$ requires
additional radiation to be present in the final state giving rise to a
decorrelation of the boson pair in the azimuthal plane. In regions far
away from back-to-back configurations with $\Delta\phi=\pi$, one can
observe large effects induced by the higher multiplicity matrix
elements.

\begin{figure}
\centering
\includegraphics[width=.46\textwidth]{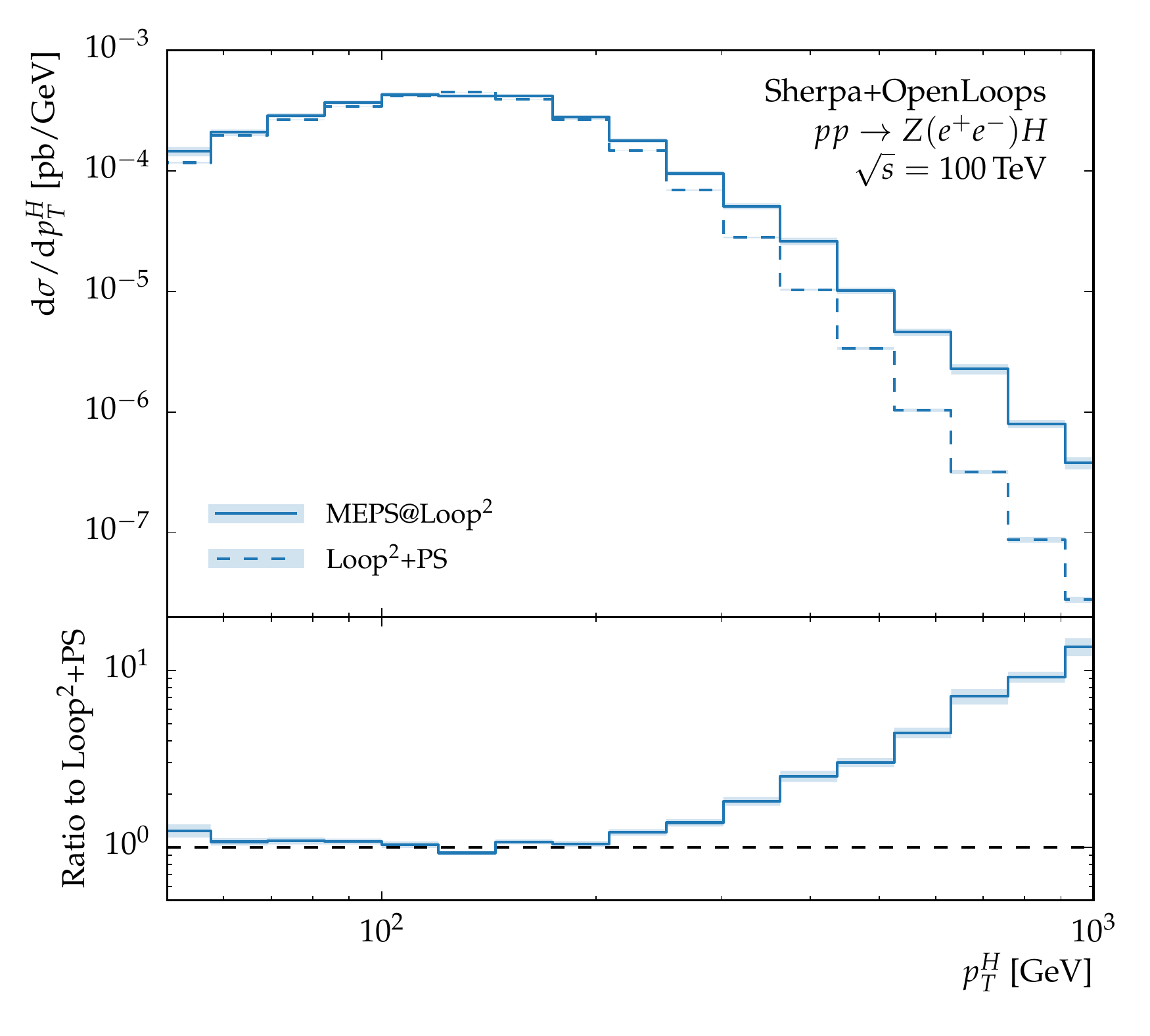}
\caption{Higgs transverse momentum distribution in loop-induced
  $Z$-associated Higgs production.We show results obtained from a
  merged calculation (solid) and results as obtained from a pure
  parton shower simulation (dashed) as well as the ratio of the two
  (lower panel).}\label{fig:l2_ptz}
\end{figure}

Large merging effects can also be observed when considering
observables that receive nontrivial contributions already from 
leading-order matrix elements. As an example, we show the Higgs 
transverse-momentum spectrum in Fig.~\ref{fig:l2_ptz} in $Z$-associated 
Higgs production. In this case, QCD real emission contributes an actual
higher-order correction to the lowest-order result since the Higgs can
recoil against the $Z$ already in a born-like kinematic configuration.
Contrary to naive expectations, these corrections are substantial. In
the tail of the distribution, the Loop$^2$+PS result undershoots the
spectrum by an order of magnitude. This indicates that, in this
regime, the Higgs-boson recoils predominantly against a jet rather
than recoiling against the $Z$ boson, despite the latter being
massive.

In conclusion, we have seen that loop-induced corrections are of
particular relevance at higher collider energies. This was
demonstrated by considering their contributions to very inclusive
event selections. When applying realistic experimental cuts, these
effects can be potentially further enhanced. Furthermore, the
description of QCD radiation patterns featuring hard QCD jet emissions
can be substantially improved using multijet-merging techniques. Due
to the increased partonic energies available at a future hadron
collider, such configurations will contribute to the characteristics
of typical events to a much larger extent than at the LHC.


\clearpage
\section{Backgrounds to searches for New Physics at 100 TeV}
\label{sec:backgrounds}
Phenomenological studies at a \SI{100}{TeV} collider will 
certainly build on the discoveries and analyses performed at 
the LHC machine. This section looks at the effects a \SI{100}{TeV} 
hadron environment has on typical analyses used to search for 
BSM physics. The search channels investigated are multilepton 
final states and monojet production.

Multilepton final states constitute a very useful signature for 
New Physics, because they are relatively rare in the SM, and the higher 
the final state lepton multiplicity the greater the suppression of 
SM backgrounds. At the LHC, these final states are used to search
for SUSY particles, in particular for chargino and neutralino 
production~\cite{Aad:2014nua,Chatrchyan:1482130}.
The dominant SM backgrounds in these searches are multi-$V$ 
production, for $V=W^\pm/Z$, $t\bar{t}$, $tZ$, $t\bar{t}V$ 
and $t\bar{t}VV$. The process classes involving top-quarks 
are particularly relevant for analyses which do not contain a 
$b$-jet veto. In principle also Higgs processes contribute to the 
multilepton background, but these shall not be considered here. 
Instead, the focus of this section is on how weak boson production 
channels as well as top quark physics can impact searches at 
\SI{100}{TeV}. Current experimental studies at the LHC can be 
easily extended to a \SI{100}{TeV} environment, by extending the 
reach in transverse momenta, missing energy and jet multiplicities.

In contrast to the multilepton analysis, monojet searches
veto all events with any leptons or multiple hard jets. The 
remaining SM processes which can contribute to this final 
state are then very few, creating a good environment for 
BSM searches. Any new, weakly interacting (meta-)stable 
particle created will leave the detector as missing energy, 
which then forms the signal for this type of search. Because 
of the veto on additional hard QCD radiation, monojet searches 
are vulnerable at \SI{100}{TeV} to cutting out large regions 
of phase-space, especially when considering the high energy 
regions. The analyses in this publication are based loosely 
on the monojet studies carried out at the LHC by both the 
ATLAS~\cite{Aad:2011xw,ATLAS:2012ky,ATLAS-CONF-2011-096,ATLAS-CONF-2012-147}
and CMS~\cite{Chatrchyan:2011nd,Chatrchyan:2012me,Khachatryan:2014rra} 
experiments, which search for hints of dark matter or extra 
dimensions. The dominant irreducible background in these searches 
is $Z\rightarrow\nu\overline{\nu}$. As well as this, there are
significant contributions from processes where leptons have been
lost in the analysis. These are $W^\pm$,
$Z\rightarrow\ell^+\ell^-$ and $t\bar{t}$ production. Two Higgs
production channels are also considered in this analysis, gluon
fusion and VBF, where in both cases the Higgs is considered to
decay to invisibles. The top-mass in the gluon fusion production
channel are considered by a reweighting of the Born matrix 
element to include the top loop~\cite{Buschmann:2014sia}, with \OpenLoops contributing
the virtual matrix  element.

\subsubsection{Set Up}
\label{SUBSEC:NewPhysicssetup}
The distributions presented in this section are based
on matrix-element plus parton-shower simulations and
do not include underlying event or hadronisation 
effects. Most processes are considered at LO
merged accuracy. The multilepton analysis includes,
for the $t\bar{t}$, $W^\pm$, $Z$ and diboson processes, the 
leading matrix element to NLO accuracy with LO matrix 
elements of higher multiplicities included via the \MENLOPS 
procedure~\cite{Hoeche:2010kg}. The renormalisation and factorisation 
scales for all processes considered are set using the CKKW 
prescription~\cite{Hoeche:2009rj,Hoeche:2012yf}, and the 
\Comix matrix element generator~\cite{Gleisberg:2008fv} was 
used for all LO calculations as well as the real 
subtraction piece of the NLO calculations. However,
for the Born-like contributions to the NLO calculation and 
the integrated subtraction terms, the \Amegic matrix element 
generator~\cite{Krauss:2001iv,Gleisberg:2007md} was employed 
with virtual matrix elements calculated through the interface to 
\OpenLoops~\cite{Cascioli:2011va}.

For the multilepton final state, single vector-boson 
production is considered off-shell, as are the 
bosons in $t\bar{t}V$ production. However, in both the 
monojet analysis and for processes with 
two or more weak bosons, the narrow width approximation is 
used and the decays are factorised from the 
production. The kinematics of the decay are then redistributed 
according to a Breit-Wigner distribution, with spin 
correlations being preserved~\cite{Hoche:2014kca}.

\subsubsection{Analyses}

The multilepton analysis is inspired by the ATLAS 
publication on trilepton searches at the
LHC at \SI{8}{TeV}~\cite{Aad:2014nua}. The analysis has been 
extended to include the single lepton channel as
well as the dilepton channel, and all lepton multiplicities 
are considered exclusive. These analyses are implemented 
within the \Rivet~\cite{Buckley:2010ar} framework.

The multilepton analysis shall provide an overview of the 
behaviour of SM multilepton processes at \SI{100}{TeV},
and the cuts are therefore kept to a minimum.
Electrons are dressed with photons within a cone of radius 
$R=0.1$ around the electron. Leptons
are required to be central, such that $|\eta_\ell|<2.5$. Jets 
are defined with the anti-$k_T$ method
with \FastJet~\cite{Cacciari:2011ma}, with $p_T>\SI{20}{GeV}$
and $|y_j|<5$. Jets which pass the jet definition
are further required to be isolated from electrons by at 
least a distance $\Delta R_{\ell j}>0.2$. Any jet
too close to an electron is rejected. Leptons are then 
required to be isolated from other leptons.
Any two leptons closer than $\Delta R_{\ell\ell}<0.1$ are 
both rejected, with the exception of two electrons,
when the harder of the pair only is retained.
For the $Z$-veto bin, any event with a pair of 
same-flavour-opposite-sign (SFOS) leptons is vetoed. 

For the monojet analysis, the cuts are based very loosely
on corresponding ATLAS and CMS studies at the LHC. One 
challenge for monojet analyses, which becomes more pronounced in the
high-energy regime that will be common at \SI{100}{TeV}, 
is the additional QCD radiation that often accompanies hard
interactions. In order to study the effect this has on
the analyses, two different regions shall be considered. 
The key difference between the two is that the first analysis, 
referred to as ``$p_T$ gap'' selection, places a strict veto on any 
secondary jet with $p_T>\SI{50}{GeV}$, but allows a softer 
second jet while the second analysis, called ``$\Delta\Phi$'' 
selection, allows one additional jet provided it satisfies 
cuts that target the dijet background. Both approaches are 
implemented for this study, along with a variant of 
the CMS analysis more in line with typical energies for 
a \SI{100}{TeV} collider. Any event with leptonic activity 
is vetoed for this analysis. Jets are defined with the 
anti-$k_T$ method with a radius parameter $R=0.4$, 
$p_T>\SI{20}{GeV}$ and $|y_j|<5$. The cuts defining the 
different regions are summarised in Tab.~\ref{TAB:missetRegions}.

\begin{table}
  \begin{tabular}{l||c|c}
  \hline
  \hline
  Observable & \;``$\Delta\Phi$'' selection\; & \;``$p_T$ gap'' selection\; \\
  \hline
  leading jet $p_T$         & \SI{0.1}{TeV}        & \SI{0.150}{TeV}         \\
  max. $p_T$ subleading jet & N/A        & \SI{50}{GeV}          \\
  min. missing $E_T$        & \SI{0.1}{TeV}        & \SI{0.1}{TeV}         \\
  max $\Delta\Phi(j_1,j_2)$ & $2$          & N/A         \\
  \hline
  \hline
  \end{tabular}
  \caption{\label{TAB:missetRegions}
  The defining cuts for the two phase-space regions considered in the monojet 
  analysis.}
\end{table}

\subsubsection{Multilepton analyses}

\begin{figure}[t]
  \begin{center}
   \setlength{\unitlength}{1cm}
     \includegraphics[width=6.8cm]{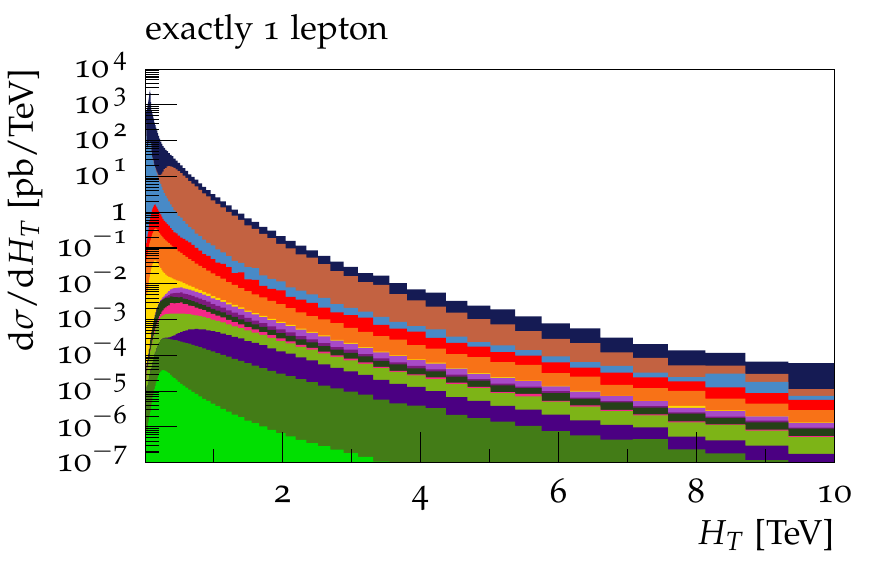}
   \includegraphics[width=6.8cm]{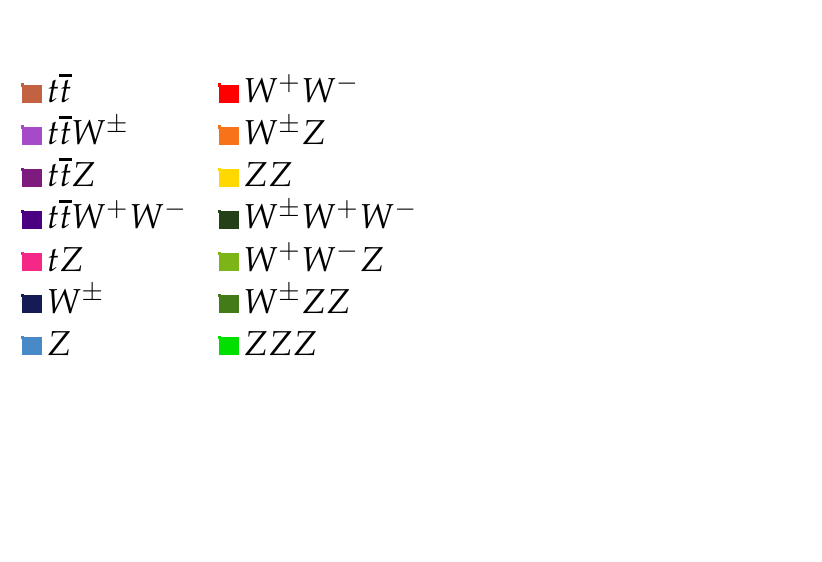}\\
     \includegraphics[width=6.8cm]{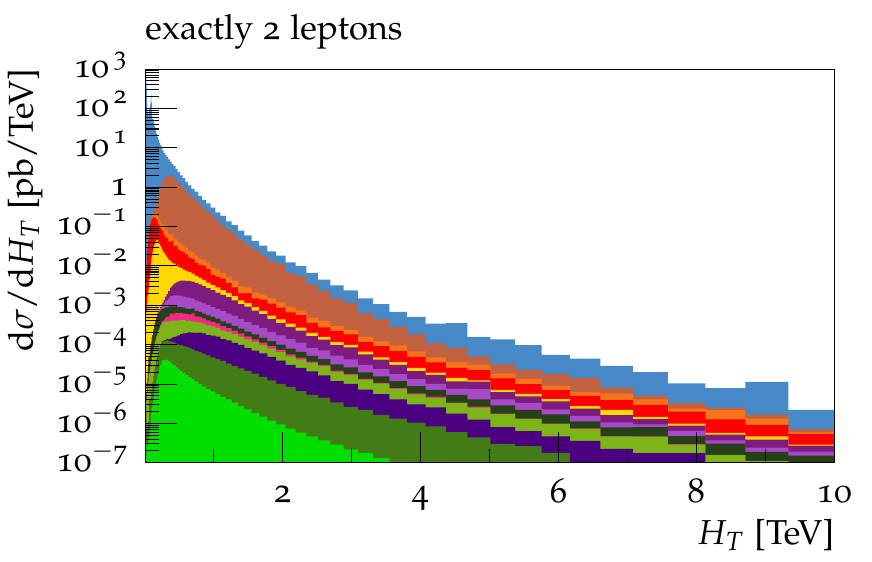}
     \includegraphics[width=6.8cm]{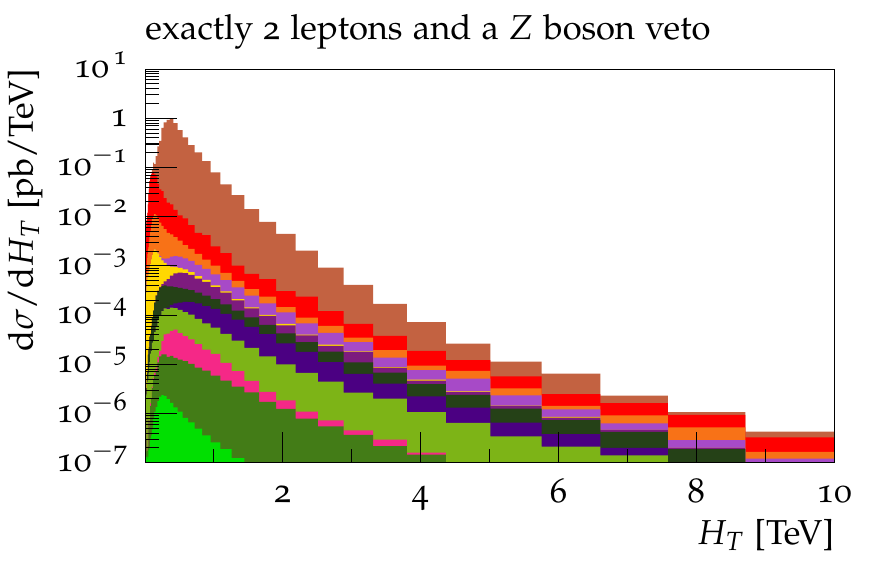}\\
     \includegraphics[width=6.8cm]{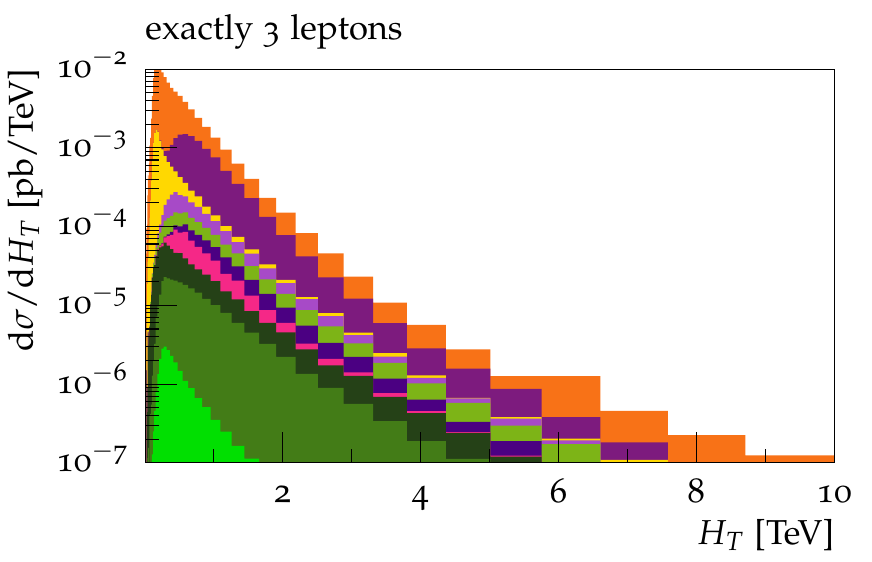}
     \includegraphics[width=6.8cm]{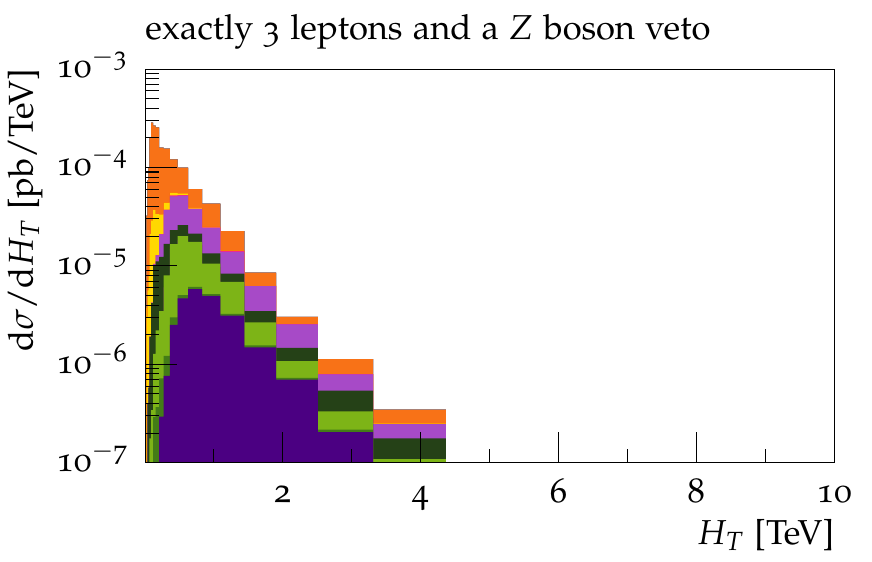}
    \caption{\label{fig:HTmultilep}The $H_T$ distributions for 
   various SM backgrounds in multilepton searches, depicted 
   for different lepton multiplicities. The right-hand plots
   on the lower two rows show the effect of including a veto
   on SFOS lepton pairs.
   The distributions are merged with up to three additional 
   light QCD partons for $V$ and $VV$ production, and up to 
   two additional light partons for all other processes 
   with a merging scale of $Q_\text{cut}=30$~GeV.}
  \end{center}
\end{figure}

\begin{figure}
  \begin{center}
   \setlength{\unitlength}{1cm}
     \includegraphics[width=6.8cm]{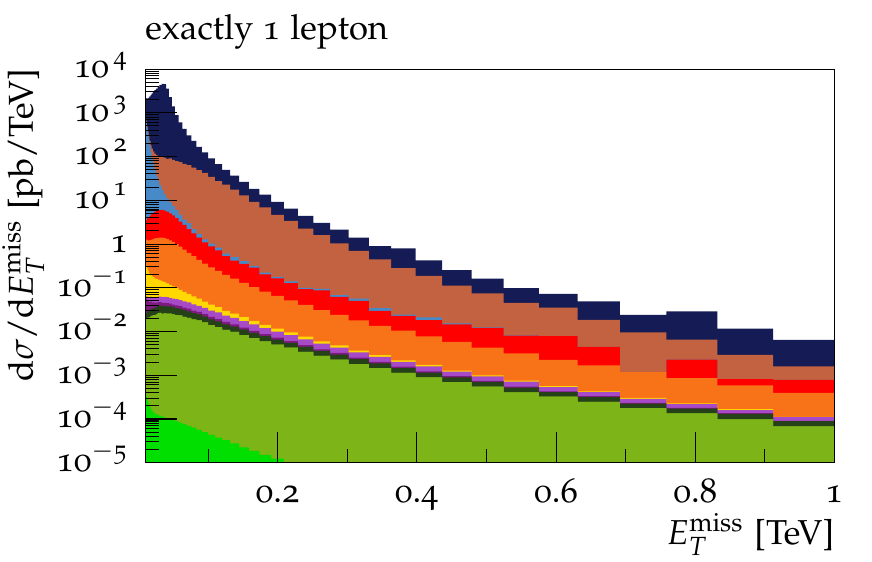}
     \includegraphics[width=6.8cm]{legend}\\
     \includegraphics[width=6.8cm]{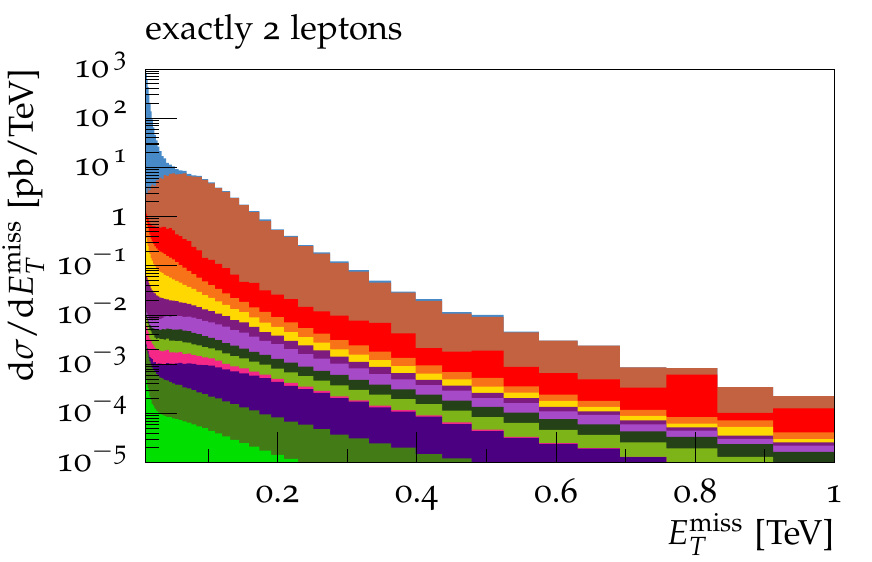}
     \includegraphics[width=6.8cm]{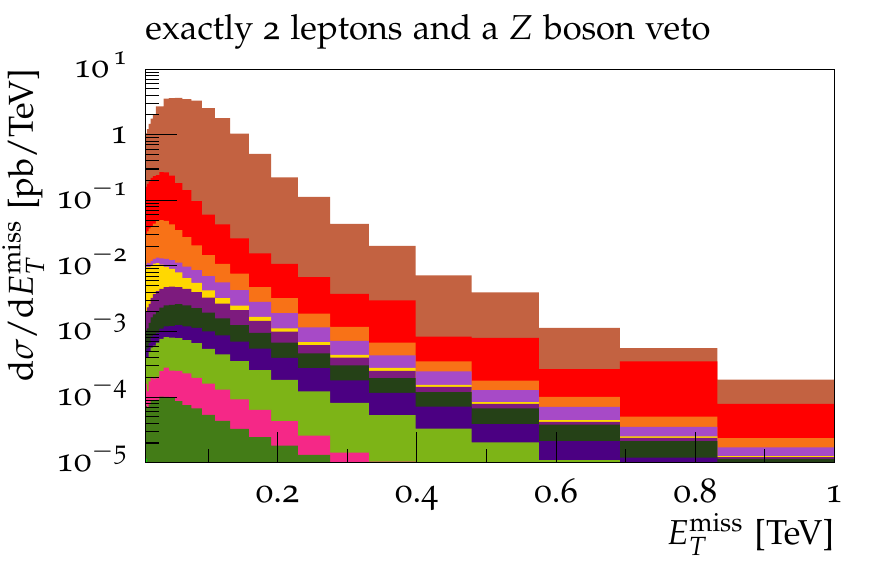}\\
     \includegraphics[width=6.8cm]{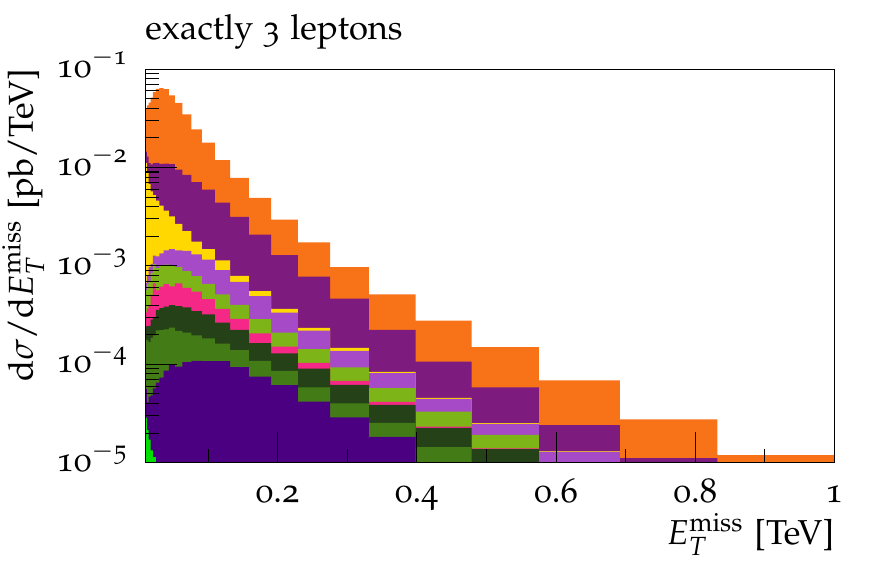}
     \includegraphics[width=6.8cm]{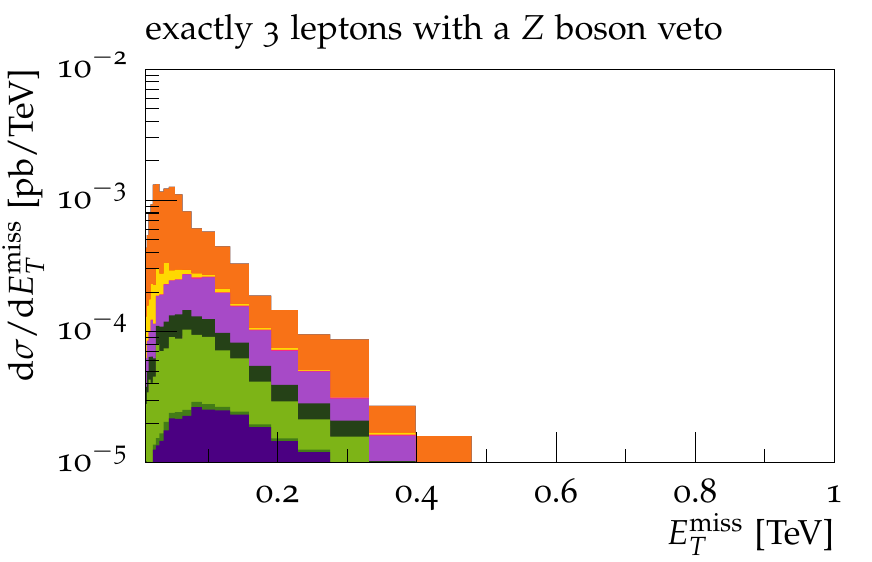}
    \caption{\label{fig:ETmissmultilep}$E_T^\text{miss}$ 
   distributions for various SM backgrounds in multilepton 
   searches, depicted for different lepton multiplicities. 
   The right-hand plots on the lower two rows show the
   effect of including a veto on SFOS lepton pairs.
   Calculational details as quoted in Fig.~\ref{fig:HTmultilep}.}
  \end{center}
\end{figure}

The $H_T$ distribution, i.e. the scalar sum of the transverse momenta of all
final-state objects in the event, including the missing $E_T$, is shown in
Fig.~\ref{fig:HTmultilep}, binned in terms of the exclusive 
lepton multiplicity. This observable provides an overview of the
relative contributions of the different processes, with 
$t\bar{t}$  dominating the one and two lepton cases and $W^\pm 
Z$ dominating the three lepton final state. Comparing the two 
plots on the middle and lower lines of Fig.~\ref{fig:HTmultilep} shows 
the effect of introducing the $Z$-boson veto. This addition 
clearly suppresses the $Z$-boson production processes in both
the two-lepton and the three-lepton channels, 
and thus significantly reduces the overall size of the 
SM background in this search channel. However, in the 
missing $E_T$ distribution shown in Fig.~\ref{fig:ETmissmultilep}, 
this veto does not have such a dramatic effect in the two-lepton
channel as in the three-lepton channel. Comparing the left-hand 
and right-hand plots of the middle (two-lepton) line as before,  
Fig.~\ref{fig:ETmissmultilep} shows a significant suppression 
of the background in the low-missing $E_T$ region, but in the 
high-energy tail of the distribution, $t\bar{t}$ production is 
dominant in the dilepton bin even before the $Z$-boson veto is 
applied, and as such there is only a small impact on the high-energy 
tail of the background processes for the missing $E_T$ 
distribution due to the $Z$-boson veto. Because the 
$t\bar{t}$ process does not contribute to the three-lepton channel,
the $Z$-boson veto has a similar effect on the missing $E_T$
distribution as it had on the $H_T$ distribution. Despite the
$Z$-boson veto, $W^\pm Z$ production continues to be the dominant
background, with $t\bar{t}W^\pm$ contributing significantly in the
higher $H_T$ or missing $E_T$ tails of the distribution. It is therefore
important in trilepton final state searches with a $Z$-boson veto to
have good control of the $t\bar{t}W^\pm$ process, at least to NLO
precision~\cite{Hirschi:2011pa,Garzelli:2012bn,Campbell:2012dh,Maltoni:2014zpa}.

\clearpage
Fig.~\ref{fig:HTmultilep} demonstrates the very large reach 
in scales the \SI{100}{TeV} FCC provides. The tails of the 
$H_T$ distributions easily stretch out to \SI{10}{TeV}. 
However, it can also be seen that the $H_T$ distribution 
becomes increasingly soft the higher the lepton multiplicity. 
Nevertheless, Fig.~\ref{fig:HTmultilep} impressively 
demonstrates the importance to have control of these 
background SM processes in the multi-TeV regime and that 
indeed many processes need to be taken into account.

Fig.~\ref{fig:ETmissmultilep} shows the missing energy 
distribution for the considered set of processes. This
observable is particularly relevant for BSM searches, where 
the signal expected is an excess at large missing energy. 
At a \SI{100}{TeV} proton-proton collider, 
Fig.~\ref{fig:ETmissmultilep} shows a significant
amount of SM background is present at these large missing 
$E_T$ values. For the three-lepton final state the dominant 
background is given by $W^\pm Z$ production. For the 
two-lepton bin with a $Z$~veto implemented, $t\bar{t}$ is the 
dominant process for most of the distribution. However, at 
high $E_T^\text{miss}$, $W^+W^-$ is also significant.
It is therefore crucial to have a very good description 
of both $t\bar{t}$ and $W^+W^-$ in these phase-space regions 
in order to claim sensitivity for New Physics at this scale.
In fact, for both channels the complete set of NNLO QCD corrections
is known~\cite{Czakon:2013goa,Gehrmann:2014fva}. For the low 
$E_T^\text{miss}$ regime at around \SI{50}{GeV} for the two-lepton 
bin with no $Z$-boson veto, both $Z$ production and $t\bar{t}$ 
production are significant, although $t\bar{t}$ still dominates 
in the tail of the distribution. For the one-lepton bin, for a 
large amount of the distribution both $W^\pm$ and $t\bar{t}$ 
production are significant.

The suppression of the SM backgrounds with increasing 
lepton multiplicity can be clearly seen in 
Figs.~\ref{fig:HTmultilep} and \ref{fig:ETmissmultilep}. 
However, even for the trilepton final state
the SM processes can reach to $\order(1\,\text{TeV})$ in 
$E_T^{\text{miss}}$ and the $H_T$ distribution can
reach very high values, up to $\order(10\,\text{TeV})$.

\subsubsection{Monojet searches}

\begin{figure}
  \begin{center}
   \setlength{\unitlength}{1cm}
     \includegraphics[width=6.8cm]{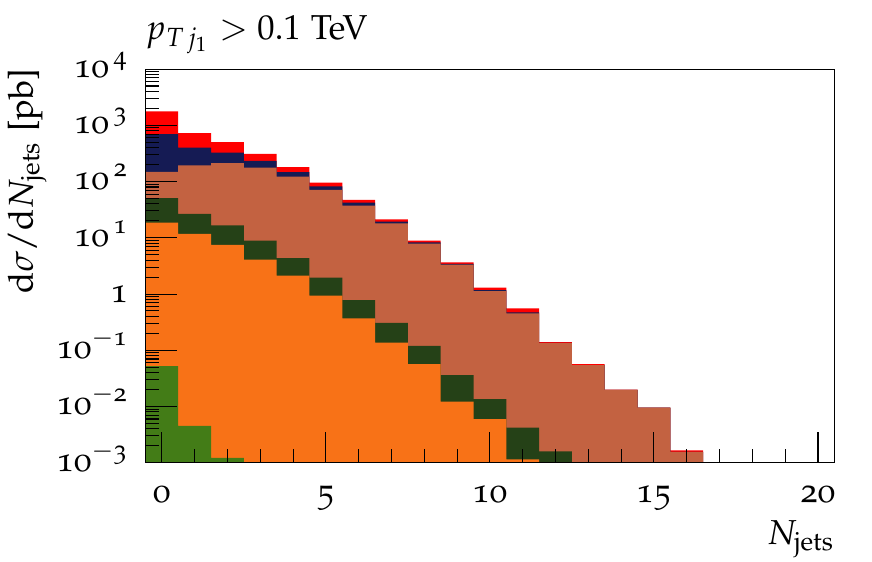}
     \includegraphics[width=6.8cm]{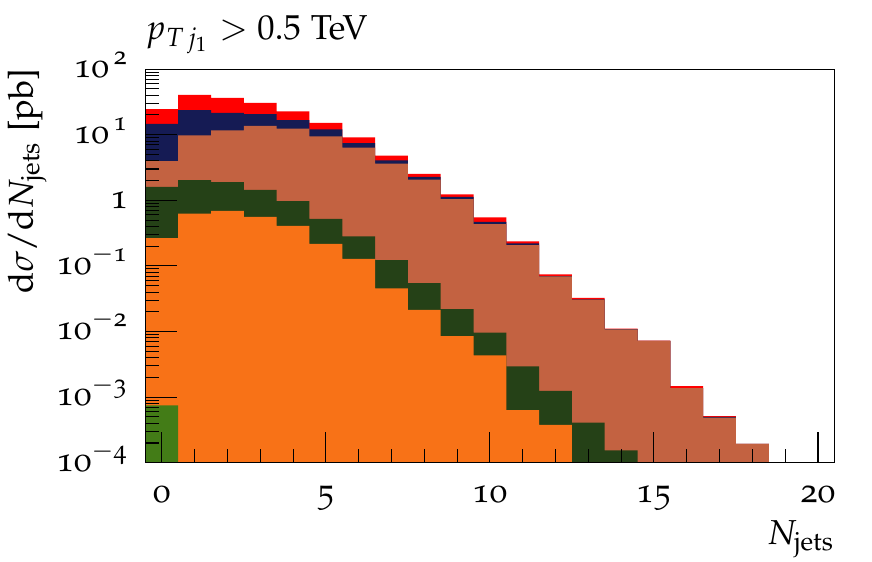}\\
     \includegraphics[width=6.8cm]{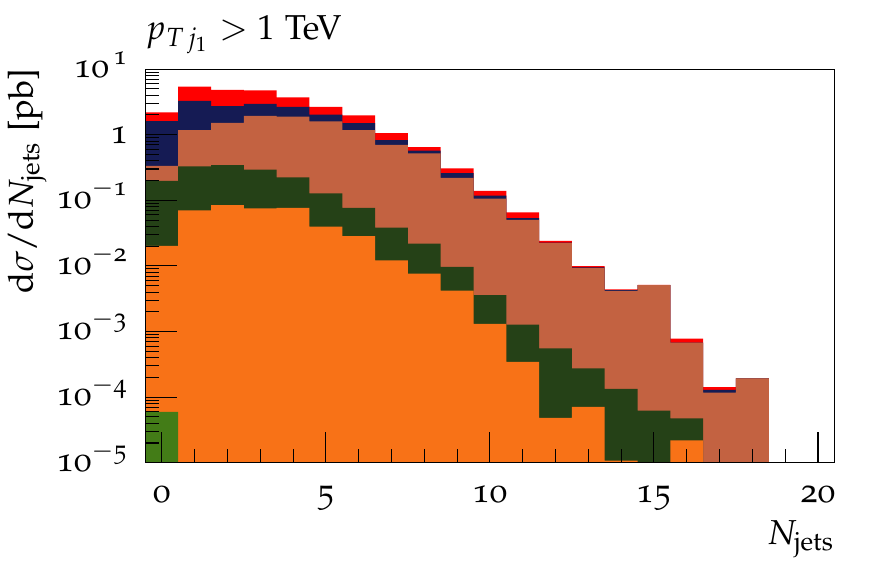}
     \includegraphics[width=6.8cm]{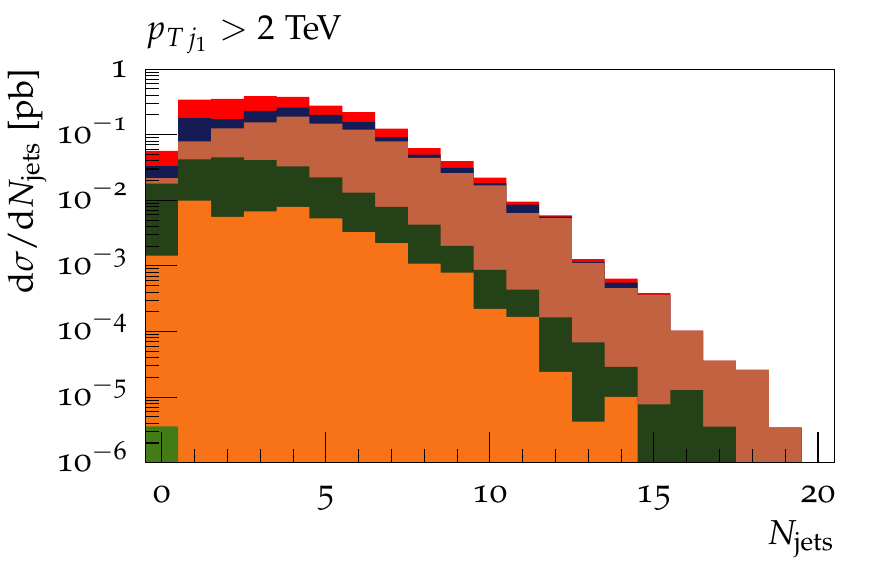}
   \begin{picture}(15,0)
     \put(13.2,2.3){\includegraphics[width=4cm]{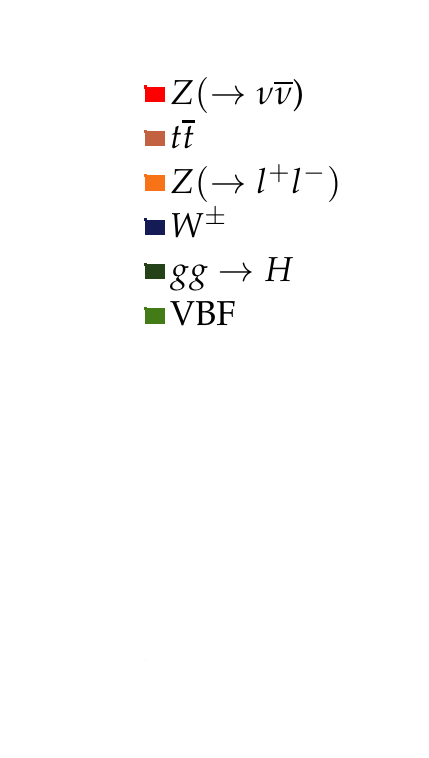}}
   \end{picture}
   \end{center}
\caption{\label{FIG:misset_ptgap}Number of jet candidates 
vetoed for the ``$\Delta\Phi$'' selection binned by leading jet 
transverse momentum. The top-left (top-right) plot shows the 
result for $p_{T\,j_1}>\SI{0.1}{TeV}$ ($p_{T\,j_1}>\SI{0.5}{TeV}$), 
the bottom-left (bottom-right) for the high $p_T$ region, i.e. 
$p_{T\,j_1}>\SI{1}{TeV}$ ($p_{T\,j_1}>\SI{2}{TeV}$). }
\end{figure}

\begin{figure}
  \begin{center}
   \setlength{\unitlength}{1cm}
     \includegraphics[width=6.8cm]{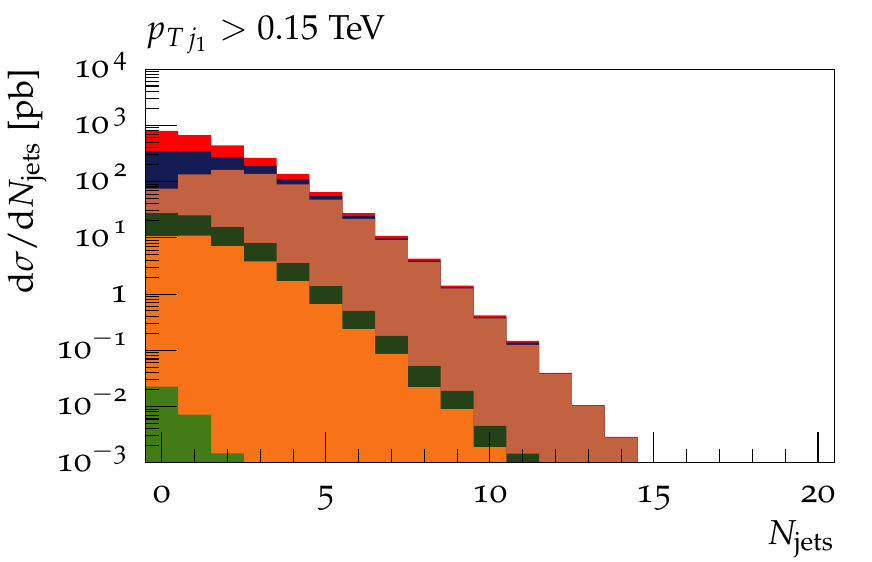}
     \includegraphics[width=6.8cm]{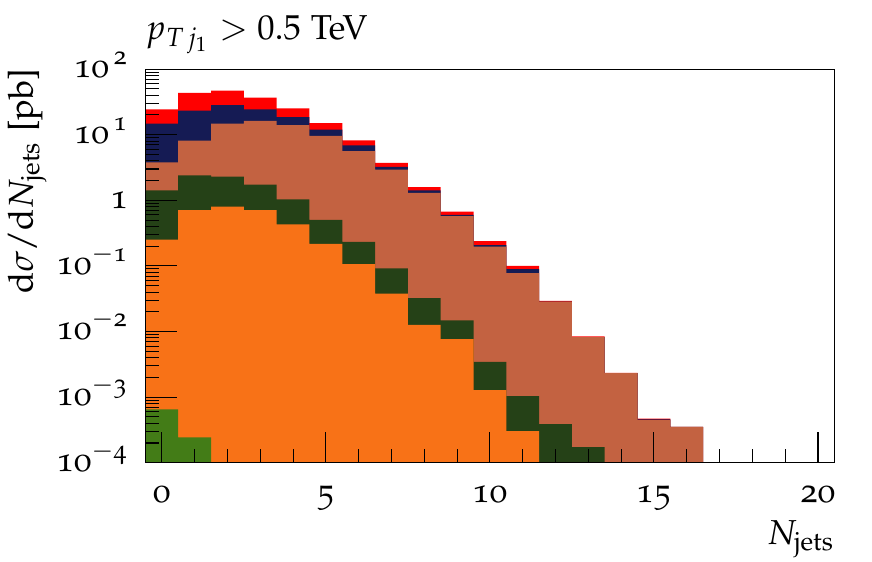}\\
     \includegraphics[width=6.8cm]{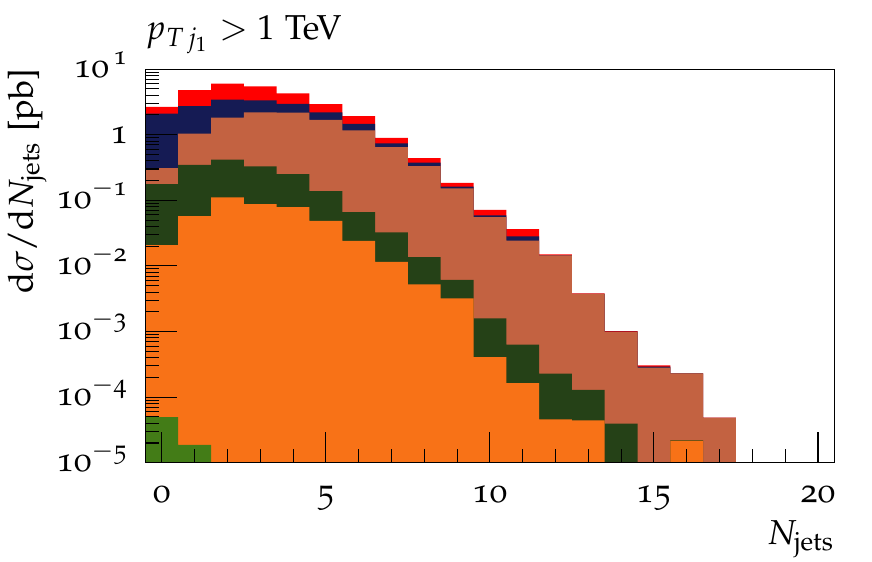}
     \includegraphics[width=6.8cm]{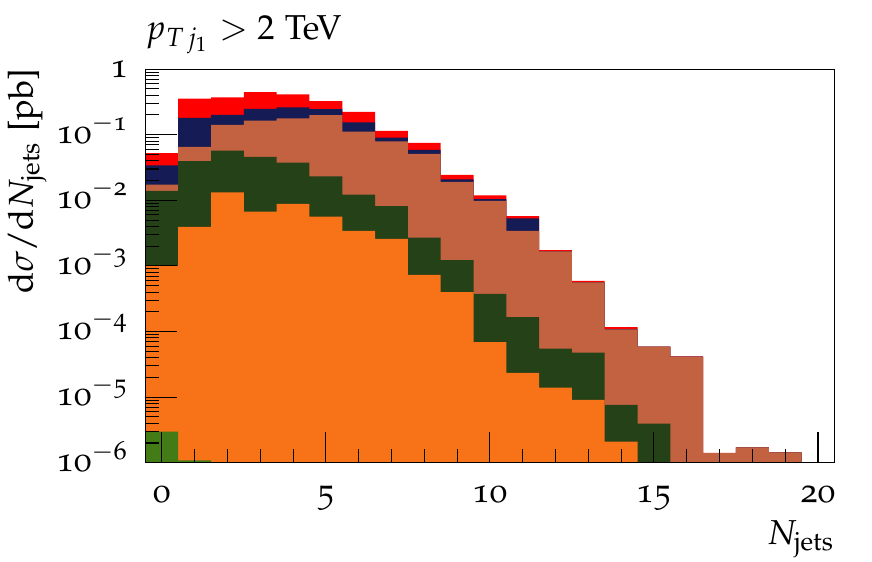}
   \begin{picture}(15,0)
     \put(13.2,2.3){\includegraphics[width=4cm]{legendmisset}}
   \end{picture}
   \end{center}
 \caption{\label{FIG:misset_ptgap50}Number of jet 
candidates vetoed for the ``$p_T$ gap'' selection binned 
by leading jet $p_T$. The top-left plot shows the result 
for $p_{T\,j_1}>\SI{0.15}{TeV}$, the top-right for 
$p_{T\,j_1}>\SI{0.5}{TeV}$ and the bottom-left (bottom-right) 
for $p_{T\,j_1}>\SI{1}{TeV}$ ($p_{T\,j_1}>\SI{2}{TeV}$).}
\end{figure}

\begin{figure}
 \begin{center}
 	\includegraphics[width=0.4\textwidth]{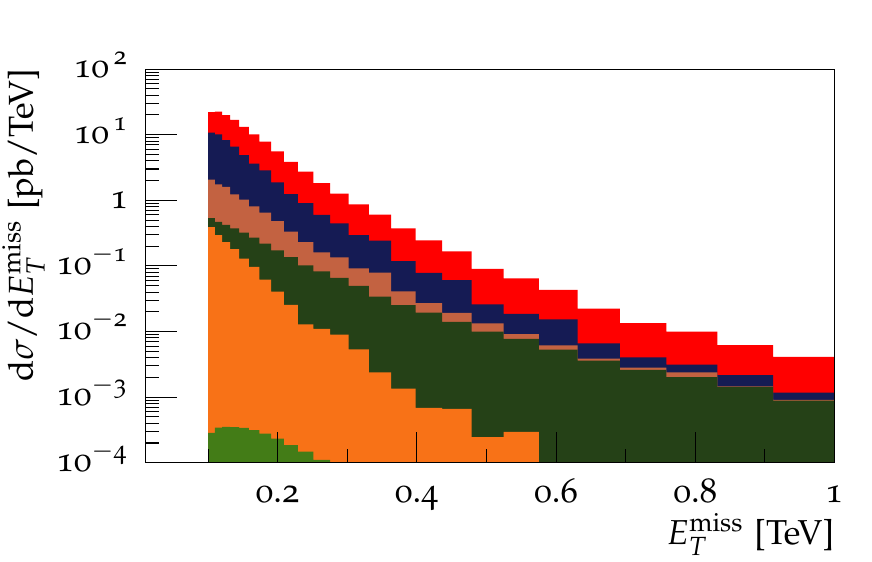}
 	\includegraphics[width=0.4\textwidth]{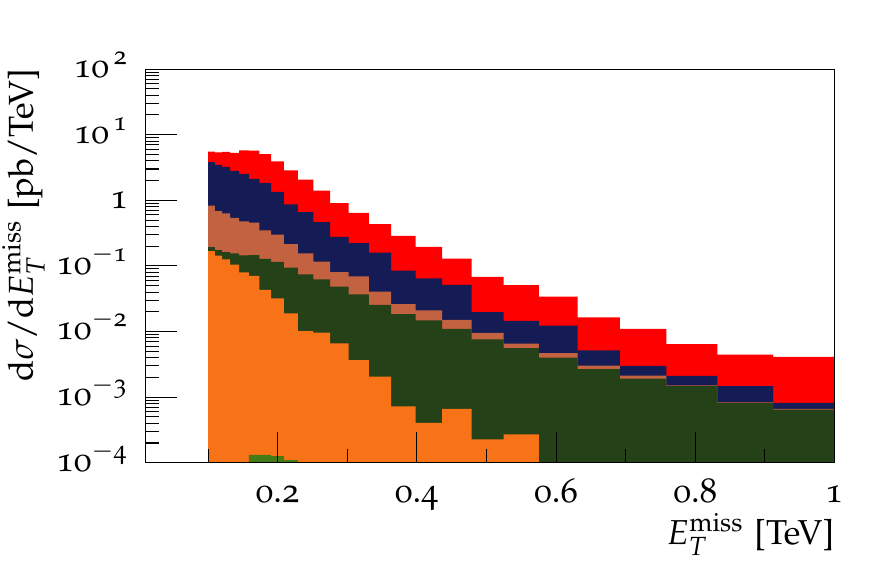}
 	\begin{picture}(15,0)
 	  \put(-30,-35){\includegraphics[width=0.18\textwidth]{legendmisset}}
 	\end{picture}
 	\caption{\label{FIG:misset_monojet} The missing energy distribution
 	for the ``$\Delta\Phi$'' selection (left panel), and for the
 	``$p_T$ gap'' selection (right panel).}
 \end{center}
\end{figure}

\begin{figure}
 \begin{center}
 	\includegraphics[width=0.4\textwidth]{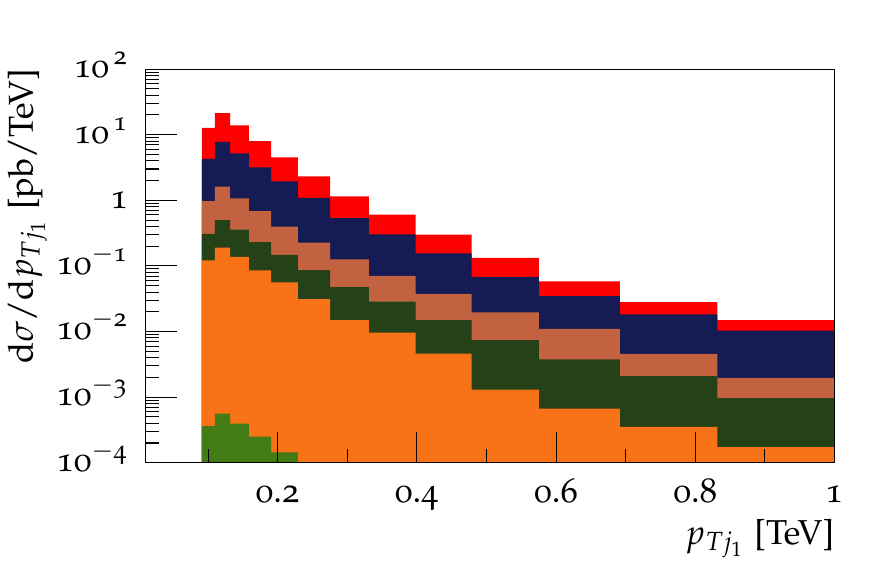}
 	\includegraphics[width=0.4\textwidth]{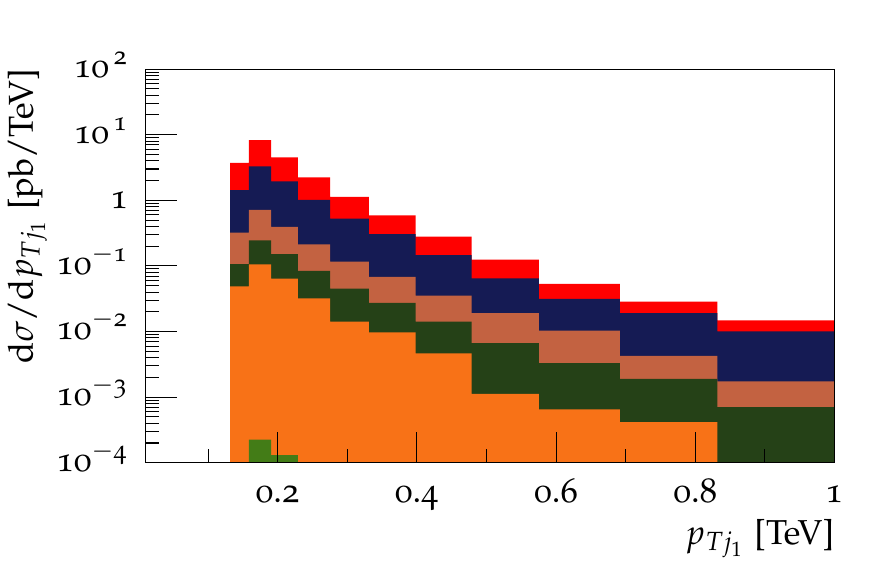}
 	\begin{picture}(15,0)
 	  \put(-30,-35){\includegraphics[width=0.18\textwidth]{legendmisset}}
 	\end{picture}
 	\caption{\label{FIG:pTlj_monojet} The leading jet $p_T$ distribution
 	for the ``$\Delta\Phi$'' selection (left panel), and for the
 	``$p_T$ gap'' selection (right panel).}
 \end{center}
\end{figure}

The current approaches to monojet studies, as outlined in the introduction, allow for
additional QCD radiation, although to different extents. This is necessary in order 
to increase the overall signal yield, and becomes increasingly important the higher 
the collider energy. Therefore, in order for monojet searches to take advantage of 
the increased energy of collisions at a \SI{100}{TeV} collider, they must be 
sufficiently inclusive to this additional QCD radiation.
Figs.~\ref{FIG:misset_ptgap} and \ref{FIG:misset_ptgap50} show the number of
jets in each event which pass all cuts but trigger the veto on QCD activity.
Accordingly, the zero-jet bin containes the total cross section of events
passing the analysis cuts. The results are given for different values of the
minimum leading jet $p_T$, and show how the average number of jets in the events
increases with increasing jet $p_T$. Fig.~\ref{FIG:misset_ptgap} shows the
results for the ``$\Delta\Phi$'' region, where a subleading jet above
$p_T=\SI{30}{GeV}$ with $\Delta\Phi_{j_1j_2}<2$ is allowed.
Fig.~\ref{FIG:misset_ptgap50} shows the results for the ``$p_T$ gap'' region,
which requires any subleading jet to have $p_T<\SI{50}{GeV}$.
Both approaches show that at low values of leading jet $p_T$, there
is a good proportion of events that pass the cuts on QCD radiation. However, as the 
leading jet becomes harder, an increasing amount of events is vetoed. 
In all of the plots, the $t\bar{t}$ background is the most suppressed by the QCD radiation
veto in the high-$p_T$ region, while the irreducible background of $Z\rightarrow\nu\overline{\nu}$
is the most dominant contribution that passes the cuts, as well as in the bins of lower
jet multiplicity.

For leading jet $p_T>\SI{0.1}{TeV}$, the average number of veto-jets in the 
``$\Delta\Phi$'' selection, cf. Fig.~\ref{FIG:misset_ptgap},  is \SI{1.206(5)}, 
compared to \SI{1.454(7)} for the ``$p_T$ gap'' selection, cf. 
Fig.~\ref{FIG:misset_ptgap50}. The corresponding jet-veto probabilities are 
$52\%$ (``$\Delta\Phi$'' selection) compared to $67\%$ (``$p_T$ gap'' selection). 
However, for $p_{T\,j_1}>\SI{0.5}{TeV}$ and above, the two different 
approaches give very similar distributions. In the highest $p_T$ region
considered here, $p_{T\,j_1}>\SI{2}{TeV}$, there is an average number of vetoed jets
of \SI{3.8(1)} (\SI{3.7(1)}) resulting in jet-veto probability of 
$99.7\%$ ($99.8\%$) for the ``$\Delta\Phi$'' (``$p_T$ gap'') analysis. Neither 
approach appears to be better suited to dealing with the typical high energies of 
a \SI{100}{TeV} collider. This implies that monojet searches as they have been 
implemented at the LHC would not be very sensitive to searches in the high $p_T$ 
phase-space regions, a more considered treatment of high levels of QCD activity 
is necessary for a monojet search at \SI{100}{TeV}.

Figs.~\ref{FIG:misset_monojet} and \ref{FIG:pTlj_monojet} show the $E_T^{\text{miss}}$ 
and leading jet $p_T$ distributions, respectively, for both the ``$\Delta\Phi$'' 
and the ``$p_T$ gap'' selections. Fig.~\ref{FIG:misset_monojet} shows very similar 
behaviour of the $E_T^{\text{miss}}$ distribution between the two regimes, beyond the 
low-$p_T$ effects from the higher leading jet $p_T$ cut in the ``$p_T$ gap'' analysis. 
Here the $Z\rightarrow\nu\overline{\nu}$ process is dominant throughout the
distribution, with $W^\pm$ production being the leading subdominant background. The 
$t\bar{t}$ process is most significant in the lower $E_T^{\text{miss}}$ bins, and does not 
contribute significantly in the tail of the distribution. Instead, the $gg\rightarrow H$ 
process becomes the most dominant subleading background at large $E_T^{\text{miss}}$.

The hierarchy of processes is similar in Fig.~\ref{FIG:pTlj_monojet} when comparing
to Fig.~\ref{FIG:misset_monojet}. The $Z\rightarrow \nu\overline{\nu}$ process is dominant 
throughout the distribution, and $W^\pm$ production is the leading subdominant contribution. 
In contrast to the $E_T^{\text{miss}}$~distribution, however, the relative contributions of 
the SM processes to the leading jet $p_T$ distribution remain more constant in the 
high-energy tail. The $t\bar{t}$ process does not disappear in the high-$p_T$ tail of 
the distribution as happens in Fig.~\ref{FIG:misset_monojet}, and similarly the 
$gg\rightarrow H$ production does not become the subleading background at large leading 
jet $p_T$. Figs.~\ref{FIG:misset_monojet} and \ref{FIG:pTlj_monojet}, show the
``$\Delta\Phi$'' and ``$p_T$ gap'' approaches leading to distributions with similar reaches in
the $E_T^{\text{miss}}$ and $p_{T\,j_1}$ distributions. Even with the large suppression 
from the veto on multiple QCD radiation, the monojet analysis could probe energies up 
to \SI{1}{TeV} at a \SI{100}{TeV} hadron collider, a reach which could be extended with 
a dedicated study into the relevant phase-space cuts for such a high-energy environment. 
A \SI{100}{TeV} collider therefore provides an excellent searching ground for New Physics, 
which could be explored with monojet analyses based on those performed at the LHC.


\section{Conclusions}
\label{sec:conclusions}
This paper presented results demonstrating the range of substantial challenges
a future \SI{100}{TeV} hadron collider will confront theory with.

In the first half of this publication we employed techniques capturing
dominant scaling patterns to extrapolate known fixed-order results for
multi-jet production in different kinematic regimes to higher multiplicities
or different regimes.  In so doing we identified regimes typical for the
emergence of Poisson scaling and Staircase scaling, with good agreement
between scaling-based predictions and simulation results in both cases.
It also addressed jet-substructure techniques at \SI{100}{TeV}.  This is
interesting from the point of view of quark--gluon discrimination, as well
as identifying the fat-jets of highly boosted particles, such as top quarks
and $W^\pm$-bosons.  Because the centre-of-mass energy is so large, the
possibility of having a highly boosted heavy particle is much larger than
at LHC energies.  The number of subjets was found to provide a sensible probe
to the initiating particle's colour charge.  It was also established that
the parton shower is in good agreement with analytical results, to NDLA
accuracy. 

At high energies, the proton begins to be probed at smaller and smaller
values of Bjorken $x$.  The gluon PDF becomes more dominant in this region,
and it was shown in this paper that the loop-induced processes for 
vector-boson production and Higgs-boson production become more significant 
in this regime.  As is well-known, HEFT is not a reliable approximation in the
high-$p_T$ tail or high-mass regime of Higgs-boson production through gluon
fusion due to top-mass effects.  Their effect becomes more relevant at a 
\SI{100}{TeV} machine, due to the increased phase space it provides.  The final
section in this paper looked at the effects the \SI{100}{TeV} environment would
have on some typical LHC analyses for BSM searches. It was seen that the
reach of these searches can extend much further in $p_T$, and that while the
multi-lepton analyses are already well suited for \SI{100}{TeV} searches, 
jet vetoes at relatively low values in monojet searches can cut out a 
lot of interesting phase space, particularly in the high-$p_T$ regions 
due to the typically large amount of QCD radiation of the \SI{100}{TeV} 
environment.

A \SI{100}{TeV} collider will provide several new challenges.  It will have 
an overwhelming amount of QCD radiation, and this paper has shown that although
some techniques used at the LHC can be used to study a \SI{100}{TeV} 
environment, others will need to be adapted or improved.  Loop-induced 
production channels will become far more significant, and will deserve 
theoretical work to improve our understanding to match the control 
requirements necessary. Moving to such a high energy comes with a 
large amount of increased complexity and challenges, but also a very 
large potential for the discovery of New Physics and tests of the 
current SM.

\smallskip

\begin{acknowledgments}
The authors are grateful to their coworkers in the \Sherpa collaboration.
We acknowledge financial support by the European Commission through the 
networks MCnetITN (PITN--GA--2012--315877) and HiggsTools 
(PITN--GA--2012--316704). EB and SS acknowledge financial support from 
BMBF under contract 05H15MGCAA.
\end{acknowledgments}

\clearpage
\bibliographystyle{amsunsrt_modp}
\bibliography{journal}

\end{document}